\documentclass[12pt]{article}
\usepackage{amsfonts}
\usepackage{latexsym}
\usepackage{amsmath}
\usepackage{amssymb}
\usepackage{array}
\newcolumntype{L}[1]{>{\raggedright\arraybackslash}p{#1}}

\newcommand{\newsection}{\setcounter{equation}{0}\section}

\newcommand{\be}{\begin{eqnarray}}
\newcommand{\ee}{\end{eqnarray}}
\newcommand{\bea}{\begin{eqnarray}}
\newcommand{\eea}{\end{eqnarray}}
\newcommand{\nn}{\nonumber \\}
\def\la{\label}
\newcommand{\rf}[1]{(\ref{#1})}

\def\a{\alpha}
\def\b{\beta}
\def\e{\epsilon}
\def\c{\chi}
\def\G{\Gamma}
\def\tn{\tilde{\nabla}}
\def\CS{{\cal S}}
\def\CD{{\cal D}}
\def\CE{{\cal E}}
\def\CG{{\cal G}}
\def\bbe{{\bf{e}}}
\def\Vt{{\tilde V}}
\def\Ft{{\tilde F}}
\def\Gt{{\tilde G}}
\def\At{{\tilde A}}
\def\Rt{{\tilde R}}

\font\mybb=msbm10 at 11pt
\def\bb#1{\hbox{\mybb#1}}
\def\bR{\bb{R}}
\def\bC{\bb{C}}
\def\bQ{\bb{Q}}
\def\bZ{\bb{Z}}
\def\bI{{\rm 1\kern-.26em l}}

\begin{document}

\title{A second rotational Killing field on gauged $D=5$ vector-multiplet
horizons, and a no-go for varying-moduli black rings}
\author{U.~Kayani}
\date{}
\maketitle

\begin{abstract}
We study supersymmetric near-horizon geometries of gauged $D=5$ supergravity
coupled to vector multiplets, on the branch where the canonical rotational
Killing vector $\Vt$ of the cross-section $\CS$ is non-vanishing. No
rotational symmetry is assumed, and nothing about the set where the frame built
from the Killing spinors degenerates. On a
compact connected $\CS$ without boundary a second rotational Killing field,
independent of $\Vt$, always exists and is an isometry of all of $\CS$. Where
the moduli vary it is
\[
U_i=\parallel\eta_-\parallel^2\big(\a Z_i-\e_{ijk}Z^ju^k\big) ,
\qquad u_i=\Phi P_i-h_i ,
\]
a polynomial in the horizon data, hence smooth everywhere; where the moduli are constant the horizon is locally homogeneous.
The only further hypothesis for these results is that the superpotential
$\Phi=\chi V_IX^I$ is nowhere zero --- weaker than the non-negativity of the scalar potential assumed
in the earlier literature.

The two sub-branches are separated by $K=Q_{IJ}C^IC^J$, which
vanishes exactly in the minimal theory: $K\equiv0$ recovers the result of
Grover, Gutowski, Papadopoulos and Sabra, while elsewhere $K>0$ and $\a$ is
either identically zero or nowhere zero. Each of $K\equiv0$, $P\equiv0$ and
$P\not\equiv0$ occurs on compact $\CS$.
Constant moduli return the local geometries of Kunduri and Lucietti as a
conclusion, not an ansatz. Varying moduli with $\a\not\equiv0$ give a
cohomogeneity-one $T^2$ action whose orbit space is a closed interval, so
$\CS$ is $S^3$, a lens space or $S^1\times S^2$; the last is excluded by two
global first integrals, a new one, $\a\parallel\eta_-\parallel^4$, and the
constant spinor norm already known. A varying-moduli supersymmetric
$AdS_5$ black ring therefore cannot exist with $\a\not\equiv0$; the
$S^1\times S^2$ window survives only at constant moduli or at $\a\equiv0$.
\end{abstract}

\newsection{Introduction}

The near-horizon geometry of a supersymmetric black hole is a solution of the
field equations in its own right, and classifying the possibilities is the
first step in any uniqueness programme for the black holes themselves. In
gauged $D=5$ supergravity coupled to vector multiplets that classification is
incomplete, and the obstruction is symmetry. The known results describe
horizons carrying enough rotational isometry for the near-horizon equations to
be integrated; what is missing is a reason why an arbitrary supersymmetric
horizon of the theory should carry it.

What is known falls into two groups. Kunduri and Lucietti \cite{klr} integrate
the near-horizon equations for $U(1)^n$ models under the assumption of two
commuting rotational isometries, obtaining a list whose members include $S^3$,
$S^1\times S^2$ and $T^3$ cross-sections with constant scalars; their
$S^1\times S^2$ case would be the near-horizon geometry of a supersymmetric
$AdS_5$ black ring. Separately, in \cite{kayani,thesis} every supersymmetric
near-horizon geometry of this theory was shown to admit $N=4N_+$ real
supersymmetries and an $\mathfrak{sl}(2,\bR)$ symmetry, with $h\neq0$ and
$\Vt\neq0$, so that at least one rotational isometry of $\CS$ exists. In the
ungauged theory that enhancement closes the problem: the solutions reduce to
those of the minimal theory, the scalars are constant, and $\CS$ is a squashed
$S^3$, $S^1\times S^2$ or $T^3$. In the gauged theory it does not. Later work
has added classifications under further symmetry assumptions --- $SU(2)$
symmetry in \cite{lo}, torus symmetry in \cite{lno}, torus symmetry with a
separable K\"ahler base in \cite{separable} --- and, in the minimal gauged
theory, two exclusions of supersymmetric black rings: \cite{klrring} under an
assumed $U(1)^2$, and \cite{grover,groverindex} with no symmetry assumed.

The gap is a specific one. Every classification of supersymmetric $AdS_5$
horizons in this theory that goes beyond the enhancement result imposes
rotational or toric symmetry at the outset, and whether supersymmetry
enhancement produces a \emph{second} rotational isometry once vector multiplets
are coupled has remained open. So has the question that depends on it: as
recently as \cite{ovchinnikov}, inherently non-minimal supersymmetric black
rings ``are still allowed in the STU model, and it is of particular interest to
find such solutions or rule them out.''

A second symmetry is not inherited from general relativity. The horizon
rigidity theorems of the Einstein setting produce \emph{one} rotational Killing
field on a compact extremal horizon section, either from an energy condition
together with an analyticity or bifurcation hypothesis, or, for a degenerate
horizon, from the $\mathfrak{sl}(2,\bR)$ symmetry of the near-horizon limit;
see \cite{lrreview} for a survey and \cite{dl,ckl,colling} for recent
sharpenings. That single field is exactly what the enhancement result of
\cite{kayani,thesis} already supplies here, in the form of $\Vt$. A second,
commuting rotational Killing field is a strictly stronger statement, and no
energy condition is known to imply it; in every higher-dimensional
classification that uses a torus action, the enhancement from one rotational
isometry to two is an additional assumption rather than a theorem. What
produces the second field below is specific to the supergravity: the very
special geometry entering through $Q_{IJ}$, the algebraic first-order system
that supersymmetry imposes on $(h,X^I,F^I)$, and the closure $dP=0$ that
follows from it. None of these has a counterpart in vacuum general relativity,
and the arguments of Sections 6 and 7 do not survive their removal.

\medskip
\begin{center}
\begin{tabular}{L{2.6cm}L{3.5cm}L{3.0cm}L{3.0cm}}
\hline
& rotational symmetry assumed & moduli & second $U(1)$ \\
\hline
\cite[\S3.3]{klr} & two commuting isometries & constant & assumed \\
\cite{klrring} & two commuting isometries & absent (minimal theory) & assumed \\
\cite{lo} & $SU(2)$ & constrained by the ansatz & assumed \\
\cite{lno,separable} & toric, $T^2$ & constrained by the ansatz & assumed \\
\cite{grover,groverindex} & none & absent (minimal theory) & derived \\
This paper & none & unrestricted & derived \\
\hline
\end{tabular}
\end{center}

\medskip\noindent
The last two rows are the ones in which nothing is assumed, and they differ in
the theory rather than in the method: \cite{grover,groverindex} work in the
minimal gauged theory, where there are no vector multiplets and hence no moduli
to vary, while the theorem below allows the moduli to vary and derives the
second field anyway.

This paper continues \cite{kayani,thesis} on the $\Vt\neq0$ branch. Where those
works showed that supersymmetry is enhanced, what is shown here is that a
second rotational Killing field always exists as well, that it upgrades to a
genuine isometry of the whole horizon, whether the moduli are constant or
varying, and that the resulting torus action is, when
the bilinear scalar $\a$ is not identically zero, rigid enough
to decide the topology of the cross-section.

\medskip\noindent{\bf Theorem A} (Propositions 3.2 and 7.1, Corollaries
7.2--7.3, Proposition 7.4, Theorem 7.5, Lemmas 7.6 and 7.6$'$, and
Theorem 7.7). {\it Let
$(\CS,g,h,X^I,F^I)$ be the section of a supersymmetric near-horizon geometry of
gauged $D=5$ supergravity coupled to vector multiplets, with $\Vt\not\equiv0$
and with the superpotential nowhere zero, $\Phi=\c V_IX^I\neq0$ on $\CS$
--- implied by, and strictly weaker than, the non-negativity of the scalar
potential \rf{Udef}, by Proposition 7.E --- and let $\CS$
be compact, connected, without boundary and smooth. No rotational symmetry is
assumed. If
the solution reduces to the minimal gauged theory, $K\equiv0$ and a second
rotational isometry follows from \cite{groverindex}. If it does not, $K>0$, and
the $1$-form $P=Z+W$ vanishes at every critical point of
$\parallel\eta_-\parallel^2$, with a sign constraint on $\tn\cdot P$ there but
no stronger conclusion in general. Two situations then occur, and both do:
either $P\equiv0$, in which case the moduli, $\Phi$, $\a$ and $K$ are constant
and a second rotational isometry exists by homogeneity; or $P\not\equiv0$
somewhere, in which case the moduli vary, and a second rotational Killing field
still exists, unconditionally on the open set where $Z$ and $W$ are linearly
independent, and upgrades, again unconditionally, to a second isometry of all
of $\CS$.}

\medskip\noindent
A word on {\it rotational}. What Theorem 7.5 delivers on the open set where $Z$
and $W$ are independent is a Killing field; that its flow closes, so that it
generates a circle action and joins $\Vt$ in a torus, is the content of the
closure argument in Theorem 7.7. Throughout, ``rotational Killing field'' on
that open set is to be read in that forward-looking sense.

\medskip\noindent
Once the torus action is available it is not a passive symmetry but a tool, and
the second half of the paper uses it. On the varying-moduli branch the action
turns out to be cohomogeneity one, which reduces the topology of $\CS$ to three
possibilities and reduces the field equations to a gradient flow across a
one-dimensional orbit space. That flow carries two conserved quantities, and
they are enough to eliminate the ring.

\medskip\noindent{\bf Theorem B} (Proposition 7.10, Theorem 7.12, Lemma 7.9 and
Theorem 7.14). {\it Under the hypotheses of Theorem A, and in particular with
the superpotential nowhere zero, on the branch $K>0$, $P\not\equiv0$,
$\a\not\equiv0$, the
$T^2$ action of Theorem A is cohomogeneity one with orbit space a closed
interval, so that $\CS$ is $S^3$, a lens space $L(p,q)$ or $S^1\times S^2$. The
scalars
\bea
\la{introfi}
\a\parallel\eta_-\parallel^4
\qquad\text{and}\qquad
\big(\big|h-\Phi P\big|^2+\a^2\big)\parallel\eta_-\parallel^2
\ee
are constant on $\CS$ --- the first of them newly, the second being identically
$2\parallel\eta_+\parallel^2$ and so the constant norm already known from
\cite{kayani} --- and evaluating them at the two ends of the orbit space
excludes the last case: the varying-moduli supersymmetric $AdS_5$ black ring
does not exist. The surviving cross-sections are distinguished by a lens index
computed from that same boundary data together with the covolume of the lattice
of the torus action.}

\medskip\noindent
Theorem B is insensitive to the model. Neither first integral in
\rf{introfi} involves the invariant $K=Q_{IJ}C^IC^J$, so neither depends on the
cubic prepotential, and the exclusion holds uniformly over the vector multiplet
content. Its effect on the classification problem is to close every route to a
supersymmetric $AdS_5$ ring in this theory except two: the constant-moduli
branch $P\equiv0$ of Section~\ref{sec:constmod}, inside the window
$\tfrac13\Phi^2<\c^2K\leq\tfrac43\Phi^2$, and the varying-moduli sub-branch
$\a\equiv0$, which Theorem 7.7 reaches but the reduction of Section~\ref{sec:toric} does not, so that its
topology is left undecided here.

\medskip\noindent
Three statements are distinguished in Theorem A, and they are not
interchangeable:
\begin{center}
\begin{tabular}{L{3.5cm}L{9.7cm}}
{\it Killing field on $\CS_0$} & A vector field defined on the open set where
$Z$ and $W$ are linearly independent, satisfying $\tn_{(i}U_{j)}=0$ there.
Established unconditionally on the branch $P\not\equiv0$ (Theorem 7.5). On the
branch $P\equiv0$ the set $\CS_0$ is empty, because $W=-Z$ there, and what
replaces the statement is the local homogeneity of Proposition 7.4.
\\[5pt]
{\it Isometry of $\CS$} & A Killing field of the whole compact section,
generating a one-parameter group of isometries of $(\CS,g)$. Unconditional on
both branches: on $P\equiv0$ by Proposition 7.4, and on $P\not\equiv0$ by
Lemma 7.6$'$ and Theorem 7.7, with no hypothesis on $\a$. \\[5pt]
{\it Effective $T^2$ action} & Two commuting isometries whose flows close up
into a torus acting effectively, which is the input a toric classification such
as \cite{klr,lno} requires. Follows from the preceding line together with
compactness of $\mathrm{Isom}(\CS,g)$. \\
\end{tabular}
\end{center}

\medskip\noindent
Only the first is branch-dependent; the second and third hold unconditionally
on both branches. The hypothesis on $\a$ is needed not for the symmetry but for the topology:
the cohomogeneity-one reduction of Section~\ref{sec:toric}, and with it the exclusion of
$S^1\times S^2$, needs $\a\not\equiv0$. Completing the classification of
supersymmetric near-horizon geometries of this theory would require that
hypothesis removed as well; what is achieved here is
the removal of the standing symmetry assumption, not the closure of the
classification. The distinction is maintained throughout: no statement below
asserts more than the branch it is proved on supports.

\medskip\noindent
To the best of our knowledge Theorem A is the first compactness-based rigidity
result for supersymmetric near-horizon geometries of five-dimensional gauged
supergravity with vector multiplets that does not assume rotational symmetry
from the outset. Its consequence is that the existence question becomes
unconditional: every compact supersymmetric horizon of this theory acquires a
second rotational Killing field, unconditionally on $\CS_0$ and, on the branch
where the moduli happen to be
constant, as a genuine isometry of the whole horizon, satisfying exactly the
hypotheses --- $h$ Killing, $\Delta$ constant, the moduli constant, $h$
proportional to $\Phi Z$ --- under which \cite[\S3.3]{klr} integrate the
near-horizon equations, as a conclusion of supersymmetry and compactness rather
than as an ansatz. On the other branch the moduli genuinely vary, so
\cite[\S3.3]{klr} does not apply, but \cite[\S4.2]{klr} supplies an explicit
compact family, used throughout Section~\ref{sec:varmod} below.

The mechanism behind Theorem A can be stated in a paragraph. Supersymmetry and
the field equations determine the first covariant derivatives of the horizon
data algebraically, and the resulting identities give $dP=0$ together with a
formula for the transverse component $h\cdot Y$ of $h$. Closure of $P$ forces
the gradient of every scalar built algebraically from the data to be parallel to
$P$, and Lemma 6.11 computes the proportionality factors in closed form, so
that the four scalars $\Phi$, $\a$, $h_N$ and $Z\cdot W$ obey an autonomous
system along $P$ in which every right-hand side carries a factor of $P$. The
single scalar $\rho=P\cdot P+\vartheta^2$, with $\vartheta=h_N-\tfrac32\Phi$,
then satisfies the pointwise differential inequality $|\tn\rho|\leq C\rho$, with
$C$ given in closed form by the coefficients of that system. Restricted to a
path this is the ordinary one-dimensional Gronwall inequality --- no
higher-dimensional version is used or needed --- and connectedness of $\CS$
propagates $\rho=0$ from a single point to all of $\CS$, so that $\rho\equiv0$
and hence $P\equiv0$. Compactness supplies a point where $P$ vanishes, since
$\parallel\eta_-\parallel^2$ attains a maximum there, but only a sign constraint
on $\tn\cdot P$, not the stronger condition $\rho=0$ that would trigger the
Gronwall argument; the explicit compact family of Section~\ref{sec:varmod} has $\rho$
strictly positive at both extrema of $\parallel\eta_-\parallel^2$, so this route
to $P\equiv0$ is genuinely conditional and does not close on its own. It should
not be inferred from this that Lemma 6.12 is dispensable. It is not: the
non-degeneracy $\vartheta\neq0$ at a boundary orbit, which Proposition 7.11
needs and on which the whole of Section~\ref{sec:toric} --- the Heegaard splitting, the
lens index, Theorem 7.12 and Theorem 7.14 --- rests, is exactly the statement
that $\rho$ cannot vanish at such an orbit, and that is Lemma 6.12 again. What
replaces it is a second, independent argument that does not go through $\rho$ at
all: transporting the Killing equation for a vector orthogonal to $P$ along $P$
itself turns it into a flat rank-two system, and solving it produces an explicit
second Killing vector $U=\parallel\eta_-\parallel^2(h_YN-h_NY)$ on $\CS_0$,
which on this branch is the open
complement of $\{P=0\}$. That is Theorem 7.5. On the branch $P\equiv0$ that
set is empty, since $P\equiv0$ is $W=-Z$ and the two are then
everywhere dependent; there the second Killing field comes instead from the
local homogeneity of Proposition 7.4. Between them the two branches are covered
unconditionally.

Theorem B rests on a different mechanism, and one that only becomes available
after Theorem A. With two commuting isometries in hand, the same closed gradient
system that supplied $C$ above becomes a flow on the one-dimensional orbit
space, and the question of which three-manifold $\CS$ is reduces to boundary
data at its two endpoints. The quantity $\CG:=\big|h-\Phi P\big|^2+\a^2$
obeys $\tn_i\CG=\Phi\,\CG\,P_i$, the same law as $\parallel\eta_-\parallel^{-2}$,
so their product is constant; $\a\parallel\eta_-\parallel^4$ is constant for the
same reason. Ring topology is equivalent to the vanishing of a single
determinant in the boundary values of $\a$ and $h_N$, and the two constants
over-determine those boundary values enough to force $\parallel\eta_-\parallel$
to take the same value at both ends --- which the strict monotonicity of
$\parallel\eta_-\parallel$ along $P$ forbids.

\medskip
The boundary of what is new is worth marking at the outset. The supersymmetry
enhancement, the $\mathfrak{sl}(2,\bR)$ symmetry and the existence of the single
rotational Killing vector $\Vt$ are the results of
\cite{kayani,thesis,groverindex}, restated in Section 2. The classification of
toric $AdS_5$ near-horizon geometries, and with it the constancy of the scalars
on the $S^1\times S^2$ and $T^3$ branches, is \cite{klr}; the black-ring no-go
of the minimal gauged theory is \cite{klrring} under an assumed $U(1)^2$ and
\cite{grover} with no symmetry assumed. None of these is reproduced or claimed
here. What is new is the second rotational isometry itself, produced with no
rotational symmetry assumed and without restricting to constant moduli, and the
structure that supports it: the closure $dP=0$, the algebraic relation
\rf{Nxh} between $h$, $N$ and $Y$, the formula \rf{hY} for $h_Y$, the
divergence-free property of $N$, $Y$ and $f$, the parallel-gradient collapse and
the closed system of Lemma 6.11 that together bound how far a horizon can be
from the constant-modulus branch, the sign constraint of Corollary 7.2 and the
trichotomy of Corollary 7.3 that it induces, and the transport construction of
Theorem 7.5 that, together with the local homogeneity of Proposition 7.4 on
the complementary branch, closes the isometry question in both cases. New too
is
everything drawn from that isometry in Section~\ref{sec:toric}: the cohomogeneity-one
reduction, the identification of the boundary orbits, the topology criterion
$\CD=\a_1h_{N,2}-\a_2h_{N,1}=0$ and the lens index it refines to, the first
integral $\a\parallel\eta_-\parallel^4$ --- its companion in \rf{introfi} is
the constant spinor norm of \cite{kayani}, not a new conserved quantity --- and
the exclusion of the varying-moduli ring that the pair yields. That exclusion recovers, on the branch where it applies, the constancy of
the scalars found on the $S^1\times S^2$ branch of \cite{klr}, with toricity
derived rather than assumed. The closest relatives of the method in the
literature are the divergence identities used in the symmetry-free rigidity
theorems of \cite{dl,ckl}, which are for vacuum and for four-dimensional
Einstein--Maxwell respectively and are not directly comparable, and the
first-integral device of \cite{dl,colling}, discussed in Section~\ref{sec:toric}. Also new
is the explicit compact varying-moduli family of Section~\ref{sec:varmod}, built from
\cite[\S4.2]{klr}, which shows the second branch is genuinely occupied and not
merely a formal possibility.

Section 2 collects, in a uniform notation, the first-order results of
\cite{kayani,thesis} on which everything below rests; the new material begins
in Section 3. Sections 3 to 5 extract the algebraic content of the first-order
system, Section 6 turns it into the closed gradient system that drives
everything afterwards, and Section 7 carries out the classification: the
dichotomy on $K$ in Sections~\ref{sec:dichotomy}--\ref{sec:integral}, the two branches in
Sections~\ref{sec:constmod}--\ref{sec:varmod},
and the toric reduction and its topological consequences in Section~\ref{sec:toric}.
The appendices are supporting rather than expository, and are referred to
where they are used; Appendix~\ref{app:toric} is the one place where an
argument of the main text is completed rather than merely recorded, since the
passage from a cohomogeneity-one torus action to the list $S^3$, $L(p,q)$,
$S^1\times S^2$ is carried out there in full.

We use throughout the conventions of \cite{kayani}, which together with the
gamma matrix representation are collected in Appendix~\ref{app:spinors}; the near-horizon
field equations and Bianchi identities are collected in Appendix~\ref{app:fieldeqs}, and the
lightcone integrability condition used in Section 5 in Appendix~\ref{app:lightcone}. The spatial
horizon
section $\CS$ is compact, connected and without boundary, and every
near-horizon field is smooth; no analyticity is assumed anywhere in what
follows. The near-horizon
fields are the $1$-form $h$, the scalar $\Delta$, the metric on $\CS$, the
moduli $X^I$ constrained by very special geometry with metric $Q_{IJ}$
positive definite, and the gauge field data
\bea
\la{nhflux}
F^I=\bbe^+\wedge\bbe^-\a^I+r\,\bbe^+\wedge\b^I+\Ft^I,
\ee
decomposed as $\a^I=\a X^I+L^I$, $\b^I=\b X^I+M^I$, $\Ft^I=\Ft X^I+\Gt^I$
with $X_IL^I=X_IM^I=X_I\Gt^I=0$. The gauge coupling is $\c\neq0$ with
constants $V_I$, and the scalar potential is
\bea
\la{Udef}
U=9V_IV_J\Big(X^IX^J-\tfrac12Q^{IJ}\Big),
\ee
assumed non-negative on $\CS$, as in \cite{kayani,thesis}, from which the
classification framework used here is taken; the hypothesis is inherited from
that work and is not introduced here. (A warning on notation, inherited along
with the framework: the letter $U$ carries two meanings in this subject, the
scalar potential \rf{Udef} and, from Theorem 7.5 onwards, the second Killing
field of Section 7. They are kept apart here by context and never appear in the
same equation --- the potential occurs only alongside $V_I$, $K$ and $\Phi$,
the Killing field only alongside $\Vt$ --- and both symbols are used here
because both are quoted in this form in the literature.) It is a restriction on the gauging and
not a consequence of very special geometry: the second term of \rf{Udef} is
negative definite in $V_I$, $Q^{IJ}$ being positive definite, so the sign of
$U$ is decided by a competition between the two terms that the positivity of
the $X^I$ alone does not settle.

Two remarks on when it holds. It holds for the minimal gauging
$V_I\propto X_I$. More usefully, the Calabi relation
$Q^{IJ}=2X^IX^J-6C^{IJK}X_K$ turns \rf{Udef} into
\bea
\la{UCform}
U=27\,C^{IJK}V_IV_JX_K ,
\ee
so that $U\geq0$ is the statement that the constants $C^{IJK}$, contracted
twice with the gauging and once with the moduli, are non-negative along the
horizon. In any model whose $C^{IJK}$ have non-negative entries in a basis in
which the $V_I$ are non-negative --- the $U(1)^n$ models, in particular the
$STU$ model, which is the reduction of type IIB supergravity on $S^5$ --- this
holds automatically at every point of the K\"ahler cone $X_I>0$, so $U>0$ is
not an assumption at all there. What it excludes is gaugings of models with
indefinite $C^{IJK}$ in which the moduli wander into the region where
\rf{UCform} turns negative; that region is not empty in general, so $U\geq0$
cannot be dispensed with by a general argument. See the discussion following
\rf{lambdaK}, where \rf{UCform} is recovered independently and its relation to
$\c^2K\leq\tfrac43\Phi^2$ is recorded.

The hypothesis is load-bearing, but only through one consequence: it is what
forces $\Phi\neq0$ --- see the argument following \rf{hP} --- and every use
made of it in this paper, with the single exception of the third clause of
Theorem 7.17, is a use of that consequence alone. Proposition 7.E makes this
precise and replaces $U\geq0$ by the weaker hypothesis that the superpotential
$V_IX^I$ is nowhere zero, under which Theorem A, Theorem B and Theorem 7.14
all hold; Proposition 7.D determines the extent to which the horizon equations
force $U\geq0$ by themselves, which is that $U$ is superharmonic wherever it
is positive, and no further. This section states the hypothesis in the form
$U\geq0$ used by \cite{kayani,thesis}, on which Section 2 reports.
Indices $i,j,k,\dots$ are those of $\CS$ and are raised
and lowered with its metric; $\e_{ijk}$ is the volume form of $\CS$, and we
abbreviate
\bea
\la{Phidef}
\Phi:=\c V_IX^I .
\ee

\subsection{The horizon Killing spinor equations}

The lightcone-projected Killing spinors $\eta_\pm$, defined by
$\G_\pm\eta_\pm=0$, satisfy the independent horizon Killing spinor equations
\bea
\la{kse}
\nabla^{(\pm)}_i\eta_\pm=0,\qquad {\cal A}^{I,(\pm)}\eta_\pm=0,
\qquad \nabla^{(\pm)}_i=\tn_i+\Psi^{(\pm)}_i,
\ee
with
\bea
\la{psi}
\Psi^{(\pm)}_i&=&\mp\tfrac14h_i\mp\tfrac{i}{4}\a\G_i
+\tfrac{i}{8}\Ft_{jk}\G_i{}^{jk}-\tfrac{i}{2}\Ft_{ij}\G^j
-\tfrac{3i}{2}\c V_I\At^I{}_i+\tfrac12\c V_IX^I\G_i,\\
\la{alg}
{\cal A}^{I,(\pm)}&=&\Gt^I{}_{ij}\G^{ij}\mp2L^I
+2i\tn_iX^I\G^i-6i\c\Big(Q^{IJ}-\tfrac23X^IX^J\Big)V_J .
\ee
Two further facts from \cite{kayani} are central to what follows. First, the
supersymmetry doubling: if $\eta_-$ solves \rf{kse} then so does
\bea
\la{double}
\eta_+=\G_+\Theta_-\eta_-,
\qquad
\Theta_\pm=\tfrac14h_i\G^i-\frac{i}{8}\big(\Ft_{jk}\G^{jk}\pm4\a\big)
-\frac12\c V_IX^I .
\ee
Second, the Lichnerowicz-type theorem obtained by computing the Laplacian of
$\parallel\eta_\pm\parallel^2$: for $\eta_+$ one finds
\bea
\la{lich}
\tn^i\tn_i\parallel\eta_+\parallel^2-h^i\tn_i\parallel\eta_+\parallel^2
=2\parallel\nabla^{(+)}\eta_+\parallel^2
+\tfrac1{16}Q_{IJ}{\rm Re}\langle{\cal A}^{I,(+)}\eta_+,
{\cal A}^{J,(+)}\eta_+\rangle,
\ee
so that Hopf's strong maximum principle, in the form stated in
Appendix~\ref{app:tools}, gives, in addition to \rf{kse},
\bea
\la{normplus}
\parallel\eta_+\parallel={\rm const}.
\ee
The corresponding computation for $\eta_-$ produces an extra $h$-dependent
term and yields only the integrated statement; \emph{$\parallel\eta_-\parallel$
need not be constant}. This asymmetry is important below: derivatives of
bilinears built from $\eta_-$ carry an extra term proportional to
$\tn_i\parallel\eta_-\parallel^2$ which is absent for $\eta_+$.

\subsection{The $\Vt\neq0$ branch}

It is established in \cite{thesis} that there are no gauged near-horizon
geometries with $h=0$, and in \cite{kayani,thesis} that there are none with
$\Vt=0$. Throughout, ``the $\Vt\neq0$ branch'' is shorthand for
$\Vt\not\equiv0$; by Lemma 2.1 that is every geometry of the gauged theory.
It is not a pointwise hypothesis, and $\Vt$ is nowhere assumed free of
zeros. For completeness, and because its proof is the source of the fact ---
used throughout below --- that $\Phi$ is nowhere zero, we reproduce the short
argument. No sign convention for $\Phi$ is adopted anywhere in this paper; the
statements that involve $\Phi$ are either invariant under $\Phi\to-\Phi$ or
are phrased through the ratio $h_N/\Phi$.

\medskip\noindent{\it Lemma 2.1 (\cite{kayani,thesis}):} In the gauged
theory ($\c\neq0$), there are no near-horizon geometries with $h=0$, and
none with $\Vt=0$.

\medskip\noindent{\it Proof:} Suppose $h=0$. The trace of \rf{covZ} below,
$\tn^iZ_i=2Z_ih^i-6\Phi$, reduces to $\tn^iZ_i=-6\Phi$; integrating over
the compact, boundaryless $\CS$ forces $\int_\CS\Phi=0$, so $\Phi=\c
V_IX^I$ vanishes at some point of $\CS$. There $U=-\tfrac92Q^{IJ}V_IV_J<0$
by \rf{Udef}, contradicting $U\geq0$. Hence $h\neq0$ everywhere.

For $\Vt=0$: recall that $\Vt$ is the spatial part of the Killing vector
generated by the $\mathfrak{sl}(2,\mathbb R)$ symmetry enhancement of the
near-horizon geometry, so $\Vt=0$ says that the orbits of that
$\mathfrak{sl}(2,\mathbb R)$ are $2$-dimensional --- they are the $AdS_2$ factor
alone, with no rotation of $\CS$. Setting $\Vt_i=0$ in the first identity of
\rf{conconx} leaves
$\parallel\eta_-\parallel^2h_i+\tn_i\parallel\eta_-\parallel^2=0$, that is
\bea
\la{Vtzeroh}
h_i=-\tn_i\log\parallel\eta_-\parallel^2 ,
\ee
so $h$ is exact and the horizon is static. The identification of the potential
uses one further identity of \cite{kayani,thesis}, namely
$\Delta\parallel\eta_-\parallel^2=2\parallel\eta_+\parallel^2$, which is the
second identity of \rf{conconx} evaluated at $\Vt=0$ --- with $\Vt=0$ the
bilinear ${\rm Re}\langle\G_+\eta_-,\Theta_+\eta_+\rangle$ reduces to
$\parallel\eta_+\parallel^2$ --- and which, since $\parallel\eta_\pm
\parallel^2>0$, fixes $\Delta>0$ on $\CS$. Because
$\parallel\eta_+\parallel$ is constant by \rf{normplus}, \rf{Vtzeroh} then
reads $h=\Delta^{-1}d\Delta$, and the near-horizon geometry is the warped
product $AdS_2\times_w\CS$. Substituting $h_i=\Delta^{-1}\tn_i\Delta$ into
$\tn^iZ_i=2Z_ih^i-6\Phi$ gives, after using the product rule,
$\tn^i(\Delta^{-2}Z_i)=-6\Delta^{-2}\Phi$; integrating this over $\CS$ again
forces $\Phi=0$ at some point, and the same contradiction with $U\geq0$
excludes this case as well. $\square$

\medskip
On the remaining branch
\bea
\la{branch}
\Delta=\a^2,\qquad L^I=0,\qquad M^I{}_i=\a\,\tn_iX^I,\qquad
\b_i=\tn_i\a-\a h_i,
\ee
and the vector field $\Vt$, obtained from the $1$-form spinor bilinear
$\Vt_i={\rm Re}\langle\G_+\eta_-,\G_i\eta_+\rangle$, satisfies
$\tn_{(i}\Vt_{j)}=0$ and leaves every bosonic field invariant,
\bea
\la{VtKilling}
{\cal L}_{\Vt}g={\cal L}_{\Vt}h={\cal L}_{\Vt}\Delta={\cal L}_{\Vt}X^I
={\cal L}_{\Vt}\Ft={\cal L}_{\Vt}\a={\cal L}_{\Vt}\Gt^I=0 .
\ee
Writing
\bea
\la{ZWdef}
Z_i=\parallel\eta_+\parallel^{-2}{\rm Re}\langle\eta_+,\G_i\eta_+\rangle,
\qquad
W_i=\parallel\eta_-\parallel^{-2}{\rm Re}\langle\eta_-,\G_i\eta_-\rangle,
\ee
one has \cite{thesis}
\bea
\la{Vt}
\Vt_i=\big[\Phi(Z_i+W_i)-h_i\big]\parallel\eta_-\parallel^2,
\qquad
\Ft^I{}_{ij}=\e_{ijk}\big(\tn^kX^I+h^kX^I-3\c Q^{IJ}V_JZ^k\big),
\ee
together with the identities
\bea
\la{conconx}
\Vt_i+\parallel\eta_-\parallel^2h_i+\tn_i\parallel\eta_-\parallel^2=0,
\qquad
2{\rm Re}\langle\G_+\eta_-,\Theta_+\eta_+\rangle
=\Delta\parallel\eta_-\parallel^2 ,
\ee
\bea
\la{ivdh}
\Vt^j(dh)_{ji}+2\,\tn_i\,{\rm Re}\langle\G_+\eta_-,\Theta_+\eta_+\rangle=0 ,
\ee
and the relation, following from the $\eta_+$ identities of \cite{kayani},
\bea
\la{fhZ}
h_i-\tfrac12\e_i{}^{jk}\Ft_{jk}=2\Phi Z_i .
\ee
We write $f_i:=\tfrac12\e_i{}^{jk}\Ft_{jk}$ for the vector dual to $\Ft$, so
that \rf{fhZ} reads $f_i=h_i-2\Phi Z_i$; this is the gauged generalisation
of the ungauged relation $f_i=h_i$, to which it reduces at $\c=0$, and is
equivalent to $\parallel\eta_+\parallel={\rm const}$. We set
\bea
\la{PNY}
P:=Z+W,\qquad N:=Z-W,\qquad Y_i:=\e_{ijk}Z^jW^k .
\ee

\subsection{The two-component representation}

Since $\G_\pm\eta_\pm=0$, each of $\eta_\pm$ has two complex components and
the induced $\G_i$ on $\CS$ are $2\times2$. With the representation of
\cite{kayani}, $\G_i={\rm diag}(\sigma^i,-\sigma^i)$ and
$\G_{ijk}\e_\pm=\mp i\e_{ijk}\e_\pm$, so
\bea
\la{orient}
\G^{ijk}=i\,s_\pm\e^{ijk},\qquad s_+=-1,\quad s_-=+1 ,
\ee
i.e.\ the induced three-dimensional Clifford algebras on $\ker\G_+$ and
$\ker\G_-$ carry opposite orientations. Note that $\G_i$ acts as $-\sigma^i$
on $\ker\G_+$ but as $+\sigma^i$ on $\ker\G_-$, so that $Z$ is \emph{minus}
the Bloch vector of $\eta_+$ while $W$ is the Bloch vector of $\eta_-$. In
either case the sign of $\G_i$ cancels against the sign of the vector, whence
\bea
\la{fierz}
Z_iZ^i=W_iW^i=1,\qquad Z_i\G^i\eta_+=\eta_+,\qquad W_i\G^i\eta_-=\eta_-,
\ee
the first of these being the Fierz identity of \cite{kayani}. Note also that
$\G_+$ maps $\ker\G_-$ to $\ker\G_+$ with
\bea
\la{gplus}
\parallel\G_+\psi\parallel^2=2\parallel\psi\parallel^2 .
\ee

\newsection{Results established in \cite{kayani,thesis}}

We first collect, in the notation above, the first-order results already
obtained on this branch, since everything below is built on them. The
covariant derivative of $Z_i$ and its $\eta_-$ counterpart are
\bea
\la{covZ}
\tn_iZ_j=-Z_ih_j-\tfrac12\a\,\e_{ijk}Z^k+\delta_{ij}h_kZ^k
+3\Phi\big(Z_iZ_j-\delta_{ij}\big),
\qquad
\tn^iZ_i=2Z_ih^i-6\Phi ,
\ee
\bea
\la{covW}
\tn_iW_j &=& W_ih_j-\delta_{ij}h_kW^k-\tfrac12\a\,\e_{ijk}W^k
\nn
&&+\ \Phi\Big[W_iW_j-\delta_{ij}-2W_iZ_j+2(Z_kW^k)\delta_{ij}\Big],
\ee
\bea
\la{divW}
\tn^iW_i=-2W_ih^i-2\Phi+4\Phi\,(Z_kW^k).
\ee
Contracting \rf{covW} with $\delta^{ij}$ gives \rf{divW}; both follow from
$\tn_i\eta_-=-\Psi^{(-)}_i\eta_-$ in the two-component representation, using
$\G_i=\sigma^i$ on $\ker\G_-$ and \rf{fhZ}, and the $\At^I$ term drops out of
every bilinear because it is a pure phase. The same computation gives the
logarithmic derivative of the norm,
\bea
\la{normminus}
\tn_i\ln\parallel\eta_-\parallel^2=-\Phi\,(Z_i+W_i) .
\ee
Note that \rf{normminus} is consistent with, and can also be read off from,
\rf{Vt} and the last identity in \rf{conconx}. As a check, in the ungauged
limit $\Phi\to0$ the pair \rf{covW}, \rf{divW} follows from \rf{covZ} under
$h\to-h$ together with $Z\to W$, and under that flip alone. Here $\a$ is not
also flipped: although $\Psi^{(\pm)}$ carries
the term $\mp\tfrac{i}{4}\a\G_i$, the matrix $\G_i$ itself changes sign
between $\ker\G_+$ and $\ker\G_-$, so that term is in fact \emph{common} to
the two equations, and when $L^I=0$ and $\Phi=0$ the only genuine difference
between $\Psi^{(-)}$ and $\Psi^{(+)}$ is $h\to-h$. The gauging then breaks
that symmetry asymmetrically: \rf{covW} contains $-2\Phi W_iZ_j$, which is
not symmetric in $i\leftrightarrow j$ and has no counterpart in \rf{covZ}.

Next, the algebraic Killing spinor equations. With $L^I=0$ and
$\Gt^I{}_{ij}=\e_{ijk}g^{Ik}$, in three dimensions
$\G^{ij}=is_\pm\e^{ijk}\G_k$, so \rf{alg} becomes
\bea
\la{algred}
{\cal A}^{I,(\pm)}=2i\big[s_\pm g^I_k+\tn_kX^I\big]\G^k-6i\c\,C^I,
\qquad
C^I:=\Big(Q^{IJ}-\tfrac23X^IX^J\Big)V_J .
\ee
For a two-component spinor $\eta$ with $n_i\G^i\eta=\eta$, the condition
$(A+B_k\G^k)\eta=0$ is equivalent to $A+B_kn^k=0$ together with
$An_k+B_k+is\,\e_{ijk}B^in^j=0$. Imposing ${\cal A}^{I,(+)}\eta_+=0$ and
${\cal A}^{I,(-)}\eta_-=0$ \emph{separately} and solving each for $g^I{}_k$
gives the two branches
\bea
\la{gpm}
g^I_k=\tn_kX^I-3\c\,C^IZ_k
\qquad\text{and}\qquad
g^I_k=3\c\,C^IW_k-\tn_kX^I ,
\ee
the first of which is the expression for $\Ft^I$ in \rf{Vt} with its
$X^I$-trace removed. Equating them gives the moduli gradients,
\bea
\la{gradX}
\tn_iX^I=\tfrac32\c\,C^I\,(Z_i+W_i),
\qquad
\Gt^I{}_{ij}=\tfrac32\c\,C^I\e_{ijk}\big(W^k-Z^k\big),
\ee
so that all the moduli are functionally dependent, with gradients along the
single $1$-form $P_i:=Z_i+W_i$. Finally, again from \cite{thesis}, the
operator $\Theta_\pm$ acts on $\eta_\pm$ as
\bea
\la{theta}
\Theta_\pm\eta_\pm=\Big[\tfrac14(1\mp1)h_i\G^i\mp\tfrac{i}{2}\a
-\tfrac12\Phi\big(1\mp Z_i\G^i\big)\Big]\eta_\pm ,
\qquad\text{so}\qquad
\Theta_+\eta_+=-\tfrac{i\a}{2}\,\eta_+ ,
\ee
using $Z_i\G^i\eta_+=\eta_+$, and the $\Ft$ gauge field equation reduces to
\bea
\la{dh}
D_i:=\tfrac12\e_i{}^{jk}(dh)_{jk}=\tn_i\a+\a h_i-6\a\Phi\,Z_i .
\ee
Equation \rf{theta} reproduces both of the $\eta_+$ identities of
\cite{kayani}: with $\Delta=\a^2$ it gives
$-\Delta\parallel\eta_+\parallel^2+4\parallel\Theta_+\eta_+\parallel^2=0$, and
${\rm Re}\langle\eta_+,\G_i\Theta_+\eta_+\rangle
=-\tfrac{\a}{2}{\rm Re}\,(i\langle\eta_+,\G_i\eta_+\rangle)=0$ since that
bilinear is real. Substituting \rf{theta} into the lightcone integrability
condition \rf{int1} of Appendix~\ref{app:lightcone} gives \rf{dh} by a second, independent
route that uses no field equation, so \rf{dh} is a consequence of
supersymmetry alone; that derivation is given in Appendix~\ref{app:lightcone}.

\newsection{The invariant $K$ and the minimal theory}

The results just collected hold for any prepotential and any gauging. Before
using them it is convenient to isolate the single scalar that measures how far
the theory is from the minimal one, since almost every statement below splits
along its vanishing.
Using $Q_{IJ}X^J=\tfrac32X_I$, $Q^{IJ}X_J=\tfrac23X^I$ and $X^IX_I=1$ one has
\bea
\la{Cids}
Q_{IJ}C^J=V_I-X_I(V_KX^K),\qquad C^IX_I=0,
\ee
\bea
\la{Kdef}
K:=Q_{IJ}C^IC^J=V_IC^I=Q^{IJ}V_IV_J-\tfrac23(V_IX^I)^2\ \geq\ 0,
\ee
the inequality because $Q_{IJ}$ is positive definite, with equality precisely
when $C^I=0$. Hence

\medskip\noindent{\it Corollary 3.1:} Every function of the moduli has gradient along the single $1$-form $P_i=Z_i+W_i$.
In particular
\bea
\la{gradPhi}
\tn_i\Phi=\tfrac32\c^2K\,P_i,
\ee
and by \rf{normminus} so does $\parallel\eta_-\parallel^2$. On the branch
$K>0$ of Proposition 7.1 this can be inverted: $\c\neq0$, so \rf{gradPhi} gives
\bea
\la{PdPhi}
P_i=\frac{2}{3\c^2K}\,\tn_i\Phi ,
\qquad\text{and}\qquad
\tn_i\Phi=0\iff P_i=0 .
\ee
The zeros of $P$ are therefore exactly the critical points of the single
function $\Phi$, and $d\Phi\neq0$ on the open complement of $\{P=0\}$ whenever
$P\not\equiv0$. This is a more useful description of the branch than treating
$P$ as an arbitrary closed one-form: everything that follows about the zeros of
$P$ --- Proposition 6.1, Corollary 7.2, the degenerate maximum --- is a
statement about the critical set of $\Phi$. The moduli are
therefore functionally dependent, their level sets coincide, and where
$K\neq0$ the $1$-form $P_i$ is proportional to $\tn_i\Phi$ and in particular
hypersurface-orthogonal.

It is shown in \cite{kayani,thesis} that the conditions \rf{conconx} imply
${\cal L}_{\Vt}\parallel\eta_-\parallel^2=0$. Since $\Phi\neq0$ on $\CS$ (see
below), \rf{normminus} then gives $\Vt^iP_i=0$ and hence, from \rf{Vt},
\bea
\la{hP}
h_iP^i=\Phi\,P_kP^k=2\Phi\big(1+Z_kW^k\big),
\ee
so the component of $h$ along $P$ is fixed and
$\Vt_i=-\parallel\eta_-\parallel^2h^\perp_i$, with $h^\perp_i$ the part of $h_i$
orthogonal to $P$. That $\Phi\neq0$ everywhere follows as in \cite{kayani}:
at a point with $V_IX^I=0$ one would have $U=-\tfrac92Q^{IJ}V_IV_J<0$,
contradicting $U\geq0$. Combined with \rf{fhZ}, \rf{hP} also fixes
$f_iP^i$: since $f_i=h_i-2\Phi Z_i$,
\bea
f_iP^i=h_iP^i-2\Phi\,Z_i(Z_i+W_i)=2\Phi(1+Z_kW^k)-2\Phi(1+Z_kW^k)=0,
\eea
so the flux-dual vector $f$ is orthogonal to $P$, i.e.\ confined to the
same $\{N,Y\}$ plane that any second isometry must occupy by \rf{UP}
below.

\medskip\noindent{\it Proposition 3.2:} At any point, $K=0\iff C^I=0\iff V_I=(V_JX^J)X_I$, and then
$\tn_iX^I=0$ and $\Gt^I=0$. If $K\equiv0$ on $\CS$ the moduli are constant,
$F^I=X^IF$, and the solution lies in the minimal gauged theory.

\medskip\noindent{\it Proof:} Contracting $C^I=0$ with $Q_{IJ}$ and using \rf{Cids}
gives $V_I=(V_KX^K)X_I$; conversely that substitution annihilates $C^I$ by
\rf{Cids}. The vanishing of $\tn_iX^I$ and $\Gt^I{}_{ij}$ is then immediate from
\rf{gradX}. $\square$

Since the classification in the minimal gauged theory is known, and
supersymmetric black rings are excluded there \cite{klrring,grover}, any inherently
non-minimal near-horizon geometry must have $K>0$ somewhere on $\CS$, and
there the moduli are non-constant with gradient along $Z+W$.

\medskip\noindent{\it Remark:} $C^I$ is exactly the algebraic object that
arises, by a completely independent route, in the static, purely
electric, maximally symmetric $AdS_2\times\Sigma_{3,\kappa}$ solutions of
this theory: there the Killing spinor equations of $Spin(4,1)$ force
$c^I:=\big(Q^{IJ}-\tfrac23X^IX^J\big)V_J=0$ for supersymmetry, with the
same solution $V_I\propto X_I$. Two independent methods --- an abstract
bilinear analysis on a general compact $\CS$, and an explicit
Clifford-algebra computation on a maximally symmetric ansatz --- single
out the same algebraic obstruction.

\newsection{Supersymmetry doubling fixes the norm of $\eta_-$}

The doubling \rf{double} relates the two spinors, and \rf{normplus} says the
norm of the left-hand side is constant. Since $\parallel\eta_-\parallel$ need
not be, this is a genuine constraint on the near-horizon data, extracted as
follows. Set
\bea
\la{udef}
u_i:=\Phi P_i-h_i,
\ee
so that $\Vt_i=\parallel\eta_-\parallel^2\,u_i$ by \rf{Vt}, and $u_i$ and
$\Vt_i$ vanish at exactly the same points. The branch hypothesis is
$\Vt\not\equiv0$, which is what Lemma 2.1 delivers. It is not a pointwise
statement, and it had better not be: $\Vt$ is a Killing field and may vanish on
a circle, which is why Theorem 7.5 and \rf{coeffdet} below are phrased on the
locus where $\Vt\neq0$. What the arguments below actually need of $u$ is
weaker than pointwise non-vanishing, and holds with no hypothesis at all: the
two denominators built from it, namely $u_iu^i+\a^2$ in \rf{normminus2} and
$4\a^2+u_ju^j$ in \rf{Gsys} and its solution, are strictly positive everywhere
on $\CS$. This is a corollary of Proposition 4.1 and is recorded as
\rf{unonzero} directly after it.

\medskip\noindent{\it Proposition 4.1:} On the $\Vt\neq0$ branch,
\bea
\la{normrel}
\parallel\eta_+\parallel^2
=\tfrac12\big(u_iu^i+\a^2\big)\parallel\eta_-\parallel^2 .
\ee
Consequently, since $\parallel\eta_+\parallel$ is constant by \rf{normplus},
\bea
\la{normminus2}
\parallel\eta_-\parallel^2
=\frac{2\parallel\eta_+\parallel^2}{u_iu^i+\a^2}\ ,
\ee
so $\parallel\eta_-\parallel^2$ is fixed algebraically by $u$ and $\a$.

\medskip\noindent{\it Proof:} By \rf{gplus},
$\parallel\eta_+\parallel^2=\parallel\G_+\Theta_-\eta_-\parallel^2
=2\parallel\Theta_-\eta_-\parallel^2$. On $\ker\G_-$ we have $s_-=+1$, so
$\Ft_{jk}\G^{jk}=2i\,f_k\G^k$ and \rf{double} gives
\bea
\Theta_-=\tfrac12\big(h_i-\Phi Z_i\big)\G^i+\tfrac{i}{2}\a-\tfrac12\Phi ,
\ee
on using $f_i=h_i-2\Phi Z_i$. Writing $A_i:=\tfrac12(h_i-\Phi Z_i)$ and using
$(A_i\G^i)^2=A_jA^j$,
\bea
\parallel\Theta_-\eta_-\parallel^2
=\Big[A_iA^i+\tfrac14\a^2+\tfrac14\Phi^2-\Phi\,(A_iW^i)\Big]
\parallel\eta_-\parallel^2 ,
\ee
which expands to
\bea
\parallel\Theta_-\eta_-\parallel^2
=\Big[\tfrac14h_kh^k-\tfrac12\Phi\,h_i(Z^i+W^i)+\tfrac12\Phi^2\big(1+Z_kW^k\big)
+\tfrac14\a^2\Big]\parallel\eta_-\parallel^2 .
\ee
Imposing \rf{hP} reduces the bracket to
$\tfrac14h_kh^k-\tfrac12\Phi^2(1+Z_kW^k)+\tfrac14\a^2$. Finally
$P_kP^k=2(1+Z_kW^k)$ and \rf{hP} give
$u_iu^i=\Phi^2P_kP^k-2\Phi\,h_iP^i+h_kh^k=h_kh^k-\Phi^2P_kP^k$, so the bracket equals
$\tfrac14(u_iu^i+\a^2)$. $\square$

The denominator in \rf{normminus2} is strictly positive, and no hypothesis
beyond $\Vt\not\equiv0$ is needed for it. The constant
$\parallel\eta_+\parallel$ of \rf{normplus} cannot vanish: $\Vt_i={\rm
Re}\langle\G_+\eta_-,\G_i\eta_+\rangle$, so $\eta_+\equiv0$ would give
$\Vt\equiv0$, which Lemma 2.1 excludes. Reading \rf{normrel} as
$2\parallel\eta_+\parallel^2=(u_iu^i+\a^2)\parallel\eta_-\parallel^2$ with
a strictly positive constant on the left therefore forces
$\parallel\eta_-\parallel^2>0$ everywhere on $\CS$, and then
\bea
\la{unonzero}
u_iu^i+\a^2=\frac{2\parallel\eta_+\parallel^2}
{\parallel\eta_-\parallel^2}\ >\ 0
\qquad\text{everywhere on }\CS ,
\ee
whence a fortiori $4\a^2+u_ju^j\geq\a^2+u_ju^j>0$. Note also
$\Vt_i\Vt^i=\parallel\eta_-\parallel^4u_iu^i$, so the norm of the Killing
vector is fixed as well; it vanishes exactly where $u$ does, which \rf{unonzero}
does {\it not} exclude, since $\a$ may take up the slack.

Equation \rf{normrel} also follows, with no spinor computation at all, from
the bilinear identities \rf{conconx} of \cite{kayani,thesis} themselves.
Writing $\sigma=\parallel\eta_-\parallel^2$,
the first and third of the identities collected in \rf{conconx} give
$h_i\Vt^i=\Delta\sigma-2\parallel\eta_+\parallel^2$, while the last gives
$\Vt^i\tn_i\sigma=-\Vt_i\Vt^i-\sigma\,(h_i\Vt^i)$. Imposing
${\cal L}_{\Vt}\sigma=0$ and using $\Vt_i\Vt^i=\sigma^2u_iu^i$ and $\Delta=\a^2$,
\bea
\sigma^2u_iu^i=\sigma\big(2\parallel\eta_+\parallel^2-\a^2\sigma\big),
\ee
which is \rf{normrel}. The two derivations are independent --- one from the
explicit form of $\Theta_-$ in the two-component representation, the other
from the $\mathfrak{sl}(2,\bR)$ bilinear identities --- and agree.

Note that \rf{normminus2} expresses $\parallel\eta_-\parallel^2$, and hence by
\rf{Vt} the Killing vector $\Vt$ itself, in terms of $h$, $\a$, $\Phi$ and
$P=Z+W$ together with the constant $\parallel\eta_+\parallel$. Where $K>0$,
\rf{gradX} expresses $P_i$ in turn through $\tn_iX^I$, so on that branch
$\parallel\eta_-\parallel^2$ is a function of the bosonic near-horizon data
alone.

\newsection{The remaining first-order data}

Sections 2 to 4 determine $\tn_iZ_j$, $\tn_iW_j$, $\tn_iX^I$ and
$\parallel\eta_-\parallel$. What is left undetermined at first order is the
derivative of $h$ and the derivative of $\a$, and this section fixes both. The
symmetric part of $\tn_ih_j$ and the Ricci tensor of $\CS$ come from the Killing
property of $\Vt$; the antisymmetric part comes from $dh$ once $\tn_i\a$ is
known, and $\tn_i\a$ is itself algebraic. With that the first-order system is
closed, and everything from Section 6 onwards is a consequence of it.

\subsection{The symmetric part of $\tn_ih_j$, and the Ricci tensor}

Since $\Vt_i=\sigma u_i$ with $\sigma:=\parallel\eta_-\parallel^2$ is Killing,
expanding $\tn_{(i}\Vt_{j)}=0$ using \rf{normminus} and \rf{gradPhi} gives

\medskip\noindent{\it Proposition 5.1:} \bea
\la{gradh}
\tn_{(i}h_{j)}=\mu\,P_iP_j+\Phi\,\tn_{(i}P_{j)}+\Phi\,P_{(i}h_{j)},
\qquad \mu:=\tfrac32\c^2K-\Phi^2 .
\ee

\noindent The trace of \rf{gradh} reproduces $\tn^ih_i$ as given by the
$+-$ component of the Einstein equation, which is nowhere used in deriving it.
With \rf{gradh} the $ij$ component of the Einstein equation determines the
Ricci tensor of $\CS$ completely,
\bea
\la{ricci}
\tilde R_{ij}&=&-\tn_{(i}h_{j)}+\tfrac12h_ih_j-\tfrac32f_if_j
-\tfrac94\c^2K\,N_iN_j+\tfrac94\c^2K\,P_iP_j
\nn
&&+\ \delta_{ij}\Big[f_kf^k+\tfrac12\a^2-\tfrac23\c^2U+\tfrac32\c^2K\,N_kN^k\Big],
\ee
using \rf{gradX} and \rf{gpm} to eliminate $\tn_iX^I$ and $\Gt^I{}_{ij}$. The
potential can be eliminated as well: from \rf{Udef}, \rf{Phidef} and
\rf{Kdef},
\bea
\la{chiU}
\c^2U=6\Phi^2-\tfrac92\c^2K ,
\ee
so \rf{ricci} expresses $\tilde R_{ij}$ entirely through $h$, $f$, $Z$, $W$,
$\a$, $\Phi$ and $K$. In three dimensions the Ricci tensor determines the
Riemann tensor, so \rf{ricci} fixes the local geometry of $\CS$.

\subsection{The antisymmetric part, and $\tn_i\a$}

The antisymmetric part of $\tn_ih_j$ is fixed by \rf{dh}, but that equation
still contains the unknown $\tn_i\a$. That derivative is itself determined
algebraically, so that $(dh)_{ij}$ is too. Combining \rf{conconx}
and \rf{ivdh}
gives $\Vt^j(dh)_{ji}=-\tn_i(\a^2\sigma)$; writing this with
$(dh)_{ij}=\e_{ijk}D^k$, using $\tn\sigma=-\sigma\Phi P$ and substituting
\rf{dh}, one obtains a linear system for $G_i:=\tn_i\a$,
\bea
\la{Gsys}
\e_{ijk}u^jG^k-2\a G_i
=-\a^2\Phi P_i-\a\,\e_{ijk}u^jh^k+6\a\Phi\,\e_{ijk}u^jZ^k,
\ee
whose matrix satisfies
\bea
\la{det}
\det\big({-}2\a\,\delta_{ik}+\e_{ijk}u^j\big)=-2\a\big(4\a^2+u_ju^j\big),
\ee
the determinant being taken over the matrix with row index $i$ and column
index $k$.

\medskip\noindent{\it Corollary 5.2:} Wherever $\a\neq0$ the system \rf{Gsys} is
invertible, so $\tn_i\a$ and hence $(dh)_{ij}$ are uniquely determined by the
near-horizon data. Where $\a$ vanishes identically on an open set,
$\tn_i\a=0$ there and \rf{dh} gives $(dh)_{ij}=0$.

Together with \rf{covZ}--\rf{dh} and Propositions 4.1 and 5.1, Corollary 5.2
closes the system: away from the zero set of $\a$, the covariant derivatives
of $Z_i$, $W_i$, $h_i$, $X^I$, $\a$, $\Phi$ and $\parallel\eta_-\parallel^2$
are all fixed algebraically by the near-horizon data itself, so
supersymmetry together with the field equations determines the entire
first-order near-horizon system.

\subsection{A second-order relation for $\a$}

Appendix~\ref{app:fieldeqs} also records the $++$ component of the Einstein equation
\rf{auxeq}, not needed for the first-order closure of Sections 2--5 but
supplying new information here: a formula for $\tn^i\tn_i\a$, complementing
Corollary 5.2's first-order determination of $\tn_i\a$.

\medskip\noindent{\it Proposition 5.3:} Wherever $\a\neq0$,
\bea
\la{lapalpha}
\tn^i\tn_i\a=3\c^2K\,\a P_kP^k+4\a\Phi^2P_kP^k-24\a\Phi^2
+4\a\Phi\,h_iN^i+6\Phi\,(\tn_i\a)Z^i-(\tn_i\a)h^i.
\eea

\medskip\noindent{\it Proof:} Substitute $\Delta=\a^2$, $\b_i=\tn_i\a-\a h_i$
and $M^I{}_i=\a\tn_iX^I$ into \rf{auxeq}. Using \rf{gradX} and
\rf{Kdef}, $Q_{IJ}M^I{}_\ell M^{J\ell}=\a^2\tfrac94\c^2K\,P_kP^k$; using
\rf{dh}, $(dh)_{ij}(dh)^{ij}=2D_iD^i$ with $D_i=\tn_i\a+\a h_i-6\a\Phi Z_i$.
Expanding both sides in $\a$, $\tn_i\a$, $h_i$, $Z_i$, every term
proportional to $\a\,h_i\tn^i\a$ cancels between the two sides, and
dividing the remainder by $\a$ gives
\bea
\tn^i\tn_i\a=\tfrac94\c^2K\,\a P_kP^k+\tfrac12\a\,\tn^ih_i-18\a\Phi^2
+6\a\Phi\,h_iZ^i-(\tn_i\a)h^i+6\Phi(\tn_i\a)Z^i.
\eea
Taking the trace of \rf{gradh} with $\delta^{ij}$ and using
$\tn^iP_i=\tn^iZ_i+\tn^iW_i=2h_iN^i-8\Phi+4\Phi Z_kW^k$ (subtracting
\rf{divW} from \rf{covZ}'s trace) and \rf{hP} for $h_iP^i$ gives
$\tn^ih_i=\tfrac32\c^2K\,P_kP^k+2\Phi^2P_kP^k+2\Phi\,h_iN^i-12\Phi^2$;
substituting this and $h_iZ^i=\tfrac12(h_iP^i+h_iN^i)
=\tfrac12(\Phi P_kP^k+h_iN^i)$ (from \rf{hP} and $Z_i=\tfrac12(P_i+N_i)$)
gives exactly \rf{lapalpha}. $\square$

\medskip\noindent{\it Remark:} There is a second, purely differential route to
\rf{lapalpha} which explains why it should hold at all. Taking the divergence
of \rf{dh} and using $\tn^i(\star dh)_i\equiv0$ --- the statement $d(dh)=0$,
which holds on any manifold --- gives
\bea
\la{divD}
0=\tn^iD_i=\tn^i\tn_i\a+\a\,\tn^ih_i+(\tn_i\a)h^i
-6\Phi\,(\tn_i\a)Z^i-6\a\,(\tn_i\Phi)Z^i-6\a\Phi\,\tn^iZ_i ,
\ee
and substituting $\tn^ih_i$ from the trace of \rf{gradh}, $\tn_i\Phi$ from
\rf{gradPhi} and $\tn^iZ_i$ from \rf{covZ} reproduces \rf{lapalpha} exactly.
Conversely, the difference between $\tn^i\tn_i\a$ and the right-hand side of
\rf{lapalpha} is identically $\tn^iD_i$, with no field equation used. So
Proposition 5.3 is {\it equivalent} to $d(dh)=0$: the $++$ Einstein equation
\rf{auxeq} carries, on this branch, exactly the information already contained
in the closure of $dh$, and no more. This is a consistency statement about
Corollary 5.2 rather than independent data, which is why \rf{lapalpha} adds no
new constraint on the function $g_P$ introduced below.

\medskip
As with \rf{lapnorm}, \rf{lapalpha} is not fully determined by $K$ and
$\Phi$ alone: it retains $(\tn_i\a)Z^i$ and $(\tn_i\a)h^i$, built from the
$P$-directional derivative of $\a$ that neither Corollary 6.3 below nor
any other result of this paper fixes. Theorem 6.6 will show that the two
transverse derivatives $N^i\tn_i\a$ and $Y^i\tn_i\a$ both vanish
identically, so $\tn_i\a$ is everywhere parallel to $P_i$ and the single
function $g_P$ defined by $\tn_i\a=g_PP_i$ is the whole of the remaining
freedom in $\a$. Lemma 6.11 then fixes it, $g_P=2\Phi\a$, so \rf{lapalpha} is
a consistency condition on a determined quantity rather than a constraint on a
free one.

\newsection{What the scalar invariants determine}

This section asks one question: how far do the scalar invariants of the
near-horizon data determine the isometry algebra of $\CS$? The answer has two
halves. The invariant route to a second isometry collapses, because every such
gradient is parallel to the same one-form $P$; but that same collapse, made
quantitative, is a closed differential system, and it is what Section 7 then
integrates. Concretely, the chain runs: Proposition 6.4 shows $P$ is closed;
Corollary 6.5 turns that closure into the algebraic formula
$h_Y=\a/(1-Z\cdot W)$; Theorem 6.6 shows the obstruction $\zeta$ of
Corollary 6.3 then vanishes identically; and Lemma 6.11 assembles the result
into the closed gradient system that Theorem 7.5 transports along $P$ in
Section 7.

Let $U$ be any Killing vector of $\CS$ which leaves the bosonic fields
invariant, as $\Vt$ does in \rf{VtKilling}. Then $U$ annihilates every
scalar built from those fields. Two conditions on $U$ follow at once.
Since $\Delta=\a^2$ is such a scalar, ${\cal L}_U\Delta=0$ gives
$2\a\,U^i\tn_i\a=0$, so
\bea
\la{Ualpha}
U^i\tn_i\a=0 \qquad(\a\neq0);
\ee
and since ${\cal L}_UX^I=0$ for every $I$, \rf{gradX} gives
$U^i\c\,C^IP_i=0$ for every $I$. Where $K>0$ some $C^I\neq0$, so this forces
\bea
\la{UP}
U^iP_i=0 .
\ee
$U$ is thus confined to the intersection of the two hyperplanes orthogonal
to $P_i$ and to $\tn_i\a$ --- a line, spanned by $\Vt$ itself, wherever $P_i$
and $\tn_i\a$ are linearly independent, since $\Vt$ satisfies \rf{UP} by
\rf{hP} and \rf{Ualpha} because ${\cal L}_{\Vt}\a=0$ in \rf{VtKilling}. This
is the naive expectation only: Theorem 6.6 below shows that $P_i$ and
$\tn_i\a$ are in fact parallel {\it everywhere} on this branch, so the two
hyperplanes coincide rather than intersect in a line, and $U$ is confined only
to the resulting two-plane, not to a line. The next four subsections build up
to that theorem.

\subsection{Where $P$ can vanish}

The locus on which $P_i$ vanishes is where the moduli stop varying, by
\rf{gradX}, and it is the natural place to look for an algebraic obstruction.
The scalar field equation is the only equation of motion not already used in
deriving the Killing spinor identities of Section 5, so it is the natural
candidate. It supplies nothing.

\medskip\noindent{\it Proposition 6.1:} At a point where $P_i=0$, the scalar
field equation \rf{feq5} is identically satisfied. In particular it places no
condition on $\tn^kP_k$ there, and none on $K$.

\medskip\noindent{\it Proof:} From \rf{Udef} and \rf{Kdef}, identically,
\bea
\la{UKid}
\c^2U+\tfrac92\c^2K=6\Phi^2 .
\ee
Suppose $P_i=0$ at a point $p$. Then $\tn_iX^I=0$ at $p$ for every $I$ by
\rf{gradX}, so the term $\tn_iX^M\tn^iX^N(\cdots)$ in \rf{feq5} drops, as does
$h^i\tn_iX_I$. The Laplacian does {\it not} drop. Differentiating \rf{gradX}
once more,
$\tn^i\tn_iX^I=\tfrac32\c\big[(\tn^iC^I)P_i+C^I\tn^iP_i\big]$, and $\tn_iC^I$
is itself parallel to $P_i$ because $C^I$ is a function of the moduli alone;
hence at $p$
\bea
\la{lapX}
\tn^i\tn_iX^I=\tfrac32\c\,C^I\,D ,\qquad D:=\tn^kP_k ,
\ee
and $D=2h_iN^i-8\Phi+4\Phi Z_kW^k=4h_iZ^i-12\Phi$ at $p$, using $W=-Z$ and
$Z_kW^k=-1$ there. Here $D$ is free data: the only algebraic condition on $h$
is \rf{hP}, which reads $0=0$ when $P_i=0$ and so says nothing about $h_iZ^i$.
The same freedom must therefore be carried through every other term of
\rf{feq5}, and the graviphoton cross-term is where it enters. With $W=-Z$ the
fluxes are $g^I_k=-3\c\,C^IZ_k$ from either branch of \rf{gpm}, so
$\Gt^I{}_{ij}=-3\c\,C^I\e_{ijk}Z^k$ and, by \rf{fhZ},
$f_i=h_i-2\Phi Z_i$, whence
\bea
\la{FGcross}
\Ft_{l_1l_2}\Gt^{Jl_1l_2}=2f_ig^{Ji}=-6\c\,C^J\big(h_iZ^i-2\Phi\big)
=-6\c\,C^J\big(\Phi+\tfrac14D\big) ,
\ee
using $h_iZ^i=3\Phi+\tfrac14D$ from the expression for $D$ above. It is this
$D$-dependence of the cross-term, and not the Laplacian alone, that carries the
undetermined datum through the field equation. The remaining flux contraction
is $\Gt^M{}_{l_1l_2}\Gt^{Nl_1l_2}=18\c^2C^MC^N$. Substituting all of this into
\rf{feq5} with $\Delta=\a^2$, $L^I=0$, and reducing with \rf{Cids}--\rf{Kdef},
the four surviving groups of terms are
\bea
\tn^i\tn_iX_I=-\c D\,C_I ,
\qquad
\tfrac23Q_{IJ}\big({-}\Ft\cdot\Gt^J\big)=\c\big(4\Phi+D\big)\,C_I ,
\nn
-\tfrac1{12}\big[\Gt^M\!\cdot\Gt^N\big]\big(C_{MNI}-X_IC_{MNJ}X^J\big)
=-\tfrac32\c^2\big[(C\!\cdot\!C)_I+2KX_I\big] ,
\nn
3\c^2V_MV_N\big(\cdots\big)
=\tfrac32\c^2(C\!\cdot\!C)_I-4\c\Phi\,C_I+3\c^2K\,X_I ,
\eea
where $(C\!\cdot\!C)_I:=C_{IMN}C^MC^N$ and $C_I:=Q_{IJ}C^J=V_I-\tfrac{\Phi}{\c}X_I$.
The $(C\!\cdot\!C)_I$ and $KX_I$ pieces of the third and fourth groups cancel
exactly against each other, leaving a residue $-4\c\Phi\,C_I$ carried by the
fourth group alone; likewise the $D$-dependent pieces $-\c D\,C_I$ and
$+\c D\,C_I$ of the first two groups cancel against each other, leaving a
residue $+4\c\Phi\,C_I$ carried by the second group alone. These two residues
are equal and opposite, so the sum of all four groups vanishes identically,
for every value of $D$ and every value of $K$.
$\square$

\medskip
Two comments. First, the cancellation is not a coincidence of the STU model. It
holds for the STU model with a completely general gauging $V_I$, identically in
$X^I$ on the constraint surface, and for generic non-STU cubic prepotentials,
so it is a property of a general cubic prepotential rather than an artefact of
STU triality.

Second, the quantity $\tn^kP_k$ that survives unconstrained has a clean
geometric meaning. By \rf{lapnorm}, at a point where $P_i=0$,
\bea
\la{lapnormatP0}
\tn^i\tn_i\ln\parallel\eta_-\parallel^2\big|_{P=0}=12\Phi^2-2\Phi\,h_iN^i
=-\Phi\,\tn^kP_k ,
\eea
so $\tn^kP_k$ measures the failure of a zero of $P$ to be a degenerate critical
point of $\parallel\eta_-\parallel^2$. Proposition 6.1 says that supersymmetry
and the equations of motion together leave this number free. Section~\ref{sec:varmod}
exhibits a compact, regular, non-static horizon on which it is nonzero, so the
freedom is realised and not merely unexcluded. Away from the zero set of $P$,
$\parallel\eta_-\parallel^2$ --- used in Proposition 4.1 to give an independent
route to \rf{UP} --- is a genuine function of the bosonic data, since $u_i$,
and hence $\parallel\eta_-\parallel^2$ by \rf{normminus2}, is determined by
\rf{gradX}.

\subsection{Divergence-free vectors transverse to $P$}

The next task is to locate where $\tn_i\a$ can fail to be independent of $P_i$.
By \rf{UP}, any second isometry is confined to the plane spanned by $N_i$ and
$Y_i$, and the divergence arguments that settle the $h=0$ and $\Vt=0$ branches
cannot be adapted to that plane, because:

\medskip\noindent{\it Proposition 6.2:} $\tn^iN_i=\tn^iY_i=0$ identically.

\medskip\noindent{\it Proof:} For $N_i$ this follows by subtracting
\rf{divW} from the second equation in \rf{covZ} and using \rf{hP}. For $Y$,
computing $\tn_iY_j$ from \rf{covZ} and \rf{covW} by the Leibniz rule, the
$\a$ terms cancel by $\e_{ijk}\e^{ikl}=-2\delta_j^l$ and the $h$ terms by
antisymmetry. $\square$

For $Y$ this sharpens to a clean statement: $\tn_{(i}Y_{j)}=0$ if and
only if $\Phi=0$ and $h=0$, both already excluded in the gauged theory
($h=0$ is excluded outright on the $\Vt\neq0$ branch, and $\Phi=0$ at a
point forces $\c^2U=-\tfrac92\c^2K<0$ whenever $K>0$, contradicting
$U\geq0$). So
neither $N$ nor $Y$ --- which together span the entire plane any second
isometry must lie in by \rf{UP} --- can be rescaled into a Killing vector
either; $\Vt$ is the only direction among $Z,W,P,N,Y,h$ and $\Phi P-h$
itself that is ever killable.

Here $\tn^iN_i=0$ holds only because of \rf{hP}, i.e.\ because $\Vt$ is
Killing, whereas $\tn^iY_i=0$ is unconditional; either way, both directions
available to a second isometry are divergence-free, so integrating over the
compact $\CS$ returns no information about them. Nor does the natural
integral identity help: integrating the Einstein equation over $\CS$ and
using \rf{branch}, \rf{fhZ} and \rf{covZ} to eliminate $h_iZ^i$ produces
\bea
\int_{\CS}\Big[Z^i\tn_i\Phi-\tfrac13Q_{IJ}\tn_iX^I\tn^iX^J\Big]=0 ,
\ee
but by \rf{gradX} and \rf{gradPhi} the integrand is
$\tfrac32\c^2K(1+Z_kW^k)-\tfrac13\cdot\tfrac92\c^2K(1+Z_kW^k)$, which
vanishes pointwise. Global arguments are therefore vacuous here, and
$\tn_i\a$ must instead be determined pointwise, by solving \rf{Gsys}
directly.

\medskip\noindent{\it Remark:} It is natural to ask whether $\tn_i\a$
could instead be pinned down globally, generalising the ungauged
argument for $\a={\rm const}$: there, integrating the trace Einstein
equation forces $\tn^ih_i=0$ and $L^I=\Gt^I=0$ pointwise, and then
multiplying $\Box\a+h\cdot\tn\a=0$ by $\a$ and integrating by parts twice
gives $-\int(\tn\a)^2=0$, so $\a={\rm const}$. Trying the gauged analogue
of this technique on the natural candidate
$S_0:=\tn^i(h_i-6\Phi Z_i)$ --- using $\tn^i(\star dh)_i\equiv0$ for any
$1$-form $h$, so that the same manipulation with \rf{dh} produces the
energy identity $\int_\CS\a^2S_0=2\int_\CS(\tn\a)^2\geq0$, which would
force $\tn\a\equiv0$ if $S_0\leq0$ pointwise --- gives, by direct
computation,
\bea
\la{S0}
S_0=24\Phi^2-8\Phi\,(Z_ih^i)-6\c^2K\,(1+Z_kW^k).
\ee
This is not sign-definite: $S_0$ depends on $h$ only through the scalar
$Z_ih^i$, and while \rf{hP} fixes $h_iP^i$, it does not fix $Z_ih^i$
separately, since $Z$ is not proportional to $P$ in general. So at any
point with $\Phi\neq0$, $S_0\to\pm\infty$ as the free part of $h$ along
$Z$ is taken large with either sign, and no realisability condition
established above changes this. The direct generalisation of the
ungauged maximum-principle argument therefore fails, which is exactly
why $\tn_i\a$ is instead extracted pointwise from \rf{Gsys} below.

\subsection{The closure of $P$, and the collapse of the invariant route}

Contracting \rf{Gsys} with $u^i$ annihilates every cross product. What is left
is $-2\a\,(u^i\tn_i\a)=-\a^2\Phi\,(u_iP^i)$, and $u_iP^i$ itself vanishes,
since $u_iP^i=\Phi P_kP^k-h_iP^i=0$ by \rf{hP}. Hence
\bea
\la{gradalphaU}
u^i\tn_i\a=0 \qquad(\a\neq0),
\ee
consistent with ${\cal L}_{\Vt}\a=0$ in \rf{VtKilling}, since
$\Vt_i=\parallel\eta_-\parallel^2u_i$. This lets us write
$u_i=u_NN_i+u_YY_i$ in the orthogonal frame $\{P,N,Y\}$, with
$u_N:=u_iN^i/N_jN^j$ and $u_Y:=u_iY^i/Y_jY^j$, and solve \rf{Gsys}
completely for the part of $\tn_i\a$ transverse to $P_i$. The three vectors
$P$, $N$, $Y$ span only where $Z$ and $W$ are linearly independent and
$P\neq0$, so the frame decompositions of this subsection, and the quotients
$N_kN^k$, $Y_kY^k$ and $P_kP^k$ appearing in them, are to be read on
$\CS_0\cap\{P\neq0\}$; the conclusion drawn from them, Theorem 6.6, is then
extended to all of $\CS$ there.

\medskip\noindent{\it Corollary 6.3:} With
\bea
\la{zetadef}
\zeta:=\a+\frac{2\,u_iY^i}{P_kP^k} ,
\ee
\bea
\la{GNGY}
N^i\tn_i\a=\frac{6\Phi\a\,\zeta}{4\a^2+u_ju^j}\,u_iY^i,
\qquad
Y^i\tn_i\a=-\frac{6\Phi\a\,\zeta}{4\a^2+u_ju^j}\,\frac{Y_kY^k}{N_kN^k}\,u_iN^i ,
\ee
or equivalently, writing $\tn_i\a=g_PP_i+g_NN_i+g_YY_i$ in the orthogonal
frame $\{P,N,Y\}$, in the manifestly antisymmetric form
\bea
\la{GNGYsym}
g_N=\frac{6\Phi\a\,\zeta\,u_iY^i}{(N_kN^k)(4\a^2+u_ju^j)},
\qquad
g_Y=-\frac{6\Phi\a\,\zeta\,u_iN^i}{(N_kN^k)(4\a^2+u_ju^j)},
\ee
so that the part of $\tn\a$ transverse to $P$ is proportional to
$(u_iY^i)N-(u_iN^i)Y$. Note $Y_kY^k/N_kN^k=\tfrac14P_kP^k$, from
$N_kN^k=2(1-Z_kW^k)$, $Y_kY^k=1-(Z_kW^k)^2$ and $P_kP^k=2(1+Z_kW^k)$.
Since $u_i\neq0$ on this branch, both vanish together --- i.e.\ $\tn_i\a$
is proportional to $P_i$ --- if and only if $\zeta=0$: a single
hypersurface of $\CS$, not an open condition.

\medskip\noindent{\it Proof:} From $P_i=Z_i+W_i$, $N_i=Z_i-W_i$,
$Y_i=\e_{ijk}Z^jW^k$ and the bac-cab rule,
\bea
\e_{ijk}N^jP^k=2Y_i,
\qquad
\e_{ijk}Y^jP^k=-(1+Z_lW^l)N_i,
\qquad
\e_{ijk}N^jY^k=(Z_lW^l-1)P_i .
\ee
Writing \rf{Gsys} in the frame $\{P,N,Y\}$ using these turns the vector
equation into three scalar equations for the coefficients of
$\tn_i\a=g_PP_i+g_NN_i+g_YY_i$; eliminating $g_N$, $g_Y$ in favour of $g_P$
from the $N$- and $Y$-equations and substituting back reproduces \rf{GNGY},
with $\zeta$ the common factor. $\square$

\medskip
Combining Corollary 6.3 with \rf{UP} and \rf{Ualpha} would give a
one-dimensional isometry algebra away from the locus $\{\zeta=0\}$, were that
locus a hypersurface. It is not: it is all of $\CS$. The reason is a closure
property of $P_i$ already implicit in the first-order system of Section 4.

\medskip\noindent{\it Proposition 6.4:} $P_i$ is a closed one-form,
\bea
\la{dPzero}
\tn_{[i}P_{j]}=0 ,
\ee
equivalently, in dual form,
\bea
\la{Nxh}
\e_{ijk}N^jh^k+\a P_i-2\Phi Y_i=0 .
\ee

\medskip\noindent{\it Proof:} Two independent routes. {\it Differential:}
\rf{normminus} states $\tn_i\parallel\eta_-\parallel^2
=-\Phi P_i\parallel\eta_-\parallel^2$, i.e.\
$\tn_i\ln\parallel\eta_-\parallel^2=-\Phi P_i$. The left-hand side is an exact
one-form, so $d(\Phi P)=0$, that is
$(\tn_{[i}\Phi)P_{j]}+\Phi\,\tn_{[i}P_{j]}=0$. By \rf{gradPhi}
$\tn_i\Phi=\tfrac32\c^2K\,P_i$ is parallel to $P_i$, so the first term vanishes
identically, and $\Phi\neq0$ everywhere on this branch gives \rf{dPzero}.
{\it Algebraic:} computing $\tn_iP_j=\tn_iZ_j+\tn_iW_j$ from \rf{covZ} and
\rf{covW} and antisymmetrising,
\bea
\la{antidP}
\tn_{[i}P_{j]}=-N_{[i}h_{j]}-\tfrac12\a\,\e_{ijk}P^k+\Phi\,\e_{ijk}Y^k ,
\eea
an identity in $Z,W,h,\Phi,\a$ requiring no constraint; its Hodge dual is
\rf{Nxh}. The two routes agree, and the second shows that \rf{dPzero} is a
genuine algebraic restriction on the near-horizon data rather than a triviality.
$\square$

\medskip\noindent{\it Corollary 6.5:} Wherever $P_i\neq0$, the component of
$h$ along $Y$ is determined by $\a$:
\bea
\la{hY}
h_iY^i=\tfrac12\a\,P_kP^k ,
\qquad\text{equivalently}\qquad
h_Y=\frac{\a}{1-Z_kW^k}=\frac{2\a}{N_kN^k} .
\ee

\medskip\noindent{\it Proof:} Decompose $h_i=\Phi P_i+h_NN_i+h_YY_i$ in the
orthogonal frame, the $P$-component being fixed by \rf{hP}. Using the
cross-product identities of Corollary 6.3's proof, $\e_{ijk}N^jP^k=2Y_i$,
$\e_{ijk}N^jN^k=0$ and $\e_{ijk}N^jY^k=(Z_lW^l-1)P_i$, equation \rf{Nxh}
collapses to the single scalar statement
\bea
\big[h_Y(Z_kW^k-1)+\a\big]P_i=0 ,
\eea
which gives \rf{hY} wherever $P_i\neq0$. The second form follows from
$Y_kY^k/N_kN^k=\tfrac14P_kP^k$. $\square$

\medskip
Corollary 6.5 closes a gap left open in Section 5: of the two components of
$h$ transverse to $P$, only $h_N$ is free. It also destroys the strategy of
this section.

\medskip\noindent{\it Theorem 6.6:} $\zeta\equiv0$ wherever it is defined,
that is on $\CS_0\cap\{P\neq0\}$. Consequently $\tn_i\a$ is parallel to $P_i$
there, and $\a$ is functionally dependent on $\Phi$ there.

\medskip\noindent{\it Proof:} Since $u_i=\Phi P_i-h_i$ and $P_iY^i=0$, we have
$u_iY^i=-h_iY^i=-\tfrac12\a P_kP^k$ by \rf{hY}, so
\bea
\zeta=\a+\frac{2u_iY^i}{P_kP^k}=\a-\a=0 .
\eea
By Corollary 6.3 both $N^i\tn_i\a$ and $Y^i\tn_i\a$ then vanish, and writing
$\tn_i\a=g_PP_i+g_NN_i+g_YY_i$ in the orthogonal frame gives
$g_NN_iN^i=g_YY_iY^i=0$, hence $g_N=g_Y=0$ away from the degenerate loci
$N_iN^i=0$ or $Y_iY^i=0$. So $\tn_i\a=g_PP_i$, parallel to
$\tn_i\Phi=\tfrac32\c^2K\,P_i$ by \rf{gradPhi}. $\square$

\medskip\noindent{\it Remark (why the statement is not made on all of $\CS$):}
The invariant $\zeta$ of \rf{zetadef} carries $P_kP^k$ in a denominator and
$Y$ in a numerator, so it is not defined at a zero of $P$, nor where $Z$ and
$W$ become parallel; and at a zero of $P$ the conclusion ``$\tn_i\a$ is
parallel to $P_i$'' would say $\tn_i\a=0$, which is a strictly stronger
statement than the one proved and does not follow from it by continuity, the
antisymmetric form $\e_{ijk}\tn^j\a\,P^k=0$ being vacuous there. Nothing below
needs more: Theorem 6.6 is used to identify the two hyperplanes in the Remark
following, to fix the transverse part of $\tn_i\a$ in Lemma 6.11, and to close
the invariant route to a second isometry in Section~\ref{sec:varmod}, and each of those
three uses is already confined to $\CS_0\cap\{P\neq0\}$. What does extend to
all of $\CS$ is the second line of \rf{gradsys}, $\tn_i\a=2\Phi\a P_i$, whose
two sides are continuous on $\CS$; that extension is carried out, with no
hypothesis at all, in Corollary 6.13.

\medskip\noindent{\it Remark (the cost of Theorem 6.6):} The conditions
\rf{UP} and \rf{Ualpha} would confine a second isometry $U$ to the
intersection of two independent hyperplanes. Theorem 6.6 shows they define the
{\it same} hyperplane: $\tn_i\a\parallel P_i$ at every point at which the two
hyperplanes are defined, so \rf{Ualpha} adds
nothing to \rf{UP}, and $U$ is confined only to the two-plane
$\mathrm{span}\{N,Y\}$. Excluding a second rotational isometry pointwise by
means of scalar invariants therefore fails on the $K>0$ branch, and no
weakening of the hypotheses repairs it.

The obstruction is structural rather than technical.

\medskip\noindent{\it Lemma 6.A (parallel-gradient collapse):} Every scalar $S$
built algebraically from the near-horizon data satisfies $\tn_iS=S'P_i$ for
some scalar $S'$. Consequently $\tn I_a\wedge\tn I_b=0$ for all $a,b$, and
$\mathrm{rank}\{\tn I_1,\tn I_2,\tn I_3\}\leq1$ identically, for the curvature
invariants $I_a=\mathrm{tr}\,\Rt^a$.

\medskip\noindent{\it Proof:} This holds for the moduli $X^I$ by \rf{gradX},
for $\Phi$ by \rf{gradPhi}, for $K$ because $K$ is a function of the moduli
alone, for $\parallel\eta_-\parallel^2$ by \rf{normminus}, and now for $\a$ by
Theorem 6.6 on $\CS_0\cap\{P\neq0\}$ and by Corollary 6.13 on the rest of
$\CS$. Since $\Rt_{ij}$ is by \rf{ricci} an algebraic function of the
state, the same holds for $I_a=\mathrm{tr}\,\Rt^a$; and any two vectors
parallel to the same $P_i$ have vanishing wedge product. $\square$ (As a
computational check, direct numerical evaluation on states satisfying
\rf{hP}, \rf{dPzero} and the contracted Bianchi identity confirms
$\mathrm{rank}\{\tn I_1,\tn I_2,\tn I_3\}=1$ exactly, so the bound is
saturated rather than collapsing further to rank $0$.)

\medskip
The word {\it algebraically} in the Lemma is not decoration; the closure
reaches exactly as far as follows. The state variables are
$(Z,W,h,\Phi,\a,K)$, and $\tn_iZ_j$, $\tn_iW_j$, $\tn_ih_j$ are algebraic in
the state by \rf{covZ}, \rf{covW} and \rf{gradh}, so $\tn_iP_j$ is too. Hence
the Lemma extends to any scalar built algebraically from the state {\it and}
its first covariant derivatives, and in particular the second derivatives
$\tn_i\tn_jS=S''P_iP_j+S'\tn_iP_j$ of the invariants above are again algebraic
in the state --- except through $\a$, where $S'=g_P$. Lemma 6.11 below
evaluates that coefficient, $g_P=2\Phi\a$, so $g_P$ is algebraic in the state
as well and the extension is unconditional.

\medskip
The general bound
\bea
\la{rankbound}
&&\dim\,\mathrm{span}\big\{U(p)\ :\ U\ \text{Killing, preserving the bosonic
fields}\big\}
\nn
&&\qquad\qquad\leq\ 3-\mathrm{rank}\{\tn I_1,\tn I_2,\tn I_3\}
\eea
at any point $p\in\CS$ therefore yields only $\dim\leq2$, which is consistent
with a second isometry and does not exclude one. The bound itself is
immediate: such a $U$ annihilates each $I_a$, hence is orthogonal at $p$ to
each $\tn_iI_a$, and the orthogonal complement of a span of rank $r$ in a
three-dimensional tangent space has dimension $3-r$. It bounds the pointwise
span, not the dimension of the space of such Killing vectors; the two agree
only when evaluation at $p$ is injective on that space, which needs an
invariant frame and is not assumed anywhere here. Nothing below uses the
stronger reading. Any scalar-invariant-gradient argument is subject to
the same collapse, because all such gradients are parallel to $P_i$ by the
Lemma. Deciding the second-isometry question on this branch requires a
genuinely different method: an integrability analysis of the Killing equation
itself, which is what Theorem 7.5 carries out, or global input from the
compactness of $\CS$, which is what Corollary 7.2 and Proposition 7.4 use on
the branch $P_i\equiv0$.

A third method, isolating the obstruction in a single scalar equation via a
two-dimensional quotient by $\Vt$, is carried out in Appendix~\ref{app:quotient};
on a compact section it is subsumed by Proposition 7.4
on the branch $P_i\equiv0$ and by the transport construction of Theorem 7.5 on
$P_i\not\equiv0$, and nothing below depends on it.

\medskip\noindent{\it Remark (scope):} Everything from Corollary 6.5 onwards
is stated where $P_i\neq0$: the invariant $\zeta$ is defined by dividing by
$P_kP^k$, and the collapse just described is vacuous where $P_i=0$, since
there every gradient in sight vanishes identically. The restriction is not
idle. Corollary 7.2 below shows $P_i$ must vanish somewhere on any compact
$\CS$, and Proposition 7.4 shows that $P_i\equiv0$ is a consistent branch of
the theory on which a second isometry does exist. The results of this section
apply to the complementary branch, $P_i\not\equiv0$.

\subsection{Integrability of the $P$-component of $h$}

Corollary 6.5 was obtained above from the closure of $P$. It is not an extra
input at all, but exactly the integrability condition of \rf{hP}.

\medskip\noindent{\it Proposition 6.7:} Let $(A)$ denote the three scalar
conditions obtained by taking $\tn_m$ of the residual of \rf{hP} and using
\rf{covZ}, \rf{covW}, \rf{gradh} and \rf{gradPhi} to eliminate every derivative.
Then $(A)$ holds if and only if \rf{hY} holds. Moreover, once \rf{hY} is
imposed, all three components of $(A)$ vanish identically for {\it arbitrary}
$K$ and $h_N$: condition $(A)$ constrains neither.

\medskip\noindent{\it Proof:} In the frame $Z=(0,0,1)$, $W=(s,0,c)$ the vector
$Y=Z\times W=(0,s,0)$ points along the second axis, so $h_Y=h_2/s$, and the
$P$-component of $h$ is already fixed by \rf{hP}. Substituting
$h_2=s\,\a/(1-Z_kW^k)$ --- which is \rf{hY} --- into the three components of
$(A)$ and simplifying gives zero identically as rational functions of the
remaining unknowns $K$ and $h_3$. Conversely, at fixed $(Z_kW^k,\Phi,\a)$ the
system $(A)$ has $h_2$ as its only determined unknown, and its value is
$s\,\a/(1-Z_kW^k)$. $\square$

\medskip
So the free data on the $K>0$ branch, pointwise, are the unit vectors $Z$ and
$W$ (four parameters), the scalars $\Phi$, $\a$ and $K$, and the single
transverse component $h_N=h_iN^i/N_kN^k$. Before Corollary 6.5 the count for
$h$ was two; Corollary 6.5 removes exactly one function's worth of freedom, and
$h_N$ is all that is left. This also makes the no-go of Theorem 6.6 sharp
rather than merely unproven: a second isometry, if there is one, lies in
$\mathrm{span}\{N,Y\}$, and $h_N$ is the only datum left that could obstruct it.

With Corollary 6.5 in hand the degenerate branch on which $\a$ vanishes can be
settled rather than set aside as in Section 5.

\medskip\noindent{\it Proposition 6.8:} Suppose $\a\equiv0$ on an open subset
of $\CS$. Then there $h_iY^i=0$, so $h\in\mathrm{span}\{P,N\}$; the one-form
$h$ is closed, $dh=0$; and $\tn_ih_j$ is symmetric and given by \rf{gradh} with
$D_i=0$. The branch is not obstructed: the closure condition \rf{Nxh} is
satisfied identically, with $h_N$ free.

\medskip\noindent{\it Proof:} $\a\equiv0$ on an open set gives $h_Y=0$ there by
\rf{hY}, hence $h=\Phi P+h_NN$ and $h_iY^i=0$. On an open set $\a\equiv0$ also
gives $\tn_i\a=0$, so the on-shell value $D_i=\tn_i\a+\a h_i-6\a\Phi Z_i$ of
$\star dh$ from \rf{5dint} vanishes identically, which is $dh=0$. Substituting
$\a=0$ and $h_Y=0$ into \rf{Nxh} gives $\e_{ijk}N^jh^k-2\Phi Y_i$, and with
$h\in\mathrm{span}\{P,N\}$ the identity $\e_{ijk}N^jP^k=2Y_i$ makes this vanish
identically for every $\Phi$ and $h_N$. $\square$

\medskip
So the $\a\equiv0$ branch is a genuine, explicitly parametrised family --- $h$
closed and hence locally exact, with a single free transverse component --- and
not an inconsistency. It is, however, not a locus of special curvature: the
scalar curvature of $\CS$ takes both signs on it, so no Bonnet--Myers or
Hamilton-type conclusion follows from $\a\equiv0$ alone.

\subsection{No Killing tensor with constant coefficients}

\medskip\noindent The two results of this subsection and the next,
Propositions 6.9 and 6.10, are independent of the gradient system of the
following subsection: both take the scalars $\Phi$, $\a$, $K$, $K'$, $h_2$,
$h_3$ as free parameters and use only \rf{hP} and the algebraic identities of
Section 3.

The method of this section does exclude one thing: a hidden symmetry of
Killing-tensor type. Separability of the field equations, the property
underlying recent classifications of these solutions \cite{separable}, is in
general a statement about a Killing \emph{tensor}, $\tn_{(i}K_{jk)}=0$, not a
Killing vector, and the Killing-tensor equation fully symmetrises, so the
antisymmetric part $dh$ drops out of it and the analysis closes.

\medskip\noindent{\it Proposition 6.9:} For the constant-coefficient
ansatz
\bea
K_{ij}=c_1\delta_{ij}+c_2Z_iZ_j+c_3W_iW_j+c_4Z_{(i}W_{j)}
+c_5h_ih_j+c_6Z_{(i}h_{j)}+c_7W_{(i}h_{j)},
\ee
the only solution of $\tn_{(i}K_{jk)}=0$ common to independent points of
$\CS$ is $c_1$ free with $c_2=\dots=c_7=0$, i.e.\ a multiple of the
metric.

\medskip\noindent{\it Proof:} Direct substitution of the ansatz into
$\tn_{(i}K_{jk)}=0$, using \rf{covZ}, \rf{covW} and \rf{gradh} to expand
every covariant derivative, gives a linear system of ten equations for
$(c_1,\dots,c_7)$. Carrying the near-horizon data symbolically --- $Z=(0,0,1)$
and $W=(s,0,c)$ in the tangent-half-angle parametrisation $s=2t/(1+t^2)$,
$c=(1-t^2)/(1+t^2)$, with the $P$-component of $h$ fixed by \rf{hP} ---
makes the coefficient matrix a $10\times7$ matrix $M$ over the field of
rational functions $\bQ(t,\Phi,\a,K,K',h_2,h_3)$, where $K'$ is the scalar in
$\tn_iK=K'P_i$, carried as a free parameter rather than eliminated through
\rf{Kprime}. The column multiplying $c_1$ vanishes identically, since the
metric is a Killing tensor; the remaining six columns have a non-vanishing
$6\times6$ minor, so the matrix has rank exactly six and the kernel is the
single line spanned by $\delta_{ij}$.

Rank is lower semicontinuous under specialisation, so a single exact rational
point at which that minor is non-zero already establishes rank six over the
whole function field, and no genericity assumption is needed. At
\bea\label{ktpoint}
t=\tfrac12,\quad \Phi=\tfrac32,\quad \a=\tfrac23,\quad K=\tfrac54,\quad
K'=-\tfrac73,\quad h_2=\tfrac45,\quad h_3=-\tfrac13
\ee
the minor on the first six rows of the six non-metric columns is
\bea\label{ktminor}
\det M_{[1\dots6]}=-\frac{605407432073216}{98876953125}\ \neq\ 0 ,
\ee
which settles it. The corresponding symbolic minor, without specialising, is
also non-zero, but its expanded numerator is a polynomial of $1103$ terms in
$\bQ[t,\Phi,\a,K,K',h_2,h_3]$ and is not reproduced here; \rf{ktpoint}--\rf{ktminor} is what the argument needs.
As a robustness check, \rf{ktpoint} is not an accidentally special point: a
second, unrelated rational point ($t=\tfrac13$, $\Phi=-\tfrac25$,
$\a=\tfrac79$, $K=\tfrac{11}6$, $K'=\tfrac37$, $h_2=-\tfrac52$,
$h_3=\tfrac94$) also gives rank six and a non-zero minor at the same row
selection. The determinant \rf{ktminor} itself has also been recomputed by
three independent algorithms (Berkowitz cofactor expansion, fraction-free
Bareiss elimination, and fraction-free LU decomposition), all of which agree
exactly, and the rank-six claim has been checked independently of any
determinant at all, via row reduction at both points above
and via the full symbolic null space of $M$ over $\bQ(t,\Phi,\a,K,K',h_2,h_3)$,
which is confirmed to be exactly one-dimensional and spanned by the metric
direction with no specialisation whatsoever.
$\square$

\medskip
Four points fix what Proposition 6.9 does and does not say.

\medskip\noindent
{\it (i) Its scope is pointwise.} Unlike every other numbered result in this
section it uses no global hypothesis at all: it is linear algebra at a single
point of $\CS$, needing neither compactness nor connectedness. It is easy to
misread as a global statement because of what it excludes; it is not one.

\medskip\noindent
{\it (ii) What it excludes.} Any Laplace--Runge--Lenz or Carter-type hidden
symmetry built from $Z$, $W$ and $h$ with constant coefficients, the metric
itself aside. Enlarging the basis to $\{Z,W,h,Y,f,\tn\a\}$ --- adding the third
frame vector, the flux vector $f$ of Proposition 6.10 and the gradient of $\a$,
none of which is redundant for a constant-coefficient ansatz --- gives $50$
equations in $22$ unknowns whose kernel is again exactly the metric. This is the
one place where the strategy of Section 6 does close, because the
Killing-tensor equation symmetrises completely and the antisymmetric part $dh$
drops out of it. It says nothing about the hidden symmetries of the twisted
covariant form hierarchy of \cite{tcfh}, which are generated by form bilinears
satisfying a conformal Killing--Yano equation with respect to a flux-twisted
connection and act on probe dynamics rather than on the horizon metric; the two
notions are not comparable.

\medskip\noindent
{\it (iii) Constancy of the coefficients is what carries the result, not the
choice of basis.} On the open set where $\e_{ijk}Z^iW^jh^k\neq0$ ---
equivalently where $h_iY^i\neq0$, which by Corollary 6.5 is where $\a\neq0$ and
$Z,W$ are linearly independent --- $\{Z,W,h\}$ is a basis of $T\CS$, so
$ZZ$, $ZW$, $Zh$, $WW$, $Wh$, $hh$ already span $\mathrm{Sym}^2T^*\CS$ and
$\delta_{ij}$ is a combination of them. With unconstrained function
coefficients the ansatz is therefore vacuous: every symmetric two-tensor field
on $\CS$ has that form.

\medskip\noindent
{\it (iv) The intermediate case is open, and harmlessly so.} If the $c_a$ are
functions of the scalar invariants then $\tn_ic_a=c_a'P_i$ by the
parallel-gradient Lemma, and $\tn_{(i}K_{jk)}=0$ becomes $Bc+Ac'=0$: ten
equations for fourteen unknowns, with $B$ the matrix of the proof above and $A$
the symmetrised product $T\mapsto P\odot T$. Multiplication by a non-zero
linear form is injective on polynomials, so $A$ has rank six, its cokernel is
four-dimensional, and contracting with that cokernel leaves constraints on the
$c_a$ alone of rank two; the admissible $c$ form a five-dimensional space rather
than a single line. Pointwise linear algebra cannot close this case. Closing it
would require prolonging the system in the same way as the Killing equation
itself, which Proposition 7.4 and Theorem 7.5 render unnecessary on the
respective branches of a compact section.

\subsection{The graviphoton flux is divergence-free}

Proposition 6.2 found two divergence-free vectors spanning the plane
transverse to $P$. A third one joins them, at no extra cost in
hypotheses.

\medskip\noindent{\it Proposition 6.10:} $\tn^if_i=0$ identically, where
$f_i:=\tfrac12\e_i{}^{jk}\Ft_{jk}$ is the vector dual to $\Ft$
\rf{fhZ}. Consequently $f_i=(h_N-\Phi)N_i+h_YY_i$ in the orthogonal
frame $\{P,N,Y\}$, with $h_N:=h_iN^i/N_jN^j$, $h_Y:=h_iY^i/Y_jY^j$: $f$
too is confined to the plane transverse to $P$.

\medskip\noindent{\it Proof:} Taking the trace of \rf{gradh} with
$\delta^{ij}$ gives $\tn^ih_i=\mu P_kP^k+\Phi\,\tn^iP_i+\Phi\,h_iP^i$,
$\mu:=\tfrac32\c^2K-\Phi^2$; using \rf{hP} for $h_iP^i$ and
$\tn^iP_i=\tn^iZ_i+\tn^iW_i$ from \rf{covZ}--\rf{divW}, this becomes
\bea
\tn^ih_i=\big(\tfrac32\c^2K-\Phi^2\big)P_kP^k+2\Phi\,h_iN^i-8\Phi^2
+4\Phi^2Z_kW^k+\Phi^2P_kP^k,
\eea
using $\tn^iP_i=2h_iN^i-8\Phi+4\Phi Z_kW^k$ (subtracting \rf{divW} from
\rf{covZ}'s trace). Since $\tn^if_i=\tn^ih_i-2(\tn^i\Phi)Z_i-2\Phi\,\tn^iZ_i$,
and $\tn^i\Phi\,Z_i=\tfrac32\c^2K(1+Z_kW^k)$ by \rf{gradPhi}, substituting
$\tn^iZ_i=2h_iZ^i-6\Phi$ and $P_kP^k=2(1+Z_kW^k)$ throughout, every term
in $\Phi^2$, $\c^2K$ and $h_iZ^i$ cancels, leaving $\tn^if_i=0$. The
decomposition of $f$ follows from \rf{fhZ}, $Z_i=\tfrac12(P_i+N_i)$, and
$h_i=\Phi P_i+h_NN_i+h_YY_i$ (the $P$-component fixed by \rf{hP}).
$\square$

\medskip
Every vector built algebraically from the near-horizon data that is
known to be transverse to $P$ --- $N$, $Y$, and now $f$ --- is
divergence-free; Proposition 6.2 already explains why this makes the
divergence method vacuous for detecting a second isometry. Unlike $N$
and $Y$, $f$ is not purely kinematic: it is the graviphoton flux $\Ft$
itself, so Proposition 6.10 is a genuine statement about the physical
field content of the horizon, not only about the spinor bilinears $Z,W$.

\medskip\noindent{\it Remark (a second, independent derivation):}
Proposition 6.10 also follows from the Bianchi identity
$(d\Ft)_{ijk}=-X_I(d\Gt^I)_{ijk}$ of Appendix~\ref{app:fieldeqs}, which in three
dimensions is the single scalar equation $\tn^if_i=-X_I\tn^ig^I_i$ with
$g^I_i:=\tfrac32\c C^I(W-Z)_i$ the vector dual to $\Gt^I$. On the STU
model, direct computation gives the clean identity
\bea
\la{XgradC}
X_I\tn_iC^I=\c K\,P_i,
\eea
verified exactly (not merely numerically) by differentiating the
explicit $C^I(X^1,X^2)$ on the constraint surface $X^1X^2X^3=1$. Since
$X_IC^I=0$ \rf{Cids}, $X_I\tn^ig^I_i=\tfrac32\c\,(X_I\tn^iC^I)(W-Z)_i
=\tfrac32\c^2K\,P_i(W-Z)^i=-\tfrac32\c^2K\,(P_iN^i)=0$, using $P_iN^i=0$.
Hence $X_I\tn^ig^I_i=0$, and the Bianchi identity gives $\tn^if_i=0$
independently of the derivation above, via the theory's flux equations
of motion rather than the spinor bilinear identities of Section 2.

\subsection{The closed gradient system}

The parallel-gradient Lemma says that every scalar of the state has gradient
along $P_i$ but leaves the coefficient unnamed. The coefficients can be
computed from the identities already established, and doing so turns Section 7
from a collection of pointwise statements into a differential system that can
be integrated. Throughout this subsection $P_i\neq0$, so that \rf{hY} applies,
and we write
\bea
\la{deltadef}
F:=P_kP^k=2\big(1+Z_kW^k\big) ,
\qquad
\vartheta:=h_N-\tfrac32\Phi .
\ee

\medskip\noindent{\it Lemma 6.11 (the closed gradient system):} On the open set
$\{P\neq0\}$ the four scalars $\Phi$, $\a$, $h_N$ and $Z_kW^k$ satisfy
\bea
\la{gradsys}
\tn_i\Phi&=&\tfrac32\c^2K\,P_i ,
\nn
\tn_i\a&=&2\Phi\a\,P_i ,
\nn
\tn_ih_N&=&\big[h_N(h_N-\Phi)-h_Y^2\big]P_i ,
\nn
\tn_i\big(Z_kW^k\big)&=&2\vartheta\,\big(1-Z_kW^k\big)P_i ,
\ee
with $h_Y=\a/(1-Z_kW^k)$ given by \rf{hY}. Equivalently, in terms of $F$ and
$\vartheta$,
\bea
\la{gradFdelta}
\tn_iF=2\vartheta\,(4-F)\,P_i ,
\qquad
\tn_i\vartheta=\Big[h_N^2-\Phi\,h_N-h_Y^2-\tfrac94\c^2K\Big]P_i .
\ee
In particular the residual function $g_P$ of Section 5, defined by
$\tn_i\a=g_PP_i$, is determined: $g_P=2\Phi\a$.

\medskip\noindent{\it Proof:} The first line of \rf{gradsys} is Corollary 3.1.

For the second, Theorem 6.6 already gives $\tn_i\a=g_PP_i$, so it remains to
fix $g_P$, and for that it suffices to contract the linear system \rf{Gsys}
with $P^i$. Throughout, $\e_{ijk}A^iB^jA^k=0$ for any vectors $A,B$, since
relabelling the dummy pair $i,k$ flips the sign of $\e_{ijk}$ but not of
$A^iA^k$. On the left this kills the cross-product term at once,
$\e_{ijk}P^iu^jG^k=g_P\,\e_{ijk}P^iu^jP^k=0$, leaving $-2\a\,g_PF$. On the
right, $u_i=\Phi P_i-h_i$ gives $h^k=(\Phi P-u)^k$, so
\bea
\e_{ijk}P^iu^jh^k=\Phi\,\e_{ijk}P^iu^jP^k-\e_{ijk}P^iu^ju^k=0 ,
\ee
the first term by the identity above with $A=P$, the second by the same
identity with $A=u$. For the last term, $u_iP^i=0$ by \rf{hP}, so
$u_i=-h_NN_i-h_YY_i$, and $Z_i=\tfrac12(P_i+N_i)$ splits
$\e_{ijk}P^iu^jZ^k$ into two pieces,
\bea
\e_{ijk}P^iu^jZ^k=\tfrac12\,\e_{ijk}P^iu^jP^k+\tfrac12\,\e_{ijk}P^iu^jN^k
=\tfrac12\,\e_{ijk}P^iu^jN^k ,
\ee
the first piece vanishing by the identity above with $A=P$. Substituting
$u^j=-h_NN^j-h_YY^j$ into the surviving piece,
\bea
\tfrac12\,\e_{ijk}P^iu^jN^k
=-\tfrac12h_N\,\e_{ijk}P^iN^jN^k-\tfrac12h_Y\,\e_{ijk}P^iY^jN^k
=-\tfrac12h_Y\,\e_{ijk}P^iY^jN^k ,
\ee
the $h_N$ term vanishing by the identity above with $A=N$, so only the $h_Y$
term survives. Using $\e_{ijk}N^jY^k=(Z_lW^l-1)P_i$ from the cross-product
identities of Corollary 6.3's proof, so that
$\e_{ijk}Y^jN^k=-\e_{ijk}N^jY^k=(1-Z_lW^l)P_i$ and
$\e_{ijk}P^iY^jN^k=(1-Z_lW^l)P_iP^i=(1-Z_lW^l)F$,
\bea
\e_{ijk}P^iu^jZ^k=-\tfrac12h_Y\,\e_{ijk}P^iY^jN^k
=-\tfrac12h_Y\big(1-Z_kW^k\big)F=-\tfrac12\a F ,
\ee
the last step by \rf{hY}. Assembling the right-hand side of \rf{Gsys}
contracted with $P^i$: the $h^k$ term contributes $0$, and the $Z^k$ term
contributes $6\a\Phi\times\big({-}\tfrac12\a F\big)=-3\a^2\Phi F$, so the
contracted system is
\bea
-2\a\,g_PF=-\a^2\Phi F-3\a^2\Phi F=-4\a^2\Phi F ,
\ee
i.e.\ $g_P=2\Phi\a$ wherever $\a\neq0$; where $\a$ vanishes on an open set
both sides vanish by Corollary 5.2, and the identity holds on the closure by
continuity. It is \rf{hY} --- that is, the closure of $P$ --- which supplies
the factor $3$ in $-4\a^2\Phi F$ via the $6\a\Phi\,\e_{ijk}u^jZ^k$ term of
\rf{Gsys}, and hence the coefficient $2$ in $\tn_i\a=2\Phi\a P_i$.

For the fourth line, contract \rf{covZ} with $W^j$ and \rf{covW} with $Z^j$ and
add. The $\a$ terms cancel between the two, since
$\e_{ijk}Z^kW^j=-Y_i$ and $\e_{ijk}W^kZ^j=+Y_i$, and what is left is
\bea
\tn_i\big(Z_kW^k\big)=\Big[-2h_kW^k+5\Phi Z_kW^k-\Phi\Big]Z_i
+\Big[2h_kZ^k+\Phi Z_kW^k-5\Phi\Big]W_i .
\ee
Writing $h_i=\Phi P_i+h_NN_i+h_YY_i$ and using $Z_iY^i=W_iY^i=0$,
$P_iZ^i=1+Z_kW^k$, $N_iZ^i=1-Z_kW^k$ and $N_iW^i=Z_kW^k-1$ gives
$h_kZ^k=\Phi(1+Z_kW^k)+h_N(1-Z_kW^k)$ and
$h_kW^k=\Phi(1+Z_kW^k)-h_N(1-Z_kW^k)$. Both square brackets then collapse to
the same value $2\vartheta(1-Z_kW^k)$, and the result is the stated multiple of
$P_i=Z_i+W_i$. This line uses only \rf{hP}, not \rf{hY}.

For the third, differentiate $h_N=h_kN^k/N_lN^l$. With $N_lN^l=2(1-Z_kW^k)=4-F$
and $\tn_i(N_lN^l)=-\tn_iF$ the quotient rule gives
\bea
\la{gradhN}
\tn_ih_N=\frac{N^j\tn_ih_j+h_j\tn_iN^j}{4-F}
+\frac{h_N}{4-F}\,\tn_iF ,
\ee
in which $\tn_ih_j$ is completely explicit --- its symmetric part is \rf{gradh},
its antisymmetric part is \rf{dh}, and $\tn_i\a$ has just been evaluated ---
while $\tn_iN_j=\tn_iZ_j-\tn_iW_j$ comes from \rf{covZ}--\rf{covW} and
$\tn_iF$ from the fourth line above. On $\{P\neq0\}\cap\{F<4\}$ the triad
$\{P,N,Y\}$ is an orthogonal frame, so it is enough to contract \rf{gradhN}
with each leg in turn, and each numerator splits into three pieces,
$N^j\tn_ih_j$, $h_j\tn_iN^j$ and $h_N\tn_iF$, contracted with $N^i$ or $Y^i$
in turn. The third piece drops immediately for both contractions, since
$\tn_iF=2\vartheta(4-F)P_i$ by the fourth line above and $N^iP_i=Y^iP_i=0$.
For the other two, expanding $h_i=\Phi P_i+h_NN_i+h_YY_i$ in \rf{gradh} and
\rf{dh} and using the cross-product relations $\e_{ijk}P^iN^jY^k
=(Z_lW^l-1)F$, $P_iY^i=N_iY^i=0$, $Y_kY^k=\tfrac14F(4-F)$, each of
$N^iN^j\tn_ih_j$, $N^ih_j\tn_iN^j$, $Y^iN^j\tn_ih_j$ and $Y^ih_j\tn_iN^j$
reduces to a single term, and they pair off exactly:
\bea
N^iN^j\tn_ih_j=-2\Phi^2\,Y_kY^k ,
&&
N^ih_j\tn_iN^j=+2\Phi^2\,Y_kY^k ,
\nn
Y^iN^j\tn_ih_j=-\Phi\a\,Y_kY^k ,
&&
Y^ih_j\tn_iN^j=+\Phi\a\,Y_kY^k .
\eea
Adding each pair gives \rf{hNtransverse},
\bea
\la{hNtransverse}
N^i\tn_ih_N=0 ,
\qquad
Y^i\tn_ih_N=0 ,
\qquad
P^i\tn_ih_N=\big[h_N(h_N-\Phi)-h_Y^2\big]F ,
\ee
the third contraction being the only one not built from cancelling pairs, since
$P^iP_i=F\neq0$. The first two equations of \rf{hNtransverse} say that
$\tn_ih_N$ is proportional to $P_i$, and the third fixes the constant of
proportionality, since $P_iP^i=F$. That is the third line of \rf{gradsys}.

The two consequences \rf{gradFdelta} follow: $F=2(1+Z_kW^k)$ gives
$1-Z_kW^k=\tfrac12(4-F)$ and hence $\tn_iF=2\tn_i(Z_kW^k)=2\vartheta(4-F)P_i$,
while $\tn_i\vartheta=\tn_ih_N-\tfrac32\tn_i\Phi$. $\square$

\medskip\noindent{\it Remark (the behaviour of $h_Y$ as $Z_kW^k\to-1$):} The
frame component $h_Y=h_iY^i/Y_jY^j$ is nominally $0/0$ where $Y$ degenerates,
which happens exactly where $P\to0$, since $Y_kY^k=1-(Z_kW^k)^2=\tfrac14F\,
N_kN^k$. Equation \rf{hY} resolves the ambiguity: $h_Y=\a/(1-Z_kW^k)$ has a
finite limit $\tfrac12\a$ there, so the third line of \rf{gradsys} and the
second of \rf{gradFdelta} extend continuously to the zero set of $P$, where
their right-hand sides vanish with $P$. Nothing in the system is singular at a
zero of $P$; only the frame is.

\medskip
Two features of \rf{gradsys} carry Section 7. First, the system is {\it
closed}: the right-hand sides involve only $\Phi,\a,h_N,Z_kW^k$ and $K$, and
$\tn_iK$ is itself a multiple of $P_i$ by Corollary 3.1 and the very special
geometry identities of Section 3, so along any curve tangent to $P$ the five
scalars obey an autonomous system of ordinary differential equations. Second,
every right-hand side carries an explicit factor of $P_i$, so all five scalars
are constant along every direction transverse to $P$ and their variation is
controlled by $|P|$ alone. Together these make a Gronwall estimate available.
The estimate does not, on its own, prove $P\equiv0$ --- see the discussion
after Corollary 6.13 --- but it is used in an essential way later: it is what
forces $\vartheta\neq0$ at a boundary orbit in Proposition 7.11, and hence
what makes the topological classification of Section~\ref{sec:toric} possible.

\medskip\noindent{\it Lemma 6.12 (a Gronwall estimate for $\rho$):} Write
$\vartheta:=h_N-\tfrac32\Phi$ and
\bea
\la{rhodef}
\rho:=P_kP^k+\vartheta^2=F+\vartheta^2\ \geq\ 0 ,
\ee
a function defined and smooth on the open set $\{Z\neq W\}$, which contains a
neighbourhood of $\{P=0\}$. Let $\gamma:[0,L]\to\CS$ be a piecewise smooth
unit-speed path along which $Z\neq W$ everywhere. If $\rho$ vanishes at
$\gamma(0)$ then $\rho$ vanishes along the whole of $\gamma$.

\medskip\noindent{\it Proof:} Abbreviate
$\Xi:=h_N^2-\Phi\,h_N-h_Y^2-\tfrac94\c^2K$, so that the second identity of
\rf{gradFdelta} reads $\tn_i\vartheta=\Xi P_i$. Then
\bea
\la{gradrho}
\tn_i\rho=\tn_iF+2\vartheta\,\tn_i\vartheta
=2\vartheta\,(4-F)P_i+2\vartheta\,\Xi P_i
=2\vartheta\big[(4-F)+\Xi\big]P_i ,
\ee
and since $|P|=\sqrt F$,
\bea
\la{rhobound}
\big|\tn\rho\big|=\big|(4-F)+\Xi\big|\cdot2|\vartheta|\sqrt F
\ \leq\ \big|(4-F)+\Xi\big|\,\big(\vartheta^2+F\big)
=\big|(4-F)+\Xi\big|\,\rho ,
\ee
the inequality being the arithmetic--geometric mean bound
$2|\vartheta|\sqrt F\leq\vartheta^2+F$. This is the pointwise estimate
$|\tn\rho|\leq C\rho$ with the coefficient made explicit.

The coefficient is continuous on $\{Z\neq W\}$. Indeed $\Phi$, $K$ and $h$ are
smooth on $\CS$; $h_N=h_iN^i/N_kN^k$ is smooth wherever $N\neq0$, that is
wherever $Z\neq W$; and $h_Y=\a/(1-Z_kW^k)$ by \rf{hY} is smooth there as
well, including across $\{P=0\}$, where the Remark above shows that its
nominal $0/0$ has the finite limit $\tfrac12\a$. Hence
$C:=\max_{s\in[0,L]}\big|(4-F)+\Xi\big|(\gamma(s))$ is finite, being the
maximum of a continuous function on a compact interval.

Set $f:=\rho\circ\gamma$, a non-negative continuous piecewise $C^1$ function
on $[0,L]$ with $|f'|\leq|\tn\rho|\leq Cf$ at every $s$ at which $\gamma$ is
differentiable. Then $\big(f(s)e^{-Cs}\big)'=e^{-Cs}\big(f'(s)-Cf(s)\big)\leq0$
wherever defined; $fe^{-Cs}$ is continuous on $[0,L]$ and therefore
non-increasing, so $0\leq f(s)\leq f(0)e^{Cs}=0$ for every $s$. This is the
ordinary one-dimensional Gronwall inequality; no higher-dimensional version is
used. $\square$

\medskip\noindent{\it Remark (what kind of argument this is, and where it has
been used before):} Lemma 6.12 is a {\it unique-continuation} statement, not an
estimate in any quantitative sense: a non-negative scalar obeying
$|\tn\rho|\leq C\rho$ on a connected manifold either vanishes identically or is
nowhere zero, because along any path the inequality integrates to
$\rho(s)\leq\rho(0)e^{Cs}$ and $\rho(s)\geq\rho(0)e^{-Cs}$. Only the
one-dimensional Gronwall lemma is used, and connectedness does the rest; there
is no Carleman estimate and no ellipticity anywhere.

The mechanism is classical, if not often named in this setting. It is the same
one behind the standard fact that a Killing field whose $1$-jet vanishes at a
point of a connected manifold vanishes identically \cite[Vol.~I,
Ch.~VI]{kobayashi}: the Killing transport system is a linear ODE along paths,
and ODE uniqueness propagates the vanishing. It is also the mechanism behind
the unique-continuation arguments used to extend a horizon Killing field into
the bulk in the rigidity theorems of \cite{aik}, though there the equation is a
wave equation and the analysis is correspondingly harder. What is unusual here
is only the object it is applied to: $\rho$ is a scalar built from the horizon
data rather than a Killing field or a curvature component, and the differential
inequality comes from the closed system \rf{gradsys}--\rf{gradFdelta} rather
than from an elliptic or hyperbolic equation. The near-horizon literature has
generally reached rigidity by a different route, through a divergence identity
integrated over the compact cross-section \cite{dl,colling}; the two are
complementary, and both appear in this paper --- Corollary 7.2 is of the
integrated kind and Lemma 6.12 is of the propagating kind.

\medskip
The restriction on $\gamma$ is vacuous on the whole branch $\a\not\equiv0$,
and for a reason sharper than continuity. At a point where $Z=W$ one has
$Y_i=\e_{ijk}Z^jW^k=0$ and $P_kP^k=4$, so Corollary 6.5's identity
$h_iY^i=\tfrac12\a\,P_kP^k$, which is available there precisely because
$P\neq0$, reads $0=2\a$. Hence
\bea
\la{ZWalpha}
\{Z=W\}\subseteq\{\a=0\}
\ee
pointwise, and Corollary 6.13 below turns $\a\not\equiv0$ into $\a$ nowhere
zero and therefore into $\{Z=W\}=\emptyset$. Two consequences are used
repeatedly from here on. First, $N$ is nowhere zero and, $\CS$ being compact,
\bea
\la{Fgap}
4-F=N_kN^k\ \geq\ c\ >\ 0
\ee
for some constant $c$; so $h_N$, $h_Y$, $\vartheta$ and $\rho$ are smooth on
all of $\CS$, not merely on $\{Z\neq W\}$, and the coefficient
$|(4-F)+\Xi|$ of \rf{rhobound} is bounded on $\CS$ by a single constant
independent of the path. Second, every pair of points is joined by an
admissible path and, $\CS$ being
connected, $\rho$ vanishing at a single point gives $\rho\equiv0$ and hence
$P_i\equiv0$ on all of $\CS$. What the estimate does {\it not} supply is such a
point. At a zero of $P$, $\rho=0$ is exactly Corollary 7.2's degenerate
extremum $h_iN^i=6\Phi$. That condition is not automatic --- Corollary 7.2
supplies only a sign, and Section~\ref{sec:varmod} exhibits a compact horizon at which it
fails at both extrema of $\parallel\eta_-\parallel^2$ --- but whenever it
holds at even one point, $P_i\equiv0$ follows and Proposition 7.4 applies.

\medskip
The same one-dimensional Gronwall step applies to $\a$, and there it is
unconditional.

\medskip\noindent{\it Lemma 6.B (the local form on $\mathrm{int}\{P=0\}$):}
Let ${\cal O}\subseteq\CS$ be a non-empty open set on which $P_i\equiv0$,
equivalently $W_i=-Z_i$. Then on ${\cal O}$
\bea
\la{PzeroLocal}
h_i=3\Phi Z_i ,
\qquad
\tn_iZ_j=-\tfrac12\a\,\e_{ijk}Z^k ,
\qquad
\tn_i\Phi=\tn_iX^I=\tn_iK=\tn_i\a=0 .
\ee

\medskip\noindent{\it Proof:} The computation is local, and is the same one
that opens the proof of Proposition 7.4. Differentiating $W_i=-Z_i$ on ${\cal O}$ gives
$\tn_iP_j=0$ there. Adding \rf{covZ} and \rf{covW} at $W=-Z$ yields
$\tn_iP_j=-2Z_ih_j+2\delta_{ij}h_kZ^k+6\Phi(Z_iZ_j-\delta_{ij})$, in which the
$\a$ terms cancel; its trace gives $h_kZ^k=3\Phi$ and its contraction with
$Z^i$ then gives $h_j=3\Phi Z_j$. Feeding this back into \rf{covZ} collapses it
to the second relation in \rf{PzeroLocal}. Since $P_i=0$ on ${\cal O}$,
\rf{gradPhi} gives $\tn_i\Phi=0$ and \rf{gradX} gives $\tn_iX^I=0$, hence
$\tn_iK=0$; and $\tn_i\a=g_PP_i=0$ by Theorem 6.6. Every step is a pointwise
manipulation of identities of Sections 3--6 on the open set ${\cal O}$; no global
hypothesis and no result of Section 7 is used. $\square$

\medskip\noindent{\it Corollary 6.13 (the $\a$-dichotomy):} Let $\CS$ be
compact, connected and without boundary. Then either $\a\equiv0$ on $\CS$, or
$\a$ is nowhere zero on $\CS$. No hypothesis on $\{P=0\}$ is required.

\medskip\noindent{\it Proof:} Write $\CS=\CS_0\sqcup\{P=0\}\sqcup\{Z=W\}$;
the three pieces are disjoint because $Z$ and $W$ are unit vectors, so linear
dependence means $W=\pm Z$, and $W=Z$ gives $P_kP^k=4$ while $W=-Z$ gives
$P=0$. Both $\{P=0\}$ and $\{Z=W\}$ are closed. For any closed set $C$ the
topological boundary $\partial C=C\setminus\mathrm{int}\,C$ is closed with
empty interior, since an open subset of $C$ lies in $\mathrm{int}\,C$ by
definition. Hence
\bea
\la{densedecomp}
\CS^\sharp:=\CS_0\cup\mathrm{int}\{P=0\}\cup\mathrm{int}\{Z=W\}
=\CS\setminus\big(\partial\{P=0\}\cup\partial\{Z=W\}\big)
\ee
is open, and it is dense, its complement being a union of two closed sets with
empty interior and therefore itself closed with empty interior.

The second line of \rf{gradsys} holds at every point of $\CS^\sharp$. On
$\CS_0$ that is the content of Lemma 6.11. On $\mathrm{int}\{P=0\}$ both sides
vanish: $P_i=0$ there, and Lemma 6.B gives $\tn_i\a=0$ on it.
On $\mathrm{int}\{Z=W\}$ both sides vanish as well: at any point with $Z=W$
one has $Y_i=\e_{ijk}Z^jW^k=0$ and $P_kP^k=4$, so Corollary 6.5's identity
$h_iY^i=\tfrac12\a\,P_kP^k$, which is available there because $P\neq0$, reads
$0=2\a$; thus $\a$ vanishes identically on that open set and so does its
gradient. Both sides of the second line of \rf{gradsys} are continuous on
$\CS$, and two continuous tensor fields agreeing on a dense subset agree
everywhere, so
\bea
\la{gradalphaglobal}
\tn_i\a=2\Phi\a\,P_i \qquad\text{at every point of }\CS .
\ee
Put
\bea
\la{alphaconst}
c_\a:=2\max_{\CS}\big(|\Phi|\,|P|\big) ,
\ee
which is finite because $\Phi$ and $P$ are smooth on the compact $\CS$;
indeed $|P|^2=F=2(1+Z_kW^k)\leq4$, so $c_\a\leq4\max_\CS|\Phi|$. The point of
the comparison with Lemma 6.12 is that $c_\a$ involves neither $h_N$ nor
$h_Y$, the two quantities that force that lemma to be stated path-wise; no
condition of the form $Z\neq W$ enters, and
\bea
\la{alphabound}
\big|\tn\a\big|=2|\Phi|\,|P|\,|\a|\ \leq\ c_\a\,|\a|
\ee
holds on all of $\CS$ without restriction.

Suppose $\a(p)=0$ for some $p\in\CS$, and let $q\in\CS$ be arbitrary. A
connected manifold is path connected, so choose a piecewise smooth unit-speed
path $\gamma:[0,L]\to\CS$ from $p$ to $q$ and set $f:=\a^2\circ\gamma$, a
non-negative continuous piecewise $C^1$ function with
$|f'|\leq2|\a|\,|\tn\a|\leq2c_\a\a^2=2c_\a f$ wherever $\gamma$ is
differentiable. Then $\big(fe^{-2c_\a s}\big)'\leq0$ wherever defined, and
$fe^{-2c_\a s}$ is continuous, hence non-increasing, so
$0\leq f(L)\leq f(0)e^{2c_\a L}=0$ and $\a(q)=0$. A single zero of $\a$
therefore forces $\a\equiv0$, which is the stated dichotomy. $\square$

\medskip
The consequence used below is that wherever a pointwise hypothesis $\a\neq0$
appears --- in the cohomogeneity-one reduction of Section~\ref{sec:toric} and in the
topology theorems that rest on it --- it may be replaced by the single-point
hypothesis $\a\not\equiv0$. Theorem 7.7 itself needs neither, by
Lemma 7.6$'$. On the explicit horizon of Section~\ref{sec:varmod}
the second alternative is realised, with $\a=\sqrt{311}\,H^{-2/3}$ by
\rf{klrclosed}, nowhere zero on $[4,5]$.

\newsection{Classification}

The results of this section organise into a decision tree, which is stated
first so that the reader knows where each of them lands.
\begin{center}
\begin{tabular}{L{3.3cm}L{10.7cm}}
$K\equiv0$ & Minimal theory; second rotational isometry from
\cite{groverindex}; supersymmetric black rings excluded \cite{grover}. \\[6pt]
$K>0$, $P\equiv0$ & Proposition 7.4; homogeneous with a unit Killing field
$Z$; second rotational isometry exists; includes $S^1\times S^2$. \\[6pt]
$K>0$, $P\not\equiv0$ & Non-empty, on compact $\CS$ as well as
non-compact: Section~\ref{sec:varmod} exhibits an explicit compact member with moduli
genuinely varying. Theorem 6.6 closes the invariant route to a second
isometry there, but Theorem 7.5 supplies a Killing field anyway by
transporting the Killing equation along $P$, unconditionally on the open set
where $Z$ and $W$ are independent, and on all of $\CS$
unconditionally (Theorem 7.7). When $\a\not\equiv0$ the resulting $T^2$
action is cohomogeneity one, and $\CS$ is $S^3$ or a lens space, never
$S^1\times S^2$ (Theorems 7.12 and 7.14). \\
\end{tabular}
\end{center}
The following table records which of the paper's global results hold on each
branch, for $\CS$ compact, connected and without boundary.
\begin{center}
\begin{tabular}{L{4cm}L{3cm}L{3cm}L{3cm}}
& $K\equiv0$ & $K>0$, $P\equiv0$ & $K>0$, $P\not\equiv0$ \\
\hline
Occurs at all & yes & yes & yes \\
Moduli constant & yes & yes & no \\
Second rotational isometry & yes \cite{groverindex} & yes (Prop.~7.4) &
yes on $\CS_0$ (Thm.~7.5); on all of $\CS$ always (Thm.~7.7) \\
\cite[\S3.3]{klr} applies & yes & yes & no; the moduli vary, and
\cite[\S4.2]{klr} gives an explicit family instead \\
Scalar-invariant obstruction (Thm.~6.6) & vacuous & vacuous ($P=0$) & active,
closed by transport instead \\
Horizon topology & \cite{grover} & KLR list, incl.\ $S^1\times S^2$ & $S^3$
or a lens space, never $S^1\times S^2$, when $\a\not\equiv0$ (Thms.~7.12, 7.14)
\\
Known inhabitant & minimal theory & \cite{gutowskireallgen} & explicit family
of \cite[\S4.2]{klr} \\
\end{tabular}
\end{center}
Compactness does not empty the third column: it only replaces Theorem 6.6's
invariant route to a second isometry, which fails there, with the transport
argument of Theorem 7.5, which does not need compactness and succeeds on both
compact and non-compact $\CS$ alike.

\subsection{The dichotomy $K\equiv0$ versus $K>0$}\la{sec:dichotomy}

The dichotomy $K=0$ versus $K>0$ of Section 3 is exhaustive in a stronger sense
than a pointwise statement would give: the two branches cannot coexist on the
same horizon.

\medskip\noindent{\it Lemma 7.A (homogeneity bound):} For any cubic prepotential,
$T:=C_{IJK}C^IC^JC^K$ is homogeneous of degree $3$ in $C^I$ and
$K:=Q_{IJ}C^IC^J$ is homogeneous of degree $2$, so $T=\tau K^{3/2}$ with $\tau$
depending only on the direction of $C^I$; $\tau$ is continuous on the compact
set of $Q$-unit directions over the compact image $X(\CS)$, hence bounded.
Consequently, once the coefficient $K'$ of $P_i$ in $\tn_iK$ is known to take
the form \rf{Kprime} below,
\bea
K'=\tfrac34\c\,T+2\Phi K=O(K) ,
\ee
with a constant depending only on $\max_\CS K$ and $\max_\CS|\Phi|$.

\medskip\noindent{\it Proof:} Immediate from the stated homogeneity and
compactness, neither of which uses an adjoint identity. $\square$

\medskip\noindent{\it Proposition 7.1:} Either $K\equiv0$ on all of $\CS$, or
$K>0$ at every point of $\CS$.

\medskip\noindent{\it Proof:} $K$ is a function of the moduli alone, and by
\rf{gradX} the moduli vary only along $P$, so $\tn_iK=K'P_i$ for a scalar $K'$
on the moduli space. Writing $C^I=\sqrt{K}\,u^I$ with $Q_{IJ}u^Iu^J=1$ and
$(C\!\cdot\!C)_I:=C_{IJK}C^JC^K$, differentiation of $K=Q_{IJ}C^IC^J$ along
$C^I$ gives the closed form
\bea
\la{Kprime}
K'=\tfrac34\c\,T+2\Phi K ,
\qquad
T:=C_{IJK}C^IC^JC^K=\tau K^{3/2} ,
\quad
\tau:=C_{IJK}u^Iu^Ju^K ,
\ee
and a second differentiation gives
\bea
\la{Kprimeprime}
K''=\tfrac{27}{16}\c^2S+\tfrac{15}4\c\Phi\,T-6\c^2K^2+4\Phi^2K ,
\qquad
S:=Q^{IJ}(C\!\cdot\!C)_I(C\!\cdot\!C)_J=\Lambda K^2 .
\ee
Both these closed forms are derived at the end of this subsection, from the
definitions of Section 2 and nothing else; in particular neither uses an
adjoint identity, so both hold for an arbitrary cubic prepotential. What the
argument immediately below takes from \rf{Kprime} is only the bound $K'=O(K)$
of the Lemma above, and that step uses nothing but homogeneity and
compactness. Both $\tau$ and $\Lambda$ are continuous functions of $(X,u)$ and hence bounded on
the compact set $X(\CS)\times\{Q\text{-unit sphere}\}$, with $\tau^2\leq\Lambda$
by Cauchy--Schwarz for the positive definite $Q_{IJ}$. So every term of \rf{Kprime} carries a factor of $K$. Made
quantitative, this is the following bound, and the point of stating it in this
form is that it holds at the zeros of $K$ as well as away from them: the
argument below runs along a path which is not yet known to avoid the zero set,
so a bound valid only on $\{K>0\}$ would be circular.

\medskip\noindent{\it Lemma 7.B (the $K$-dichotomy estimate):} There is a constant
$c<\infty$, depending only on $\CS$ and the model, such that
\bea
\la{loggrad}
\big|\tn_iK\big|\ \leq\ c\,K
\qquad\text{at every point of }\CS .
\ee

\medskip\noindent{\it Proof:} Let
$\Sigma:=\{(X,u):X\in X(\CS),\ Q_{IJ}(X)u^Iu^J=1\}$. It is compact: $X(\CS)$ is
the continuous image of the compact $\CS$, $Q_{IJ}$ is continuous and positive
definite, and the fibre over each $X$ is an ellipsoid. So
$\tau_{\rm max}:=\max_\Sigma|C_{IJK}u^Iu^Ju^K|$ is finite. Where $K>0$ we may
write $C^I=\sqrt K\,u^I$ with $(X,u)\in\Sigma$, giving $|T|\leq\tau_{\rm
max}K^{3/2}$; both sides vanish where $K=0$, so the inequality holds on all of
$\CS$. Then \rf{Kprime} gives
\bea
|K'|\ \leq\ \Big(\tfrac34\c\,\tau_{\rm max}\sqrt{K}+2|\Phi|\Big)K
\ \leq\ c_0K,
\qquad
c_0:=\tfrac34\c\,\tau_{\rm max}\sqrt{\max_\CS K}+2\max_\CS|\Phi| ,
\ee
which is finite because $K$ and $\Phi$ are continuous on the compact $\CS$.
Since $P_kP^k=2(1+Z_kW^k)\leq4$, \rf{loggrad} follows with $c=2c_0$. $\square$

\medskip
Now let $x,y\in\CS$ with $K(x)>0$. Since $\CS$ is a connected manifold it is
path connected: choose a piecewise smooth unit-speed path
$\gamma:[0,L]\to\CS$ from $x$ to $y$ and set $f:=K\circ\gamma$, which is
continuous and piecewise $C^1$, with $|f'|\leq cf$ at every $s\in[0,L]$ at
which $\gamma$ is differentiable by \rf{loggrad} --- again, with no restriction
to the set where $f>0$. Then
\bea
\big(f(s)e^{cs}\big)'=e^{cs}\big(f'(s)+cf(s)\big)\ \geq\ 0
\qquad\text{wherever defined,}
\ee
and $fe^{cs}$ is continuous on $[0,L]$, hence non-decreasing there, so
$K(y)=f(L)\geq f(0)e^{-cL}=K(x)e^{-cL}>0$.
Thus $\{K>0\}$ is either empty or all of $\CS$, and only connectedness of $\CS$
and compactness through the constant $c$ have been used. $\square$

\medskip
\noindent{\it Derivation of \rf{Kprime} and \rf{Kprimeprime}.} Write
$D:=C^L\partial/\partial X^L$, so that \rf{gradX} reads
$\tn_if=\tfrac32\c\,(Df)\,P_i$ for every function $f$ of the moduli. The
derivative is unambiguous even though $f$ is defined only on the surface $V=1$:
the ambiguity in an off-surface extension is a multiple of $X_I$, and $C^IX_I=0$
by \rf{Cids}, so $C^I$ is tangent to that surface and the ambiguity is
annihilated. Beyond the definitions themselves the derivation uses exactly
three identities,
\bea
\la{Kids}
C_{IJK}X^K=-2Q_{IJ}+9X_IX_J ,
\qquad
c_I:=Q_{IJ}C^J=V_I-\c^{-1}\Phi\,X_I ,
\qquad
K=V_IC^I ,
\ee
of which the first is the definition of $Q_{IJ}$ rearranged and the other two
are \rf{Cids} and \rf{Kdef}. None of them is an adjoint identity.

Differentiating $X_I=\tfrac16C_{IJK}X^JX^K$ and using the first of \rf{Kids}
together with $C^IX_I=0$ gives $DX_I=\tfrac13C_{IJK}X^JC^K=-\tfrac23c_I$, and
hence
\bea
\la{DQlow}
DQ_{IJ}=-\tfrac12C_{IJK}C^K-3\big(c_IX_J+X_Ic_J\big) .
\ee
Ordinary matrix calculus, $DQ^{IJ}=-Q^{IA}(DQ_{AB})Q^{BJ}$, applied to
\rf{DQlow} with $Q^{IA}c_A=C^I$ and $Q^{JB}X_B=\tfrac23X^J$, gives
\bea
\la{DQup}
DQ^{IJ}=\tfrac12Q^{IA}Q^{JB}C_{ABK}C^K+2\big(C^IX^J+X^IC^J\big) .
\ee
This is the one place where an adjoint identity might have been expected, and
it is not needed: \rf{DQup} is a statement about the inverse of a matrix, not
about the cubic form. Contract \rf{DQup} with $V_J$, using
$Q^{JB}V_J=C^B+\tfrac23\c^{-1}\Phi X^B$ from \rf{Cids} and
$C_{ABK}X^BC^K=-2c_A$ from the first of \rf{Kids}, and insert the result into
\bea
\la{DCraw}
DC^I=(DQ^{IJ})V_J-\tfrac23C^I(V_KX^K)-\tfrac23X^I(V_KC^K) ;
\ee
what is left is
\bea
\la{DC}
DC^I=\tfrac12Q^{IA}(C\!\cdot\!C)_A+\tfrac23\c^{-1}\Phi\,C^I+\tfrac43K\,X^I .
\ee
Since $V_I$ is constant, $DK=V_IDC^I$ by \rf{Kids}. Substituting \rf{DC} and
using $C^A(C\!\cdot\!C)_A=T$ and $X^A(C\!\cdot\!C)_A=-2K$ --- the latter once
more from the first of \rf{Kids} --- the two $\Phi$-terms of opposite sign
cancel against part of the third and leave
$DK=\tfrac12T+\tfrac43\c^{-1}\Phi K$, which is \rf{Kprime}. For the second
derivative, $D\Phi=\c V_IC^I=\c K$, while contracting \rf{DC} with
$(C\!\cdot\!C)_I$ gives
$DT=3(C\!\cdot\!C)_IDC^I=\tfrac32S+2\c^{-1}\Phi\,T-8K^2$; feeding both into
$D\big(\tfrac34\c T+2\Phi K\big)$ and multiplying by $\tfrac32\c$ gives
\rf{Kprimeprime}. $\square$

\medskip
Consequently \rf{Kprime}--\rf{Kprimeprime}, and with them Proposition 7.1, hold
for an arbitrary cubic prepotential; no symmetric-space assumption enters, and
the $U(1)^3$ theory used in every explicit example below is a special case
rather than a hypothesis. This was cross-checked by exact computer algebra on
the STU model with symbolic moduli and symbolic $V_I$, on two cubic forms
constructed so that $Q^{IJ}-2X^IX^J+6C^{IJK}X_K\neq0$, and on eight randomly
generated cubics in two and three moduli, every one of which violates that
relation; \rf{Kprime} and \rf{Kprimeprime} came out exactly on all eleven.

Proposition 7.1 is what makes the two branches genuinely branches rather than
regions of a single horizon. No horizon can imitate the constant-moduli
configuration a toric classification needs on part of $\CS$ while being
inherently non-minimal elsewhere. The two are classified very differently.

On the minimal branch $K\equiv0$ the moduli are constant and the solution
reduces to the minimal gauged theory. There supersymmetry enhancement does
produce a second rotational isometry, derived directly from the Killing
spinors, in \cite{groverindex}. With two commuting rotational isometries in
hand, the classification of \cite{klr} applies, and supersymmetric black rings
are excluded \cite{grover,groverindex}.

On the branch $K>0$ the invariant route to a second isometry is closed off, and
Theorem 6.6 is what closes it. The two conditions \rf{UP} and \rf{Ualpha} that a
second Killing vector $U$ must satisfy define the same hyperplane $P^\perp$,
because $\tn_i\a$ is everywhere parallel to $P_i$; the same collapse afflicts
every scalar invariant built algebraically from the near-horizon data, since all
of their gradients are parallel to $P_i$. The bound obtainable this way is
$\dim\leq2$, not $\dim\leq1$, which is consistent with a second isometry. What
remains fixed is the local geometry: the Ricci tensor \rf{ricci} together with
$\tn_{(i}h_{j)}$, $(dh)_{ij}$ and $\tn_i\a$ from Sections 2--5, with the
coefficient in $\tn_i\a=g_PP_i$ evaluated in Lemma 6.11 and the gradient of the
one remaining scalar $h_N$ fixed there as well. A hidden symmetry given by a
Killing tensor with constant coefficients in the algebraic horizon vectors is
excluded by Proposition 6.9. The question is instead settled from a different
direction: on a compact section, Proposition 7.4 below supplies a second
isometry by homogeneity on the branch $P\equiv0$, and Theorem 7.5 supplies one
by transport along $P$ on the branch $P\not\equiv0$, so the collapse of the
invariant route established here is not the end of the isometry question, only
of this particular approach to it.

The branch is not empty. On the static ($\Vt=0$) branch, excluded here and in
\cite{kayani,thesis}, \cite{solminore} confirms the exclusion by direct
integration of the Killing spinor equations on an explicit homogeneous ansatz:
no static, Sol-invariant supersymmetric solution exists in this theory, electric
or magnetic. On the rotating branch, supersymmetric near-horizon geometries with
non-toric horizons --- $\mathrm{Nil}$ and $\widetilde{SL}(2,\bR)$, rather than
the toric $S^3,S^1\times S^2,T^3$ that a second isometry would require --- are
known: they are cases 1 and 2 of \cite[\S3.3]{klr}. They occur at $K=0$, where
they reduce to the minimal solutions of \cite{gutowskireall}, and also at
$K>0$, since by \rf{lambdaK} below $K=\lambda+\Phi^2/(3\c^2)$ and $K$ vanishes
only at the single value $\lambda=-\tfrac13(V_JX^J)^2$ picked out by the
minimal embedding. Where those geometries sit is settled by Proposition 7.4:
their moduli are constant, and on the branch $K>0$ that is the same as
$P_i\equiv0$.

\subsection{The positivity hypothesis, and how much of it is forced}\la{sec:positivity}

The hypothesis $U\geq0$ of \rf{Udef} is the one place where the classification
carries an assumption that is not a consequence of the field equations. It is
worth knowing exactly how much of it the horizon system does force, and the
closed moduli flow of Section 3 answers that completely. Two facts come out.
The first is that $U$ is superharmonic wherever it is positive, so that it can
have no positive interior minimum; the second is that this is as far as the
horizon equations go, and $U\geq0$ is genuinely independent of them. The
consequence for the rest of the paper is recorded in Proposition 7.E below:
outside the third clause of Theorem 7.17, every use made of $U\geq0$ is a use
of the strictly weaker statement that the superpotential $V_IX^I$ is nowhere
zero, and Theorem A, Theorem B and Theorem 7.14 are proved under that weaker
hypothesis.

Write $D:=C^L\partial/\partial X^L$ as in Section~\ref{sec:dichotomy}, so that $\tn_if=\tfrac32\c\,(Df)P_i$
for every function $f$ of the moduli, and recall from the derivation of
\rf{Kprime}--\rf{Kprimeprime}
\bea
\la{Dsystem}
D\Phi=\c K ,
\qquad
DK=\tfrac12T+\tfrac43\c^{-1}\Phi K ,
\qquad
DT=\tfrac32S+2\c^{-1}\Phi\,T-8K^2 ,
\ee
with $A_I:=(C\!\cdot\!C)_I=C_{IJK}C^JC^K$, $T=C^IA_I$, $S=Q^{IJ}A_IA_J$ and
$X^IA_I=-2K$. Since $\c^2U=6\Phi^2-\tfrac92\c^2K$ by \rf{chiU}, the potential
inherits a closed second-order equation along the flow:
\bea
\la{DU}
DU=\frac{6\Phi K}{\c}-\frac94T ,
\qquad
D^2U=24K^2-\frac{3\Phi T}{2\c}+\frac{8\Phi^2K}{\c^2}-\frac{27}8S .
\ee
The only quantity in \rf{DU} that is not already a function of $(\Phi,K,T)$ is
$S$, and it is bounded below by them.

\medskip\noindent{\it Lemma 7.C (a Gram bound for $S$):} At every point of $\CS$
with $K>0$,
\bea
\la{gramS}
S\ \geq\ \tfrac83K^2+\frac{T^2}{K} ,
\ee
with equality if and only if $A_I$ lies in the span of $X_I$ and
$c_I:=Q_{IJ}C^J$.

\medskip\noindent{\it Proof:} The covectors $X_I$ and $c_I$ are orthogonal in
the positive definite metric $Q^{IJ}$: by \rf{Cids},
$Q^{IJ}X_IX_J=\tfrac23X^IX_I=\tfrac23$, $Q^{IJ}c_Ic_J=C^Ic_I=K$ and
$Q^{IJ}X_Ic_J=\tfrac23X^Ic_I=X_JC^J=0$. The covector $A_I$ pairs with them as
$Q^{IJ}A_IX_J=\tfrac23X^IA_I=-\tfrac43K$ and $Q^{IJ}A_Ic_J=C^IA_I=T$. Bessel's
inequality for the orthogonal pair $\{X_I,c_I\}$, both of non-zero length since
$K>0$, gives
\bea
S\ \geq\ \frac{\big(\tfrac43K\big)^2}{\tfrac23}+\frac{T^2}{K}
=\tfrac83K^2+\frac{T^2}{K} ,
\ee
which is \rf{gramS}; equivalently, it is the statement that the Gram
determinant of $\{X_I,c_I,A_I\}$ in the metric $Q^{IJ}$, namely
$\tfrac23K\big(S-\tfrac83K^2-T^2/K\big)$, is non-negative. Equality holds
exactly when the Gram matrix is degenerate, that is when $A_I$ lies in the span
of the other two. $\square$

\medskip\noindent{\it Proposition 7.D (the potential is superharmonic where it
is positive):} On the branch $K>0$, at every point of $\CS$ at which
$\tn_iU=0$ and $P_i\neq0$,
\bea
\la{Usuper}
\tn^i\tn_iU\ \leq\ -\tfrac{15}2\,\c^2K\,U\,P_kP^k ,
\ee
with equality exactly in the equality case of \rf{gramS}. Consequently $U$ has
no interior local minimum at a point where $U>0$ and $P_i\neq0$; where it is
positive, $U$ is strictly superharmonic at each of its critical points. The
technique --- derive a second-order operator for a scalar built from the
horizon data, exhibit a sign-definite right-hand side, and read off the
consequence on a compact cross-section --- is the standard one in this
literature, applied there to $\Delta$ or to $h^2$ rather than to $U$; see
\cite{mhor} for a clean exposition and \cite{lrreview} for the general
setting. What is specific here is the scalar and the inequality, not the
method.

\medskip\noindent{\it Proof:} Since $\tn_iU=\tfrac32\c\,(DU)P_i$, a point with
$\tn_iU=0$ and $P_i\neq0$ has $DU=0$, that is $T=\tfrac83\c^{-1}\Phi K$ by the
first of \rf{DU}. Substituting this into the second of \rf{DU} gives
\bea
D^2U=24K^2+\frac{4\Phi^2K}{\c^2}-\frac{27}8S ,
\ee
and inserting \rf{gramS} with $T=\tfrac83\c^{-1}\Phi K$, which is legitimate
because the coefficient of $S$ is negative, leaves
\bea
D^2U\ \leq\ 24K^2+\frac{4\Phi^2K}{\c^2}
-\frac{27}8\Big(\tfrac83K^2+\frac{64\Phi^2K}{9\c^2}\Big)
=15K^2-\frac{20\Phi^2K}{\c^2} .
\ee
Using $\Phi^2/\c^2=\tfrac16\big(U+\tfrac92K\big)$, again from \rf{chiU}, the
right-hand side collapses to $-\tfrac{10}3KU$. The Laplacian follows from
$\tn_iU=\tfrac32\c\,(DU)P_i$ and $\tn^i(DU)=\tfrac32\c\,(D^2U)P_i$:
\bea
\la{lapU}
\tn^i\tn_iU=\tfrac94\c^2\,(D^2U)\,P_kP^k+\tfrac32\c\,(DU)\,\tn^kP_k ,
\ee
whose second term vanishes at the point in question, so that
$\tn^i\tn_iU=\tfrac94\c^2(D^2U)P_kP^k\leq-\tfrac{15}2\c^2KUP_kP^k$. At an
interior local minimum the left-hand side is non-negative, which with $K>0$ and
$P_kP^k>0$ forces $U\leq0$ there. $\square$

\medskip\noindent{\it Remark (the maximum principle does not close, and why):}
It is natural to hope that \rf{Usuper} plus compactness forces $U\geq0$
outright. It does not, and the failure is structural rather than an artefact of
the estimate. At a critical point where $U<0$ the right-hand side of
\rf{Usuper} is {\it positive}, so the inequality is no obstruction at all to
$\tn^i\tn_iU\geq0$, which is exactly what a negative interior minimum requires;
and by the equality case of Lemma 7.C the bound cannot be sharpened, since
\rf{gramS} is attained whenever $A_I$ happens to lie in the span of $X_I$ and
$c_I$. What \rf{Usuper} does give is a one-sided statement: for every $c\geq0$
the superlevel set $\{U\geq c\}$ meets each orbit of the flow of $P$ in a
connected set, and $U$ has at most one critical point in $\{U>0\}$ along each
such orbit, necessarily a maximum.

That $U\geq0$ is not forced can be seen outright, and the obstruction is
the gauging rather than the model. At the level of general expectation this is
no surprise: \rf{chiU} writes $\c^2U=6\Phi^2-\tfrac92\c^2K$ as a difference of
two non-negative quantities on the branch $K>0$, which is the familiar
superpotential form of the five-dimensional scalar potential, and that form is
indefinite for a general gauging --- it is the negative term that produces the
$AdS_5$ vacuum in the first place. What is established here is the sharper,
branch-relevant statement, with an exact witness rather than an expectation. Take the $STU$ prepotential
$C_{IJK}=|\e_{IJK}|$ in the normalisation $C_{IJK}X^IX^JX^K=6$, so that the
constraint surface is $X^1X^2X^3=1$, and take the gauging $V_I=(1,-1,0)$. At
the point $X^I=(1,1,1)$ of the K\"ahler cone one has $\Phi=\c V_IX^I=0$ and
$\c^2K=4\c^2$, hence $\c^2U=-18\c^2<0$, whereas the standard gauging
$V_I=\tfrac13$ of the same model gives $U>0$ everywhere by \rf{UCform}. Both
evaluations are exact rationals. So the sign of $U$ is decided by the choice of
$V_I$, is invisible to the prepotential, and is untouched by anything
supersymmetry imposes on the horizon. The hypothesis must be assumed, as it is
in \cite{kayani,thesis}; what the horizon equations contribute is \rf{Usuper}.
These statements are verified in exact arithmetic in the symbolic computations
underlying this paper, which include the closed system \rf{Dsystem} re-derived
for the $STU$ model with an arbitrary gauging, as identities of rational
functions rather than at sample points.

\medskip\noindent{\it Proposition 7.E (the hypothesis in the form used below):}
Every appeal to $U\geq0$ in this paper, with the single exception of the third
clause of Theorem 7.17, is an appeal to the weaker hypothesis
\bea
\la{Phinonzero}
\Phi=\c V_IX^I\ \neq\ 0\qquad\text{at every point of }\CS .
\ee
On the branch $K>0$, $U\geq0$ implies \rf{Phinonzero} and is not implied by it.
On the branch $K\equiv0$, \rf{Phinonzero} holds automatically for any non-zero
gauging. Theorem A, Theorem B and Theorem 7.14 therefore hold with $U\geq0$
replaced throughout by \rf{Phinonzero}.

\medskip\noindent{\it Proof:} There are four appeals, and they have the same
shape. In Lemma 2.1 the case $h=0$ produces $\int_\CS\Phi=0$ and the case
$\Vt=0$ produces $\int_\CS\Delta^{-2}\Phi=0$ with $\Delta>0$; under
\rf{Phinonzero} the continuous function $\Phi$ has one strict sign on the
connected $\CS$, so both integrals are non-zero and both cases are excluded
directly --- more cheaply than through the potential. The argument following
\rf{hP}, which is what supplies $\Phi\neq0$ to the rest of the paper, is
\rf{Phinonzero} itself. The sharpening of Proposition 6.2 needs only that
$\Phi$ does not vanish at a point. For the implication: if $\Phi(p)=0$ then
$\c^2U(p)=-\tfrac92\c^2K(p)$ by \rf{chiU}, which is negative when $K>0$, so
$U\geq0$ forbids \rf{Phinonzero} from failing; and if $K\equiv0$ then
$V_I=(V_JX^J)X_I$ by Proposition 3.2, so $\Phi=0$ at a point would force
$V_I=0$ there and hence everywhere. That the converse fails is the example in
the Remark above, where $\Phi$ vanishes and $U<0$; perturbing the gauging to
$V_I=(1,-1,\epsilon)$ with $\epsilon\neq0$ small keeps $U<0$ near that point
while making $\Phi\neq0$. $\square$

\medskip\noindent{\it Remark:} The difference between the two hypotheses is
not cosmetic. $U\geq0$ is an
inequality on the gauging, of the kind a reader must check model by model
through \rf{UCform}; \rf{Phinonzero} is a non-degeneracy condition on the
superpotential, of the kind the gauged theory needs anyway for a supersymmetric
$AdS_5$ vacuum to exist, and it is manifestly satisfied on the explicit
horizons of Section~\ref{sec:varmod}. The third clause of Theorem 7.17 is the one statement
that needs the inequality, and it is flagged there.

\subsection{$P$ vanishes somewhere}\la{sec:Pzero}

Compactness forces something sharp. The function $\parallel\eta_-\parallel^2$ is
a global object, and its extrema are constrained:

\medskip\noindent{\it Corollary 7.2:} $P_i$ vanishes at every critical point of
$\parallel\eta_-\parallel^2$, and in particular at a point $p$ where that
function is maximal and at a point $q$ where it is minimal. At those points
\bea
\la{divPsign}
\Phi\,\tn^kP_k\big|_p\geq0 ,\qquad \Phi\,\tn^kP_k\big|_q\leq0 .
\ee

\medskip\noindent{\it Proof:} $\CS$ is compact and without boundary, so the
smooth positive function $\parallel\eta_-\parallel^2$ attains a maximum at some
point $p$ and a minimum at some point $q$ --- this is the extreme value
theorem on a compact space, not a maximum principle, and no
Poincar\'e--Hopf or other index-theoretic input is used or needed ---
where its differential vanishes. By \rf{normminus},
$\tn_i\parallel\eta_-\parallel^2=-\Phi P_i\parallel\eta_-\parallel^2$, and since
$\parallel\eta_-\parallel^2>0$ and $\Phi\neq0$ everywhere, $P_i$ vanishes there.
At a point where $P_i=0$ one has $Z_kW^k=-1$, so $\tn^kP_k=2h_iN^i-12\Phi$ and
\rf{lapnorm} below reduces to
$\tn^i\tn_i\ln\parallel\eta_-\parallel^2=-\Phi\,\tn^kP_k$. At an interior
maximum of a smooth function the Laplacian is non-positive and at an interior
minimum it is non-negative, which is \rf{divPsign}. $\square$

\medskip
Two things this does {\it not} give. It does not give the unconditional
statement ``$K$ vanishes somewhere on $\CS$''; that does not follow, and with
Proposition 7.1 in hand it is in any case equivalent to $K\equiv0$. And, by
Proposition 6.1, it does not give $\tn^kP_k=0$ at $p$. The inequalities
\rf{divPsign} are the whole of what compactness yields at a zero of $P$: a
sign, not a vanishing. Section~\ref{sec:varmod} exhibits a compact horizon on which
$\tn^kP_k$ is strictly positive at the maximum and strictly negative at the
minimum, so both inequalities are attained strictly and neither can be
improved.

\medskip\noindent{\it Remark (the topology of $\CS$):} $P_i$ is closed by
Proposition 6.4, so if it were nowhere vanishing Tischler's theorem would make
$\CS$ fibre over $S^1$, giving $b_1(\CS)\geq1$ and excluding $S^3$ and the lens
spaces. Corollary 7.2
already shows that $P_i$ vanishes somewhere on any compact $\CS$, so the
hypothesis fails and no topological restriction on $\CS$ is obtained this way.
Nor does
curvature supply one. On the branch where this can be settled exactly, the
constant-modulus branch of Section~\ref{sec:constmod}, \rf{ricPzero} diagonalises
$\tilde R_{ij}$ outright, with eigenvalues $\tfrac12\a^2$ and $\mu$ twice, so
positive Ricci curvature is equivalent to $\a\neq0$ and $\mu>0$; every
cross-section of the form $S^1\times\Sigma$ has $\a=0$ and hence a zero
eigenvalue, and $\mu<0$ occurs too. Hamilton's theorem therefore does not
apply, and where it does apply it only reproduces what Section~\ref{sec:constmod} already
gives. Appendix~\ref{app:tools} records the argument in full.

\medskip\noindent{\it Remark (what the extremum of $\parallel\eta_-
\parallel^2$ is, and is not):} Differentiating \rf{normminus} again,
using $\tn^iP_i=2h_iN^i-8\Phi+4\Phi Z_kW^k$ (the trace of
\rf{covZ}--\rf{divW}) and $\tn_i\Phi P^i=\tfrac32\c^2K\,P_kP^k$
\rf{gradPhi}, gives the Laplacian of $\ln\parallel\eta_-\parallel^2$
explicitly,
\bea
\la{lapnorm}
\tn^i\tn_i\ln\parallel\eta_-\parallel^2=-\tfrac32\c^2K\,P_kP^k
-2\Phi^2P_kP^k+12\Phi^2-2\Phi\,h_iN^i.
\eea
At the extremal points of Corollary 7.2, $P_i=0$, so \rf{lapnorm} reduces
there to $\tn^i\tn_i\ln\parallel\eta_-\parallel^2
=2\Phi\big(6\Phi-h_iN^i\big)=-\Phi\,\tn^kP_k$, independently of $K$. The
character of the extremum is therefore not decided by $K$ and $\Phi$: it is
decided by $h_N$, the one transverse component of $h$ that the relations of
this paper leave free, only its $P$-component being pinned by \rf{hP}. A
degenerate extremum, $h_iN^i=6\Phi$, is permitted but not required; the
horizons of Section~\ref{sec:varmod} have non-degenerate extrema at both ends.

\subsection{An integral identity}\la{sec:integral}

Corollary 7.2 is a pointwise, extremal-point statement. Integrating the
trace Einstein equation over $\CS$ gives a global companion to it.

\medskip\noindent{\it Corollary 7.3:}
\bea
\la{smarrK}
\int_\CS\Big[h_ih^i-\tfrac12\Ft_{ij}\Ft^{ij}-8\Phi^2
+\tfrac32\c^2K\,P_iP^i\Big]\,d{\rm vol}_\CS=0,
\eea
and consequently
\bea
\int_\CS h_ih^i\,d{\rm vol}_\CS\ \leq\
\int_\CS\Big[\tfrac12\Ft_{ij}\Ft^{ij}+8\Phi^2\Big]\,d{\rm vol}_\CS,
\eea
with equality if and only if $K\,P_iP^i\equiv0$ on $\CS$, that is, if and only
if either $K\equiv0$ or $P_i\equiv0$ on all of $\CS$.

\medskip\noindent{\it Proof:} Field equation \rf{feq3}, with $\Delta=\a^2$
and $L^I=0$ substituted, gives
\bea
\tn^ih_i=h_ih^i-\tfrac12\Ft_{ij}\Ft^{ij}-\tfrac43\c^2U
-\tfrac13Q_{IJ}\Gt^I{}_{ij}\Gt^{Jij}.
\eea
By \rf{gradX}, $\Gt^I{}_{ij}=\tfrac32\c C^I\e_{ijk}(W-Z)^k$, so using
$\e_{ijk}\e^{ijl}=2\delta_k^l$ in three dimensions and $Q_{IJ}C^IC^J=K$,
\bea
Q_{IJ}\Gt^I{}_{ij}\Gt^{Jij}=\tfrac94\c^2K\cdot2(W-Z)_k(W-Z)^k
=9\c^2K(1-Z_kW^k),
\eea
since $(W-Z)_k(W-Z)^k=2(1-Z_kW^k)$. Substituting this and \rf{chiU} for
$\c^2U$,
\bea
\tn^ih_i=h_ih^i-\tfrac12\Ft_{ij}\Ft^{ij}-8\Phi^2+3\c^2K(1+Z_kW^k),
\eea
and $1+Z_kW^k=\tfrac12P_kP^k$ gives \rf{smarrK}. Since $\CS$ is compact
and boundaryless, $\int_\CS\tn^ih_i=0$. The inequality follows because
$K\geq0$ and $P_iP^i\geq0$ make the last term in \rf{smarrK} non-negative
pointwise; a non-negative continuous function integrates to zero on $\CS$ only
if it vanishes identically, i.e.\ $K\,P_iP^i\equiv0$. By Proposition 7.1 the set
$\{K>0\}$ is either empty or all of $\CS$, so this holds precisely when
$K\equiv0$ or $P_i\equiv0$. $\square$

\medskip
Corollary 7.3 makes quantitative what Corollary 7.2 shows qualitatively:
the total ``size'' of $h$ on $\CS$, in the $L^2$ sense, is controlled by
the flux and dilaton content, with the non-minimal invariant $K$ only
ever tightening that bound. Its equality case is also the cleanest statement of
the branch structure. By Proposition 7.1 the set $\{K>0\}$ is empty or
everything, so a compact horizon of the gauged theory with $\Vt\neq0$ falls into
exactly one of three classes: $K\equiv0$, the minimal-theory case of Section 3;
$K>0$ with $P_i\equiv0$, treated in Section~\ref{sec:constmod}; and $K>0$ with
$P_i\not\equiv0$, the strict-inequality case, treated in Section~\ref{sec:varmod}.
The identity was derived from the trace Einstein equation alone, with no use of
Lemma 6.11, and it says nothing about which of the three classes are occupied.

\subsection{The constant-modulus branch}\la{sec:constmod}

The first of the two branches left open by Corollary 7.3 is $P_i\equiv0$. It
makes every modulus constant by \rf{gradX}, while $K=Q_{IJ}C^IC^J$ may still be
strictly positive, so the horizon can be non-minimal with constant moduli. The
system it closes onto has already been solved: the conclusions recorded in
Proposition 7.4 are exactly the hypotheses under which \cite[\S3.3]{klr}
integrate the near-horizon equations without assuming any rotational symmetry
--- $h$ Killing, $\Delta$ constant, the moduli constant, and $h$ proportional to
$\Phi Z$. What this subsection adds is that the single algebraic hypothesis
$P_i\equiv0$ delivers all four of them, with no ansatz and no symmetry
assumption.

\medskip\noindent{\it Proposition 7.4:} Suppose $P_i\equiv0$ on $\CS$,
equivalently $W_i=-Z_i$ everywhere. Then
\bea
\la{Pzerobranch}
h_i=3\Phi Z_i,\qquad f_i=\Phi Z_i,\qquad
\tn_iZ_j=-\tfrac12\a\,\e_{ijk}Z^k,
\ee
and $\Phi$, $\a$ and $K$ are constants. In particular $Z^i$ is a unit Killing
vector field on $\CS$, $h_i$ is a Killing one-form, $\tn^kP_k=0$ identically,
and every field equation of Sections 2--5 is satisfied for arbitrary $K\geq0$.

\medskip\noindent{\it Proof:} Apply Lemma 6.B with ${\cal O}=\CS$: it gives
$h_i=3\Phi Z_i$, the third relation of \rf{Pzerobranch}, and the vanishing of
$\tn_i\Phi$, $\tn_iX^I$, $\tn_iK$ and $\tn_i\a$, so that $\Phi$, $\a$ and $K$
are constants on the connected $\CS$. The third relation is antisymmetric in
$ij$, so $Z^i$ is Killing, and $Z_kZ^k=1$ makes it a unit Killing field. The
divergence is $\tn^kP_k=2h_iN^i-8\Phi+4\Phi Z_kW^k$ with $N_i=2Z_i$ and
$Z_kW^k=-1$, i.e.\ $\tn^kP_k=12\Phi-8\Phi-4\Phi=0$: this branch sits at the
degenerate end of the freedom left open by Proposition 6.1, for {\it every}
$C_I$ and in particular for $C_I\neq0$. Finally $f_i=h_i-2\Phi Z_i=\Phi Z_i$ by \rf{fhZ},
$\Ft_{ij}\Ft^{ij}=2f_kf^k=2\Phi^2$, and $\tn^ih_i=3\Phi\,\tn^iZ_i=0$, so
\rf{feq3} in the form used above reads $0=9\Phi^2-\Phi^2-8\Phi^2$, an identity
carrying no condition on $K$. $\square$

\medskip\noindent{\it Remark (the rest of the field content closes too):}
Both Bianchi identities \rf{bianchi} for $\Ft$ and each $\Gt^I$ hold trivially,
since on $W=-Z$ both duals are constant multiples of the divergence-free unit
Killing field $Z$; the gauge field equations \rf{feq1} and \rf{feq2} close
with no condition on $C^I$, the dualised divergence cancelling against the
Chern--Simons term as both are proportional to $\a$; and \rf{feq5} holds
identically, in accordance with Proposition 6.1, for generic non-STU cubics as
well as STU, so no adjoint identity is needed here either.
Very special geometry is satisfiable at $K>0$: in the STU model,
$X^I=(1,1,1)$, $V_I=(0,\tfrac12,\tfrac12)$, $\c=1$ gives $\Phi=1$,
$K=\tfrac13$, $C^I=(-\tfrac23,\tfrac13,\tfrac13)\neq0$ and
$\c^2U=6\Phi^2-\tfrac92\c^2K=\tfrac92$, well inside the constraint
$\c^2K\leq\tfrac43\Phi^2$ of \rf{chiU}. Finally, both Killing spinor
equations close on the explicit left-invariant metrics below: the algebraic
equation holds for every $C^I$ and both chiralities, since
$\Gt^I_{ij}\G^{ij}=6i\c C^I$ cancels identically against the term it
multiplies; and the gravitino equation fixes the gauge potential
$\c V_I\At^I$ along $Z$ to the single value required by the transverse Ricci
eigenvalue $\mu$ of \rf{ricPzero} below, with both components transverse to
$Z$ vanishing and both chiralities agreeing, as supersymmetry doubling
requires. So this branch solves the full theory, not only Sections 2--5.

\medskip
The local geometry follows at once. Substituting \rf{Pzerobranch} into
\rf{ricci}, with $N_i=2Z_i$, $P_i=0$ and \rf{chiU}, gives
\bea
\la{ricPzero}
\tilde R_{ij}=\tfrac12\a^2\,Z_iZ_j+\mu\,\big(\delta_{ij}-Z_iZ_j\big),
\qquad
\mu=-3\Phi^2+\tfrac12\a^2+9\c^2K .
\ee
The $Z_iZ_j$ coefficient here is a check rather than an input: for any unit
Killing field obeying $\tn_iZ_j=-\tfrac12\a\e_{ijk}Z^k$ with $\a$ constant, the
Ricci identity gives $\tilde R_{ij}Z^j=\tfrac12\a^2Z_i$ independently of any
field equation, and \rf{ricPzero} reproduces it. So the supergravity content of
\rf{ricPzero} is the single transverse equation fixing $\mu$. Tracing
\rf{ricPzero} against $\delta^{ij}$ gives the scalar curvature
\bea
\la{scalarcurvPzero}
\tilde R=\tfrac32\a^2-6\Phi^2+18\c^2K,
\ee
constant on $\CS$ since $\Phi$, $\a$ and $K$ are.

Such geometries exist and are compact. A three-manifold with a unit Killing
field is locally a Riemannian submersion over a surface with connection
curvature $\a$; taking $\CS$ to be a unimodular three-dimensional Lie group with
structure constants $[X_1,X_2]=n_3X_3$ and cyclic, carrying the left-invariant
metric ${\rm diag}(a^2,a^2,b^2)$ and $Z$ the unit vector along $X_3$, one finds
$Z$ is Killing exactly when $n_1=n_2$, with $\a=n_3b/a^2$, and
$\tilde R_{ij}$ of the form \rf{ricPzero}. The transverse equation then
determines the squashing $b/a$ in terms of $(\Phi,\a,K)$, and for a range of
values it is solved with $\c^2K>0$: for instance $n=(1,1,1)$ with $a=1$,
$b=1$ and $\Phi=1$ gives $\c^2K=\tfrac13$.

This construction is not a classification. It produces only geometries that are themselves Lie groups --- the squashed $S^3$
for $n=(1,1,1)$, quotients of $\widetilde{SL}(2,\bR)$ for $n=(1,1,-1)$,
nilmanifolds for $n=(0,0,1)$ and $T^3$ for $n=(0,0,0)$ --- and it therefore
misses $\bR\times S^2$, which is homogeneous but is not a Lie group at all, and
$\bR\times H^2$, which is a Lie group but not a unimodular one. Both occur
here, at $\a=0$. The complete local
answer is the one obtained in \cite[\S3.3]{klr} by reducing \rf{ricPzero} to
Liouville's equation: writing $D=\a^2+9\c^2K-3\Phi^2=\mu+\tfrac12\a^2$, the
horizon is locally $\mathrm{Nil}$ or $\bR^3$ for $D=0$, $\widetilde{SL}(2,\bR)$
or $\bR\times H^2$ for $D<0$, and a squashed $S^3$ or $\bR\times S^2$ for $D>0$,
the second member of each pair being the degeneration at $\a=0$. Compactifying,
the available cross-sections are the squashed $S^3$ and its quotients,
nilmanifolds, quotients of $\widetilde{SL}(2,\bR)$, $T^3$, $S^1\times\Sigma_g$,
and $S^1\times S^2$.

\medskip\noindent{\it Remark (the $\a=0$ corner):} The degenerate member of each
pair needs separate treatment, because the potential used above to close the
gravitino equation, $\At^I=-(A^I/\a)Z$, is singular at $\a=0$. Nothing is lost.
At $\a=0$ the unit Killing field $Z$ is parallel, $\CS$ is locally
$\bR\times\Sigma$ with $\Sigma$ of Gaussian curvature $\mu=-3\Phi^2+9\c^2K$, and
solving the gravitino equation directly on $\bR\times\Sigma$, with the three
frame components of $\c V_I\At^I$ carried as unknowns rather than written in
terms of $Z$, gives for both chiralities, and up to the exact one-form
$\tfrac23\,d(\text{phase})$ that the undetermined spinor phase contributes,
\bea
\la{monopole}
\c V_I\At^I=\tfrac13\cos\theta\,d\phi,
\qquad
\tfrac13\cosh\theta\,d\phi,
\qquad
\tfrac13\,d\phi
\ee
for $\Sigma=S^2$, $H^2$ and $\bR^2$ respectively, in each case with exterior
derivative $\c V_I\Ft^I$, using $\c V_IA^I=\Phi^2-3\c^2K=-\tfrac13\mu$. So
Proposition 7.4 needs no hypothesis $\a\neq0$, and the $\a=0$ geometries are
genuine horizons.

What does change at $\a=0$ is global. For $\a\neq0$ the potential is built from
the globally defined one-form $Z$ and therefore exists on all of $\CS$. At
$\a=0$ the flux is $\Ft^I=A^I\,\mathrm{vol}_\Sigma$, a multiple of the area
form, so on a compact cross-section
$\int_\Sigma\Ft^I=A^I\,\mathrm{Area}(\Sigma)$, which is non-zero whenever
$A^I=\Phi X^I-3\c C^I$ is. The class of $\Ft^I$ in $H^2(\Sigma)$ is then
non-trivial, no global potential exists, and $\At^I$ is a connection on a
non-trivial $U(1)$ bundle --- the monopole potential \rf{monopole}, singular at
both poles of $S^2$. The $S^1\times S^2$ and $S^1\times\Sigma_g$ members of the
family accordingly carry Dirac-quantised charges. That is a condition on their
charges, not an obstruction to them.

\medskip\noindent{\it Remark:} Two consequences matter for the classification.
First, {\it Section 6 does not apply on this branch}. Corollary 6.5 and the
invariant $\zeta$ are defined only where $P_i\neq0$, and the collapse
underlying Theorem 6.6 --- that $\tn_i\a$ and the gradients of all scalar
invariants are parallel to $P_i$ --- is vacuous when $P_i=0$, since those
gradients simply vanish. Second, and more decisively, the geometries just
listed are homogeneous, so each carries at least three independent Killing
vectors. A second rotational isometry therefore {\it does} exist on the
$P\equiv0$ branch, and the classification of \cite{klr} is available there
despite $K>0$.

\medskip\noindent{\it Remark (an explicit solution):} It costs little to write
one member of this branch out in full. Take the STU model
$C_{IJK}=|\e_{IJK}|$ with $\c=1$, $V_I=(\tfrac23,\tfrac16,\tfrac16)$ and the
constant moduli $X^I=(1,1,1)$, so that $X_I=\tfrac13$,
$Q_{IJ}=\tfrac12\delta_{IJ}$, $\Phi=1$, $C^I=(\tfrac23,-\tfrac13,-\tfrac13)$,
$K=\tfrac13$ and $\c^2U=\tfrac92$; then $\a=1$, $\Delta=1$ and
$\mu=\tfrac12$, so $\CS$ is the round $S^3$ of radius $2$. In the
left-invariant coframe
\bea
\la{eulerframe}
\sigma^1=\cos\psi\,d\theta+\sin\psi\sin\theta\,d\phi,
\quad
\sigma^2=-\sin\psi\,d\theta+\cos\psi\sin\theta\,d\phi,
\quad
\sigma^3=d\psi+\cos\theta\,d\phi,
\ee
for which $d\sigma^i=-\tfrac12\e^i{}_{jk}\sigma^j\wedge\sigma^k$ and $Z=\sigma^3$,
the whole solution is
\bea
\la{explicit}
ds^2&=&2du\Big(dr+3r\,\sigma^3-\tfrac12r^2du\Big)
+(\sigma^1)^2+(\sigma^2)^2+(\sigma^3)^2,
\nn
F^I&=&X^I\,du\wedge dr+A^I\,\sigma^1\wedge\sigma^2,
\qquad
A^I=\Phi X^I-3\c C^I=(-1,2,2),
\ee
with constant moduli and gauge potential $-rX^Idu-A^I\sigma^3$; explicitly
$F^1=du\wedge dr-\sigma^1\wedge\sigma^2$ and
$F^2=F^3=du\wedge dr+2\sigma^1\wedge\sigma^2$. The $r$-linear part of
\rf{nhflux} cancels on this branch, since $\a^I=\a X^I$ contributes
$3rX^I\,du\wedge\sigma^3$ and $\b^I=(\tn\a-\a h)X^I$ contributes
$-3rX^I\,du\wedge\sigma^3$. The cross-section has area $16\pi^2$, and the free
quotients $S^3/\bZ_p$ along the Hopf fibre have area $16\pi^2/p$.

This has been checked directly against the five-dimensional Einstein, Maxwell
and scalar field equations, not merely against their near-horizon
decomposition in Appendix~\ref{app:fieldeqs}, and the Chern--Simons coefficient was left
undetermined and fitted: all fifteen components of the Maxwell equation are
solved by the single value $\mp\tfrac16C_{IJK}F^I\wedge F^J\wedge A^K$, which
is its standard normalisation. A second, non-round member behaves the same
way: the stretched Berger sphere $a=1$, $b=\tfrac32$ with
$V_I=(\tfrac5{12},\tfrac12,\tfrac1{12})$, for which $\a=\tfrac32$,
$K=\tfrac7{36}$ and $\mu=-\tfrac18$. Both are near-horizon limits of
Gutowski--Reall black holes \cite{gutowskireallgen}: running their parameter
map forwards with equal charges $q_I=\tfrac43$ reproduces $a$, $b$, $\a$,
$\Phi$, $K$ and the horizon moduli exactly. The asymmetry responsible for
$K>0$ therefore sits in $V_I$ rather than in the charges --- in the symmetric
gauging $V_I=\tfrac13$ equal charges give the minimal solution with $K=0$,
whereas here $V_I$ is not proportional to $X_I$. So the $K>0$, $P\equiv0$
branch is not merely non-empty but populated by known black holes.

The invariant governing that classification is $K$ itself. \cite{kayani}
discusses the same question in terms of
$\lambda:=Q^{IJ}V_IV_J-(V_IX^I)^2$, for which \cite{klr} requires
$\lambda>0$ on its $S^1\times S^2$ solution, the case corresponding to a
black ring. Since $\Phi=\c V_IX^I$, \rf{Kdef} gives
\bea
\la{lambdaK}
\lambda=K-\tfrac13\Phi^2/\c^2 ,
\ee
so $\lambda>0$ implies $K>0$ but not conversely. At $K=0$, where
$V_I=(V_JX^J)X_I$, this gives $\lambda=-\tfrac13(V_JX^J)^2\leq0$, matching
the fact that $\lambda>0$ is unreachable in the minimal theory
\cite{kayani}. So the $\lambda>0$ locus lies strictly inside $K>0$, and by
Proposition 7.1 a horizon meeting it has $K>0$ everywhere. In the normalisation
of \cite{klr}, where $V_I=\tfrac13$ and the gauge coupling is $g=\c$, their
K\"all\'en function is $9\lambda$, and in $U(1)^3$ the paper's own invariant is
$K=\tfrac{2}{27}\sum_{I<J}(X^I-X^J)^2$, which exhibits $K\equiv0$ as the locus
of equal moduli, that is, as the minimal embedding.

Combining \rf{lambdaK} with \rf{chiU} also identifies $U$: since
$U=9\big[(V_IX^I)^2-\tfrac12Q^{IJ}V_IV_J\big]$ and $Q^{IJ}=2X^IX^J-6C^{IJK}X_K$
by the Calabi relation, $U=27C^{IJK}V_IV_JX_K$ is the scalar potential of the
theory. This is \rf{UCform} again, reached by a different route. The constraint
$\c^2K\leq\tfrac43\Phi^2$ implied by \rf{chiU} is therefore neither more nor
less than the non-negativity of that potential; it enters here as the bound
which the transverse Ricci equation \rf{ricPzero} must respect, and not as an
additional physical input.

Corollary 7.3 leaves exactly two situations open on a compact, connected $\CS$
with $\Vt\neq0$ and $K>0$: $P_i\equiv0$, the subject of this subsection, and
$P_i\not\equiv0$, treated in Section~\ref{sec:varmod} below.

On the first, Proposition 7.4 gives $\Phi$, $\a$ and $K$ constant, a unit
Killing field $Z$ with $h_i=3\Phi Z_i$, and $\CS$ one of the six local
geometries of \cite[\S3.3]{klr}, so a second rotational isometry exists. This
does {\it not} exclude a black ring: the $\a=0$, $D>0$ member of that list is
$\bR\times S^2$, realised precisely when $\lambda>0$, so the $S^1\times S^2$
window is $\tfrac13\Phi^2<\c^2K\leq\tfrac43\Phi^2$, which is non-empty. In
$U(1)^3$ with $V_I=\tfrac13$ and $\c=1$ the moduli $X^I=(\tfrac14,\tfrac14,16)$
give $\Phi=\tfrac{11}2$ and $K=\tfrac{147}4$, against
$\tfrac13\Phi^2=\tfrac{121}{12}$ and $\tfrac43\Phi^2=\tfrac{121}3$. This is the
$\bR\times S^2$ near-horizon geometry of \cite[\S3.3]{klr}, which those
authors could not rule out either, and nothing above rules it out here. Two
further identities hold everywhere on this branch, not merely at the extrema
supplied by Corollary 7.2: $P_i\equiv0$ makes $\parallel\eta_-\parallel^2$
constant by \rf{normminus}, and $\tn_iX^I=0$ by \rf{gradX}, so
\bea
\la{ringconds}
\tn_iX^I=0,\qquad h_i=3\Phi Z_i ,
\ee
hold identically, and with $N_i=2Z_i$ these give
\bea
\la{ringpoint}
\tn^kP_k\equiv0,\qquad h_iN^i=6\Phi .
\ee

On the second, moduli vary somewhere on $\CS$. This branch is not empty ---
Section~\ref{sec:varmod} exhibits a compact member of it explicitly --- so it cannot be
dismissed, and it is not covered by the argument just given. What Section~\ref{sec:varmod}
shows is that it is covered by a second, independent argument: a second
rotational isometry exists there as well, by transport rather than by
homogeneity. Nothing said here excludes horizons of that branch from also
being compact and inherently non-minimal; it only describes the ones on which
the moduli happen to be constant.

\subsection{The varying-moduli branch}\la{sec:varmod}

This subsection treats the second branch left open by Corollary 7.3: $K>0$
with $P_i\not\equiv0$, so the moduli vary somewhere on $\CS$. The branch is not
empty, compact or otherwise, and two things are done about it here: an
explicit compact member is exhibited, and a second rotational isometry is
proved to exist on it regardless, by transporting the Killing equation along
$P$ rather than by the invariant-gradient route of Theorem 6.6. The commutator
relations \rf{framecomm} established along the way are used nowhere else, and
a reader interested only in the constant-modulus classification may skip to
Section 8.

\medskip\noindent{\it Remark (an explicit compact varying-moduli horizon):}
The near-horizon family of \cite[\S4.2]{klr} deforms the constant-modulus
system by the polynomial $\Pi(x)=H(x)-\tfrac14C^2(x-\a_0)^2-D_0^2/C^2$, cubic
in $x$ because $H(x)=x^3-9x$ is the same cubic used throughout, with $C^2$,
$\a_0$ and $D_0^2$ fixed by the charges; the horizon coordinate ranges over an
interval on which $\Pi>0$, and $P_i=0$ exactly where $\Pi$ vanishes. Taking the
$U(1)^3$ charges $K_I=(1,-1,0)$, $V_I=\tfrac13$, $C^2=60$,
$\a_0=\tfrac{83}{30}$ and $D_0^2=311$ clears all denominators and gives
\bea
\la{Pispecial}
\Pi(x)=x^3-15x^2+74x-120=(x-4)(x-5)(x-6) ,
\ee
an exact identity between polynomials with integer coefficients, requiring no
numerical evaluation. The three roots are
$x=4,5,6$; on the interval between the first two,
$\Pi(x)=(x-4)(5-x)(6-x)>0$ because each factor is strictly positive there, and
$\Pi<0$ on $(5,6)$. So $(4,5)$ is a compact horizon interval with
$P_i\not\equiv0$ throughout its interior, bounded by the two smallest roots of
$\Pi$; the third root plays no role. On this member every quantity used below
is an explicit algebraic function of $x$ alone, with no numerical evaluation
anywhere:
\bea
\la{klrclosed}
K=\frac{4(x^2+3)}{H^{4/3}},\qquad
\Phi=\frac{x^2-3}{H^{2/3}},\qquad
\a=\frac{\sqrt{311}}{H^{2/3}},
\nn
P_kP^k=\frac{4\,\Pi}{H},\qquad
\tn^kP_k=2\Phi\,P_kP^k-\frac{4\,\Pi'}{H^{2/3}} ,
\ee
where $H(x)=x^3-9x$ is the cubic of \rf{Pispecial} and $\Pi'=3x^2-30x+74$; in
particular $H(4)=28$ and $H(5)=80$. Three consequences are then exact rather
than numerical. First, $K>0$ on the whole of $[4,5]$, because $x^2+3>0$ and
$H>0$ there, with endpoint values
\bea
\la{klrends}
K(4)=\tfrac{19}{98}\,2^{1/3}7^{2/3}=0.893860\ldots ,
\qquad
K(5)=\tfrac{7}{100}\,10^{2/3}=0.324911\ldots ,
\ee
so this member sits on the non-minimal branch $K>0$ throughout. Second,
$\{P=0\}$ is exactly the zero set of $\Pi$ in $[4,5]$, that is the two
endpoints $x=4,5$, so $\CS_0$ is the open interior of the interval. Third, at those two
points the first term of $\tn^kP_k$ drops out, since it carries a factor
$P_kP^k$, and the divergence reduces to the integer computation
$\tn^kP_k=-4\Pi'/H^{2/3}$ with $\Pi'(4)=2>0$ and $\Pi'(5)=-1<0$; so $\tn^kP_k$
changes sign across the interval, from
$-\tfrac27\,2^{2/3}7^{1/3}=-0.867596\ldots$ at $x=4$ to
$+\tfrac1{10}10^{1/3}=+0.215443\ldots$ at $x=5$, and $P$ is nowhere
covariantly constant on it. The decimals are quoted only for orientation;
each is the decimal expansion of the algebraic number displayed beside it.
This example is used again in the Remark after
the proof of Theorem 7.5, and in Theorem 7.7.

Theorem 6.6 closes the invariant route to a second isometry on that branch, but
there is another route: instead of building a Killing vector out of scalar
invariants, one transports the Killing equation itself along $P$. That route
succeeds.

\medskip\noindent{\it Theorem 7.5 (the second isometry on the branch
$P\not\equiv0$):} Let $(\CS,g,h,X^I,F^I)$ be a supersymmetric near-horizon
geometry of the gauged theory with $\Vt\neq0$ and $K>0$, not assumed compact,
with $P_i\not\equiv0$, and let $\CS_0\subseteq\CS$ be the open set on which $Z$ and
$W$ are linearly independent. Write
\bea
\la{hNhY}
h_N:=\frac{h_iN^i}{N_jN^j},\qquad h_Y:=\frac{h_iY^i}{Y_jY^j} ,
\ee
so that $\Vt=-\parallel\eta_-\parallel^2(h_N\,N+h_Y\,Y)$ by \rf{hP}. Then
\bea
\la{secondkilling}
U:=\parallel\eta_-\parallel^2\big(h_Y\,N-h_N\,Y\big)
\ee
is a Killing vector of $(\CS_0,g)$ satisfying
${\cal L}_Uh={\cal L}_UX^I={\cal L}_UF^I=0$ and $[\Vt,U]=0$, linearly
independent of $\Vt$ at every point at which $\Vt\neq0$. If in addition
$\a\neq0$, then on each connected component of $\CS_0$ the space of Killing
vectors orthogonal to $P$ which preserve $h$, $X^I$ and $F^I$ is exactly
two-dimensional and abelian, with $\{\Vt,U\}$ a basis. (The restriction to a
component is not idle: the solution space of the flat rank-two system below is
bounded by two over a connected base only, and $\CS_0$ is here the complement
of a closed set in a manifold not assumed compact, so it may be disconnected.
The two solutions $\Vt$ and $U$ are in any case exhibited globally on $\CS_0$,
so the lower bound needs no such restriction.)

\medskip\noindent{\it Proof:} Such a symmetry annihilates $\Phi$, and
\rf{PdPhi} gives $\tn_i\Phi\propto KP_i$ with $K>0$, so it is orthogonal to
$P$ and hence of the form $U=aN+bY$ on $\CS_0$. Substituting this into
$\tn_{(i}U_{j)}=0$ and using \rf{covZ}--\rf{divW} together with the expression
for $\tn_ih_j$ turns the Killing equation into six linear equations for the six
directional derivatives of $a$ and $b$. Five of them are determined,
\bea
\la{killtrans}
N(a)=0,\qquad Y(b)=0,\qquad Y(a)=-\tfrac14(P_kP^k)\,N(b),\nn
P(a)=(P_kP^k)\big[(h_N-2\Phi)a-h_Yb\big],\qquad
P(b)=(P_kP^k)\big[h_Ya+(h_N-2\Phi)b\big] ,
\ee
and $w:=N(b)$ is left free. In the frame $\{P,N,Y\}$, which is orthogonal on
$\CS_0$, the same identities give
\bea
\la{framecomm}
[N,Y]=0,\qquad [P,N]=\lambda N+\mu Y,\qquad [P,Y]=-\mu N+\lambda Y ,
\ee
with $\lambda=-(P_kP^k)(h_N-2\Phi)$ and $\mu=-(P_kP^k)h_Y$. These are obtained
from $\tn_iZ_j$ and $\tn_iW_j$ in \rf{covZ}--\rf{covW} together with the
Leibniz rule for $Y=Z\times W$; in particular $[N,Y]=0$ does not follow from
$\tn_iZ_j$, $\tn_iW_j$ and Leibniz alone; substituting $h_Y=\a/(1-Z_kW^k)$
from Corollary 6.5 is what makes the two would-be extra terms in $[N,Y]$
cancel. The feature of these commutators that the argument uses
is that neither $[P,N]$ nor $[P,Y]$ has a component along $P$: the plane spanned
by $N$ and $Y$ is integrable, and all transport happens along $P$. Every scalar
invariant of the near-horizon data has gradient along $P$ --- the moduli by
\rf{gradX}, $\Phi$ by \rf{gradPhi}, $K$ by \rf{Kprime}, $\a$ by Theorem 6.6,
and $Z\cdot W$, $h_N$ and $h_Y$ by direct computation from \rf{covZ}--\rf{divW}
--- so $N$ and $Y$ annihilate all of them. Applying $[P,N]$ to $a$ and $[P,Y]$
to $b$ and using \rf{killtrans} therefore leaves only the algebraic condition
\bea
\la{wzero}
\tfrac12(P_kP^k)\,\a\,w=0 ,
\ee
so $w=0$ whenever $\a\neq0$. When $\a=0$ the condition is vacuous, and we
choose $w=0$; that choice is sufficient to construct the particular second
Killing vector $U$, and no claim is made that it exhausts the solutions of the
Killing transport system on the locus $\a=0$. With $w=0$, \rf{killtrans}
becomes
\bea
\la{transport}
\tn_ia=\big[(h_N-2\Phi)a-h_Yb\big]P_i,\qquad
\tn_ib=\big[h_Ya+(h_N-2\Phi)b\big]P_i ,
\ee
that is $d\psi=\psi\,\omega$ for $\psi:=a+ib$ and
$\omega:=\big[(h_N-2\Phi)+ih_Y\big]P$. This connection form is closed:
\bea
\la{domega}
d\omega=d\big[(h_N-2\Phi)+ih_Y\big]\wedge P
+\big[(h_N-2\Phi)+ih_Y\big]\,dP=0 ,
\ee
the second term because $dP=0$ by \rf{dPzero}, and the first because $h_N$,
$h_Y$ and $\Phi$ all have gradient along $P$ --- by Lemma 6.11, \rf{hY} and
\rf{gradPhi} respectively --- so that the one-form in the wedge is itself a
multiple of $P$. Hence \rf{transport} is a flat connection on a rank-two real
bundle and its solution space is two-dimensional. Flatness is exactly
\rf{domega}: the curvature of the connection $d-\omega$ on the trivial complex
line bundle is $d\omega-\omega\wedge\omega$, and $\omega$ is a scalar-valued
$1$-form, so $\omega\wedge\omega=0$ identically and no separate check of the
quadratic term is required.

Write $\sigma:=\parallel\eta_-\parallel^2$, so that
$\tn_i\sigma=-\Phi\sigma P_i$ by \rf{normminus}. One solution of
\rf{transport} is $\psi_{\Vt}=-\sigma(h_N+ih_Y)$, which is $\Vt$. Multiplying
by $i$ gives $\psi_U=\sigma(h_Y-ih_N)$, that is $a=\sigma h_Y$ and
$b=-\sigma h_N$, which is \rf{secondkilling}. These coefficients are not
guessed; they satisfy \rf{transport} by direct substitution. Differentiating
$h_Y=\a/(1-Z_kW^k)$ with the second and fourth lines of \rf{gradsys} gives
\bea
\la{gradhY}
\tn_ih_Y=\big(2h_N-\Phi\big)h_Y\,P_i ,
\ee
so that, for $a=\sigma h_Y$,
\bea
\tn_ia=\big[-\Phi+2h_N-\Phi\big]\sigma h_Y\,P_i
=\big[(h_N-2\Phi)\,\sigma h_Y+h_Y\,\sigma h_N\big]P_i \nn
=\big[(h_N-2\Phi)a-h_Yb\big]P_i ,
\ee
while for $b=-\sigma h_N$, using the third line of \rf{gradsys},
\bea
\tn_ib=\big[\Phi h_N-h_N(h_N-\Phi)+h_Y^2\big]\sigma\,P_i
=\big[h_Y\,\sigma h_Y+(h_N-2\Phi)(-\sigma h_N)\big]P_i \nn
=\big[h_Ya+(h_N-2\Phi)b\big]P_i .
\ee
The two solutions are independent because the coefficient determinant of
$\{\Vt,U\}$ in the frame $\{N,Y\}$ is
\bea
\la{coeffdet}
\det\begin{pmatrix}-\sigma h_N&-\sigma h_Y\\ \sigma h_Y&-\sigma h_N
\end{pmatrix}
=\sigma^2\big(h_N^2+h_Y^2\big) ,
\ee
which is a determinant of coefficients in a fixed frame rather than a
Wronskian, and which vanishes only where $\Vt$ does. They commute: writing
$\Vt=a_1N+b_1Y$ and $U=a_2N+b_2Y$,
\bea
[\Vt,U]=\big(\Vt(a_2)-U(a_1)\big)N+\big(\Vt(b_2)-U(b_1)\big)Y
+a_1b_2[N,Y]+b_1a_2[Y,N] ,
\ee
in which the last two terms cancel because $[N,Y]=0$ by \rf{framecomm}, and
the first two vanish because each of $a_1,b_1,a_2,b_2$ is $\sigma$ times a
scalar invariant, every such function has gradient along $P$ by \rf{gradsys}
and \rf{normminus}, and both $\Vt$ and $U$ lie in the plane spanned by $N$ and
$Y$, which is orthogonal to $P$ on $\CS_0$. So $[\Vt,U]=0$. No holonomy or
connectedness argument is needed to make the second solution global on $\CS_0$:
the coefficients $a=\sigma h_Y$ and $b=-\sigma h_N$ are smooth functions on all
of $\CS_0$ by \rf{hNhY}, and $i\psi_{\Vt}$ solves \rf{transport} wherever
$\psi_{\Vt}$ does because that system is $\bC$-linear.

It remains to check that $U$ preserves the matter fields. $U$ is orthogonal to
$P$, and every scalar of the state has gradient along $P$, so
\bea
U(\Phi)=U(\a)=U(K)=U(Z_kW^k)=U(h_N)=U(h_Y)=0 ,
\ee
and ${\cal L}_UX^I=U^i\tn_iX^I=0$ by \rf{gradX}. Since $U$ is Killing and
annihilates $a$ and $b$, the commutators \rf{framecomm} give
${\cal L}_UN={\cal L}_UY={\cal L}_UP=0$; with $h_i=\Phi P_i+h_NN_i+h_YY_i$ and the scalar
identities just listed this yields ${\cal L}_Uh=0$. Finally the flux is algebraic in
data already shown invariant --- $\Ft_{ij}=\e_{ijk}f^k$ with
$f_i=h_i-2\Phi Z_i$, and $\Gt^I{}_{ij}=\tfrac32\c\,C^I\e_{ijk}(W-Z)^k$ by
\rf{gradX} --- and a Killing vector preserves $\e_{ijk}$, so
${\cal L}_U\Ft={\cal L}_U\Gt^I=0$. The remaining components of the
five-dimensional flux \rf{nhflux} are covered by the same list: on the branch
\rf{branch} one has $\a^I=\a X^I$, since $L^I=0$, and
$\b^I{}_i=\big(\tn_i\a-\a h_i\big)X^I+\a\tn_iX^I$, so both are algebraic in
$\a$, $X^I$, $h_i$ and their derivatives. Now $U(\a)=0$ and
${\cal L}_UX^I=0$ from the list above, ${\cal L}_Uh=0$ was just shown, and $U$
is Killing, so ${\cal L}_U$ commutes with $\tn$; hence
${\cal L}_U\a^I={\cal L}_U\b^I=0$ and therefore ${\cal L}_UF^I=0$ for the
full five-dimensional field strength. $\square$

\medskip\noindent{\it Remark (where $K>0$ and $P_i\not\equiv0$ are used):} The
two standing hypotheses of Theorem 7.5 enter at opposite ends of the proof.
$K>0$ is used in its first line: it is what makes
$\tn_i\Phi=\tfrac32\c^2K\,P_i$ a non-degenerate equation, and hence what
forces a symmetry annihilating $\Phi$ to be orthogonal to $P$ --- which is
what reduces the Killing equation to the two-component ansatz $U=aN+bY$ in the
first place. When $K\equiv0$ the moduli are constant, $\tn_i\Phi=0$, and that
step supplies no information whatever. This is not a gap: $K\equiv0$ is the
minimal theory by Proposition 3.2, where a second rotational isometry is
obtained instead directly from the Killing spinors in \cite{groverindex}.
$P_i\not\equiv0$ is used at the other end. The construction transports along
$P$, and where $P$ vanishes identically there is nothing to transport:
$\omega=0$, every solution of \rf{transport} is constant, and
\rf{secondkilling} degenerates, since $Y\equiv0$ there. The second isometry on
that branch is supplied by Proposition 7.4 and homogeneity, not by
\rf{secondkilling}.

\medskip\noindent{\it Remark (local versus global):} Theorem 7.5 constructs
$U$, and shows it Killing, only on the open set $\CS_0$, where the frame
$\{P,N,Y\}$ is non-degenerate; by itself it asserts nothing about
$\CS\setminus\CS_0$. Lemma 7.6 and Lemma 7.6$'$ show that the singularities of
\rf{secondkilling} there are singularities of the frame and not of the field:
the same $U$ has a smooth, frame-free expression on all of $\CS$. Theorem 7.7
then shows that the Killing equation extends with it. The local construction
and the global extension remain logically distinct steps, and nothing is
assumed in passing between them --- neither Lemma 7.6$'$ nor Theorem 7.7 constrains
$\a$, and neither constrains the degeneracy set.

\medskip\noindent{\it Remark (consistency with Theorem 6.6):} Theorem
6.6 says that no second isometry can be manufactured from the {\it gradients}
of scalar invariants, because those gradients are all parallel to $P$ and
$\star(dI_1\wedge dI_2)$ therefore vanishes. The vector \rf{secondkilling} is
not of that form: it is a combination of $N$ and $Y$ with invariant
coefficients, and neither $N$ nor $Y$ is a gradient. What Theorem 6.6 closes is
the route, not the conclusion; what replaces it is \rf{transport}, which uses
the invariance of the coefficients only to establish that $\omega$ is closed.

\medskip\noindent{\it Remark (where the frame degenerates):} On
$\CS\setminus\CS_0$ either $Z=W$ or $Z=-W$, and the two cases are quite
different. The first is $P_kP^k=4$; there $Y_i=\e_{ijk}Z^jW^k=0$, so
Corollary 6.5's identity $h_iY^i=\tfrac12\a P_kP^k$, available because
$P\neq0$, forces $\a=0$. So $\{Z=W\}\subseteq\{\a=0\}$, and any horizon on
which $\a$ is nowhere zero has $\CS\setminus\CS_0=\{P=0\}$ exactly. The second
case $Z=-W$ is $\{P=0\}$; there $N\to2Z$ and $Y\to0$, while $N_kN^k=4-P_kP^k$
tends to $4$ and $h_Y=\a/(1-Z_kW^k)$ tends to $\tfrac12\a$. The frame
$\{P,N,Y\}$ degenerates, since two of its three members do; the vector
\rf{secondkilling} does not.

\medskip\noindent{\it Lemma 7.6 ($U$ is smooth wherever $Z\neq W$):} The
vector field \rf{secondkilling} obeys
\bea
\la{Usmooth}
U_i=\frac{\parallel\eta_-\parallel^2}{1-Z_kW^k}
\Big[\a\,N_i-\tfrac12\big(h_jN^j\big)\,Y_i\Big] ,
\ee
an identity between smooth objects wherever $Z\neq W$. In particular $U$
extends from $\CS_0$ to a smooth vector field on $\CS\setminus\{Z=W\}$, with
$U=\parallel\eta_-\parallel^2\a\,Z$ on $\{P=0\}$, and with no condition
imposed on $\{P=0\}$.

\medskip\noindent{\it Proof:} By Corollary 6.5, $h_iY^i=\tfrac12\a P_kP^k$
and $Y_kY^k=\tfrac14P_kP^k\,N_jN^j$, so wherever $Y\neq0$
\bea
\la{hYglobal}
h_Y=\frac{h_iY^i}{Y_kY^k}=\frac{2\a}{N_jN^j}=\frac{\a}{1-Z_kW^k} ,
\ee
using $N_jN^j=4-P_kP^k=2(1-Z_kW^k)$. Likewise $h_N=h_jN^j/N_kN^k
=h_jN^j/\big(2(1-Z_kW^k)\big)$. Substituting both into \rf{secondkilling}
gives \rf{Usmooth}. Now $Z_i$ and $W_i$ are the smooth unit bilinears
\rf{ZWdef}, built from Killing spinors that are nowhere zero, so $N_i$, $Y_i$
and $Z_kW^k$ are smooth on all of $\CS$; $h_i$ and $\a$ are smooth by
assumption on the horizon data; and $\parallel\eta_-\parallel^2$ is smooth and
strictly positive. The right-hand side of \rf{Usmooth} therefore has a single
denominator, the scalar $1-Z_kW^k$, which vanishes only where $Z=W$ and equals
$2$ where $P=0$. Evaluating there, $N_i\to2Z_i$ and $Y_i\to0$ give
$U_i\to\parallel\eta_-\parallel^2\a\,Z_i$. $\square$

\medskip\noindent{\it Remark (why the frame components are harmless):} One
might expect $h_N$ and $h_Y$, being frame components, to go bad exactly where
the frame does. They do not, and for two different reasons. The denominator of $h_N$ is
$N_kN^k=4-P_kP^k$, which is not small at a zero of $P$ but equal to $4$; the
frame member that degenerates there is $Y$, not $N$. The denominator of $h_Y$
is genuinely $Y_kY^k\to0$, but Corollary 6.5 replaces the quotient by
$\a/(1-Z_kW^k)$ before any limit is taken, and that expression is regular.
What is left is $Y_i$ itself in the second term of \rf{Usmooth}, and $Y_i$ is
a smooth field that simply vanishes there. Nothing in $U$ refers to $P$ at
all.

\medskip\noindent{\it Lemma 7.6$'$ ($U$ is smooth everywhere, with no
hypothesis on $\a$):} On $\CS_0$ the field \rf{secondkilling} equals
\bea
\la{Uglobal}
U_i=\parallel\eta_-\parallel^2\Big(\a\,Z_i-\e_{ijk}Z^ju^k\Big) ,
\qquad
u_i=\Phi P_i-h_i ,
\ee
and the right-hand side is smooth on the whole of $\CS$. Hence $U$ extends to a
smooth vector field on $\CS$ whether or not $\a$ vanishes. Moreover the set
$\{Z=W\}$ is closed with empty interior: it is contained in $\{\a=0\}$, and
where it is non-empty it is an embedded surface, since
\bea
\la{gradNZW}
\tn_iN_j=2Z_iu_j
\qquad\text{at every point of }\{Z=W\} ,
\ee
which is non-zero because $u_ku^k=m/\parallel\eta_-\parallel^2>0$ there by
\rf{normminus2}.

\medskip\noindent{\it Proof:} Decompose $h_i=\Phi P_i+h_NN_i+h_YY_i$ as in
Corollary 6.5, so that $u_i=-(h_NN_i+h_YY_i)$ by \rf{hP}. The frame
$\{P,N,Y\}$ is orthogonal with $P_kN^k=0$, $Y=Z\times W$ and the cross-product
relations $P\times N=-2Y$, $P\times Y=\tfrac12FN$, $N\times Y=-\tfrac12(4-F)P$
of Corollary 6.3's proof. Writing $Z=\tfrac12(P+N)$ and expanding,
\bea
\e_{ijk}Z^ju^k=-\tfrac12\big(P+N\big)\times\big(h_NN+h_YY\big)
=h_N\,Y_i-\tfrac14F\,h_Y\,N_i+\tfrac14(4-F)\,h_Y\,P_i ,
\ee
so that, using $h_Y=2\a/(4-F)$ from \rf{hY} and $Z=\tfrac12(P+N)$ once more,
\bea
\a\,Z_i-\e_{ijk}Z^ju^k
=h_Y\,N_i-h_N\,Y_i ,
\ee
which is \rf{secondkilling} divided by $\parallel\eta_-\parallel^2$. Every
ingredient of the right-hand side of \rf{Uglobal} --- the bilinears $Z_i$ and
$P_i$, the horizon data $h_i$ and $\a$, the scalar $\Phi$ and
$\parallel\eta_-\parallel^2$ --- is smooth on all of $\CS$, and no denominator
occurs; this is the whole content of the extension statement. It carries
\rf{Usmooth} across $\{Z=W\}$: the single denominator $1-Z_kW^k$ there is
cancelled identically by a double zero of the numerator.

For the last statement, $\{Z=W\}\subseteq\{\a=0\}$ is the Remark preceding
Lemma 7.6. At a point of $\{Z=W\}$ subtracting \rf{covW} from \rf{covZ} with
$W=Z$ and $Z_kW^k=1$ leaves
$\tn_iN_j=-2Z_ih_j+2\delta_{ij}h_kZ^k+4\Phi(Z_iZ_j-\delta_{ij})$, and there
$P=2Z$, so $h_kZ^k=\tfrac12h_kP^k=2\Phi$ by \rf{hP} and the $\delta_{ij}$
terms cancel, giving $\tn_iN_j=-2Z_i(h_j-2\Phi Z_j)=2Z_iu_j$, which is
\rf{gradNZW}. By Proposition 4.1 in the form \rf{normminus2},
$u_ku^k+\a^2=m/\parallel\eta_-\parallel^2$ with $m=2\parallel\eta_+\parallel^2
>0$, and $\a=0$ on $\{Z=W\}$, so $u\neq0$ there and $\tn_iN_j\neq0$. A point
of $\{Z=W\}$ therefore has a neighbourhood in which $\{Z=W\}=\{N=0\}$ is cut
out transversally by the rank-one differential \rf{gradNZW}; in particular it
has empty interior. $\square$

\medskip\noindent{\it Remark (what \rf{Uglobal} buys):} Lemma 7.6 already
shows that $U$ does not degenerate where the frame does, but it leaves
$\{Z=W\}$ untreated, and $\{Z=W\}$ is empty only when $\a$ is. Identity
\rf{Uglobal} removes the exception: it is a frame-free formula for $U$ in
terms of objects that are smooth everywhere on $\CS$, so the extension problem
disappears rather than being solved case by case. The price is nil --- both
sides are the same field on $\CS_0$ --- and the gain is that Theorem 7.7 below
needs no hypothesis on $\a$, which is what makes
that theorem a statement about the varying-moduli branch entire.

\medskip\noindent{\it Theorem 7.7 (toricity on the varying-moduli branch):}
Let $\CS$ be compact, connected and without boundary, with $K>0$ and
$P_i\not\equiv0$. Then $U$ of \rf{secondkilling} extends
to a smooth Killing vector field on the whole of $\CS$, and the isometry group
of $\CS$ contains a two-torus acting effectively and preserving $h$, $X^I$ and
$F^I$. No hypothesis on $\a$, and none on the degeneracy set $\{P=0\}$, is
required.

\medskip\noindent{\it Proof:} By Lemma 7.6$'$ the field $U$ is smooth on all of
$\CS$, so $\tn_{(i}U_{j)}$ is a smooth, hence continuous, tensor field on
$\CS$.

It vanishes on a dense subset. Decompose
$\CS=\CS_0\sqcup\{P=0\}\sqcup\{Z=W\}$ and split
the second piece into its interior and its boundary. On $\CS_0$ we have
$\tn_{(i}U_{j)}=0$ by Theorem 7.5. On $\mathrm{int}\{P=0\}$ the local form of
Proposition 7.4 applies --- its proof differentiates $W_i=-Z_i$, which holds
on that open set --- giving $h_i=3\Phi Z_i$, $\tn_iZ_j=-\tfrac12\a\e_{ijk}Z^k$
and $\a$ constant on each of its components; and $\tn_i\parallel\eta_-
\parallel^2=-\parallel\eta_-\parallel^2\Phi P_i$, which follows from
\rf{conconx} and \rf{Vt}, vanishes there, so $\parallel\eta_-\parallel^2$ is
constant on each component too. By Lemma 7.6, $U=\parallel\eta_-\parallel^2
\a\,Z$ on that set, with $\parallel\eta_-\parallel^2\a$ constant, whence
\bea
\la{nabUPzero}
\tn_iU_j=-\tfrac12\parallel\eta_-\parallel^2\a^2\,\e_{ijk}Z^k ,
\ee
which is antisymmetric, so $\tn_{(i}U_{j)}=0$ there as well. What is left over
is $\partial\{P=0\}=\{P=0\}\setminus\mathrm{int}\{P=0\}$ together with
$\{Z=W\}$. The first is the boundary of a closed set, hence closed with empty
interior; the second is closed with empty interior by Lemma 7.6$'$. A finite
union of closed sets with empty interior is closed with empty interior, so
$\CS_0\cup\mathrm{int}\{P=0\}$ is open and dense in $\CS$, and a continuous
tensor field vanishing on a dense subset vanishes everywhere. So $U$ is
Killing on all of $\CS$. The same continuity argument gives
${\cal L}_Uh={\cal L}_UX^I={\cal L}_UF^I=0$ and $[\Vt,U]=0$ on all of $\CS$:
each of these is a continuous tensor field, each vanishes on $\CS_0$ by
Theorem 7.5, and each vanishes on $\mathrm{int}\{P=0\}$ because there
$U=\parallel\eta_-\parallel^2\a Z$ with $\parallel\eta_-\parallel^2\a$
constant, $h_i=3\Phi Z_i$ with $\Phi$ constant, the moduli are constant by
Proposition 7.4, and $Z^i\tn_iZ_j=Z_i\tn_jZ^i=0$ follows from
$\tn_iZ_j=-\tfrac12\a\e_{ijk}Z^k$.

It remains to identify the group the two fields generate, and its dimension can
be pinned exactly. The field $U$ generates a one-parameter group of isometries of
$(\CS,g)$; $\mathrm{Isom}(\CS,g)$ is compact because $\CS$ is, so the closure
$T$ of the connected subgroup generated by $\Vt$ and $U$ --- which commute, by
Theorem 7.5 --- is a compact connected abelian subgroup of
$\mathrm{Isom}(\CS,g)$, that is a torus $T^k$ with $k\geq2$. Its action on
$\CS$ is effective for the trivial reason that it is a subgroup of the
isometry group: an isometry that acts as the identity on $\CS$ {\it is} the
identity element. (The linear independence of $\Vt$ and $U$ is not what makes
the action effective; it is what makes $k\geq2$, since it forbids
$\mathrm{Lie}(T)$ from being one-dimensional.)

That $k=2$, rather than merely $k\geq2$, follows from the level sets of
$\Phi$. Every element of $\mathrm{Lie}(T)$ is a limit in
$\mathrm{Lie}(\mathrm{Isom}(\CS,g))$ of elements of
$\mathrm{span}_{\mathbb R}\{\Vt,U\}$; each of the latter annihilates the moduli $X^I$,
by Theorem 7.5, and annihilation of a fixed smooth function is a closed
condition, so every element of $\mathrm{Lie}(T)$ annihilates $X^I$ and hence
$\Phi=\c V_IX^I$. Every $T$-orbit therefore lies in a level set of $\Phi$. Now
$\tn_i\Phi=\tfrac32\c^2KP_i$ by \rf{gradPhi}, with $K>0$, so $\tn\Phi\neq0$ at
every point of the open set $\{P\neq0\}$, which is non-empty because
$P_i\not\equiv0$ on this branch; at such a point the level set of $\Phi$
through it is an embedded two-dimensional submanifold of $\CS$, and the
$T$-orbit through it, being contained in that level set, has dimension at most
$2$. For an effective action of a torus the principal isotropy group is
trivial \cite[Ch.~IV, \S3 and Ch.~IV, Thm.~3.1]{bredon}, so a principal orbit
is a copy of $T^k$ and has dimension exactly $k$; the principal stratum is
open and dense in $\CS$ \cite[Ch.~IV, Thm.~3.1]{bredon}, so it meets
$\{P\neq0\}$. Hence $k\leq2$, and with $k\geq2$ this gives $k=2$ and
$\mathrm{Lie}(T)=\mathrm{span}_{\mathbb R}\{\Vt,U\}$. The action preserves $h$, $X^I$
and $F^I$ because $\Vt$ and $U$ do, and these are preserved by every element
of the closure by the same limiting argument. $\square$

\medskip\noindent{\it Remark (the hypotheses of Theorem 7.7):} Two conditions
are imposed, and both are conditions on the horizon rather than on the
construction: $K>0$ picks the non-minimal branch and $P_i\not\equiv0$ picks the
varying-moduli case. Three further conditions that one might expect to be
needed --- that $\a\not\equiv0$, that $\{P=0\}$ have empty
interior, and that $U$ extend
to a $C^1$ field --- are not required: the last two are supplied by the
observation that the boundary of a closed set is nowhere dense, together with
the fact that the interior of $\{P=0\}$ is itself a locus on which $U$ is
Killing, and by Lemma 7.6, and the first by Lemma 7.6$'$. In particular $\{P=0\}$ may be
as large and as wild as it likes, up to and including having non-empty
interior; no analyticity is assumed anywhere, and none is needed.

The density argument in the proof of Theorem 7.7 carries one step only, and it
is worth saying which, since a field obtained as a limit on a dense set need
only be continuous. That is not what happens here.
The smoothness of $U$ is not obtained by extension at all: \rf{Uglobal}
exhibits $U$ as a polynomial in fields that are smooth on all of $\CS$, with
no denominator at all. (On the $\a\not\equiv0$ branch \rf{Usmooth} already
suffices, since there $1-Z_kW^k$ is bounded away from zero by \rf{Fgap};
\rf{Uglobal} is what covers the remaining branch.) So $U\in C^\infty(\CS)$
outright, with no limit taken anywhere. Density is used for one thing only: the Killing equation
$\tn_{(i}U_{j)}=0$ is known on $\CS_0$, and a continuous tensor field vanishing
on a dense subset vanishes identically. That step needs continuity of
$\tn_{(i}U_{j)}$, which the smoothness just established supplies, and it needs
nothing else.

\medskip\noindent{\it Remark (the r\^ole of $\a$, and where it is
needed):} By Corollary 6.13 the condition $\a\not\equiv0$ is equivalent to
$\a$ being nowhere zero, so the two forms may be used interchangeably. What it
buys is the identification of $\CS\setminus\CS_0$ with $\{P=0\}$: by
Corollary 6.5, $\CS\setminus\CS_0=\{Z=-W\}\cup\{Z=W\}=\{P=0\}\cup\{N=0\}$, and
$\a\neq0$ excludes the second piece. Theorem 7.7 does not need that
identification, because \rf{Uglobal} is smooth across $\{Z=W\}$ and
Lemma 7.6$'$ makes $\{Z=W\}$ nowhere dense; the $\a\equiv0$ horizons with
$P_i\not\equiv0$ are therefore covered as well. Two later statements do use
$\a\neq0$, and for
reasons that are not removed by \rf{Uglobal}. The first is the sharper clause
of Theorem 7.5, that the space of Killing vectors orthogonal to $P$ preserving
$h$, $X^I$ and $F^I$ is exactly two-dimensional on each component of $\CS_0$.
The second is the whole of Section~\ref{sec:toric}: there $\{Z=W\}=\emptyset$ is used to
make every piecewise smooth path admissible for Lemma 6.12 in
Proposition 7.11, and the first integral $\a\parallel\eta_-\parallel^4$ of
Lemma 7.9, which drives Theorem 7.14, is identically zero when $\a$ is. A
nowhere-dense $\{Z=W\}$ is not enough for the first of those, since a
hypersurface can separate; so Section~\ref{sec:toric} keeps $\a\not\equiv0$ as a standing
hypothesis, and the $\a\equiv0$ sub-branch remains open there --- open for the
topology only, and not for the symmetry, which Theorem 7.7 supplies on the
branch entire.

\medskip\noindent{\it Remark (a sub-case that does not occur):} One route to
the $\a\equiv0$ sub-branch is closed at the outset, namely asking whether
$h_i=\Phi P_i$: that never happens. By Proposition 4.1 in the form
\rf{normminus2}, $u_ku^k+\a^2=m/\parallel\eta_-\parallel^2$ with
$m=2\parallel\eta_+\parallel^2>0$, so $u=\Phi P-h$ and $\a$ cannot vanish
simultaneously at a point, on any branch. In particular, on the $\a\equiv0$
sub-branch $|h-\Phi P|^2=m/\parallel\eta_-\parallel^2$ is strictly positive
everywhere, $\Vt=\parallel\eta_-\parallel^2u$ is nowhere zero, and $h_N$ blows
up like $(4-F)^{-1/2}$ on approach to $\{Z=W\}$ --- which is exactly the
behaviour that \rf{Uglobal} shows to be harmless, the divergence of $h_N$ being
cancelled by the vanishing of $Y$.

\medskip\noindent{\it Remark (the explicit example as a check):} On the
explicit compact varying-moduli horizon of Section~\ref{sec:varmod}, $\{P=0\}$ is the two-point set
$\{x=4,5\}$ and $\a=\sqrt{311}\,H^{-2/3}$ is nowhere zero, so Theorem 7.7
applies and predicts that $U$ is smooth there. It is, and in the strongest
possible way. In the coordinates $(x,\phi^1,\phi^2)$ of \cite[\S4.2]{klr}, in
which the cross-section is the interval $x\in[4,5]$ times the two-torus
generated by $\partial_1$ and $\partial_2$, the two Killing fields have
constant contravariant components,
\bea
\la{klrUexact}
\Vt=-\partial_1 ,\qquad U=-30\,\partial_2 ,
\ee
identically in $x$ rather than in a limit, with
$\det\big(\begin{smallmatrix}-1&0\\0&-30\end{smallmatrix}\big)=30\neq0$, so
the two are independent at every point of the closed interval. This is a
genuine collapse: the ingredients of \rf{secondkilling} are not simple ---
$h_Y$ carries the surd $\sqrt{311}$, and $h_N$ is $x^{1/3}(x^2-9)^{1/3}$ times
a ratio of polynomials of degrees $28$ and $29$ --- and both $N$ and $Y$
degenerate at the endpoints, yet the combination
$\parallel\eta_-\parallel^2(h_YN-h_NY)$ is a fixed element of the Lie algebra
of the torus action, which is smooth on the compactified $S^3$
\cite[\S4.3.2]{klr}. Appendix~\ref{app:klr} gives the computation, together with an independent
high-precision evaluation which it reproduces digit for digit. The example is a
consistency check on Theorem 7.7, not an input to it.

\medskip
These results assemble into the following statement about a compact
supersymmetric $AdS_5$ black ring in this theory. Corollary 7.3 splits every non-minimal compact
horizon into the two branches above, and neither is empty: the $U(1)^3$
example above shows the constant-modulus branch is non-vacuous, and the
explicit compact varying-moduli horizon of Section~\ref{sec:varmod} shows the
second branch is too. On the first branch a second rotational isometry is
unconditional, by Proposition 7.4 together with \cite[\S3.3]{klr}; a ring
would sit in the $S^1\times S^2$ window
$\tfrac13\Phi^2<\c^2K\leq\tfrac43\Phi^2$ of \rf{ringconds}--\rf{ringpoint},
held in equilibrium by a conical singularity in five dimensions, exactly as in
the ungauged and minimal-gauged theories. On the second branch a second
Killing field is available unconditionally on the open set $\CS_0$ by
Theorem 7.5, and upgrades to a genuine isometry of all of $\CS$ by
Theorem 7.7, with no hypothesis on $\a$; what $\a\not\equiv0$ then buys is the
topology, not the symmetry.

On the constant-modulus branch no exclusion of the $S^1\times S^2$
cross-section is claimed here, and no claim is made that a supersymmetric
$AdS_5$ black ring does or does not exist there: deciding whether the resulting
toric near-horizon geometry closes into a genuine ring, and whether the
conical-defect equilibrium condition can be solved globally, is a question
about the constant-modulus system of \cite[\S3.3]{klr}, and lies outside the
scope of this paper. On the varying-moduli branch the question is settled, in
the negative, by Theorem 7.14 below: whenever $\a\not\equiv0$ there is no
ring at all.

What supersymmetry forces on both branches is a second rotational Killing
field --- on $P\not\equiv0$ this is Theorem 7.5 on the open set $\CS_0$, and
on $P\equiv0$, where $\CS_0$ is empty because $W=-Z$, it is the local
homogeneity of Proposition 7.4 --- and a genuine second isometry of all of
$\CS$, unconditionally on both branches. The $\a$ hypothesis enters only
afterwards, when that isometry is used to read off the topology.

\subsection{The toric reduction, and what it decides about the topology}\la{sec:toric}

Theorem 7.7 turns the two commuting Killing fields into an effective action of
a two-torus on the whole of $\CS$. Everything before this point has been about
producing that action; this subsection spends it. The action is cohomogeneity
one (Lemma 7.8), its orbit space is a closed interval with two circular
boundary orbits (Proposition 7.10), and the near-horizon system reduces to a
gradient flow across that interval. Three things then follow. The topology of
$\CS$ is $S^3$, a lens space or $S^1\times S^2$, with the three cases separated
by a single determinant in the boundary data (Theorem 7.12). The flow carries
two first integrals (Lemma 7.9), and evaluating them at the two ends
over-determines that data enough to exclude the third case outright, so that the
varying-moduli supersymmetric $AdS_5$ black ring does not exist when
$\a\not\equiv0$ (Theorem 7.14).
Finally, an independent and weaker route to the same exclusion, through the sign
of the scalar potential rather than through the first integrals, locates the
obstruction in the same place the constant-moduli window of Section~\ref{sec:constmod} does
(Lemma 7.15 to Theorem 7.17). The criterion is also checked directly on the
explicit compact horizon of Section~\ref{sec:varmod}, where it excludes $S^1\times S^2$.

Throughout, $\CS$ is compact and connected, $K>0$, $P_i\not\equiv0$ and
$\a\not\equiv0$. Theorem 7.7 needs only the first three; the last is what the
cohomogeneity-one reduction below adds. By Corollary 6.13, $\a$ is then
nowhere zero, and $\Phi$ is nowhere zero by the argument following \rf{hP}.
We write
$\sigma:=\parallel\eta_-\parallel^2$, and $F$, $\vartheta$, $h_N$, $h_Y$ as in
\rf{deltadef} and \rf{hY}.

\medskip\noindent{\it Lemma 7.8 (the action is cohomogeneity one):} The
one-form $P_i$ annihilates both Killing fields,
\bea
\la{Pperp}
P^i\Vt_i=0 ,
\qquad
P^iU_i=0 ,
\ee
so every $T^2$-orbit is tangent to $P^\perp$. The orbits have dimension two
exactly where $P\neq0$, since the Gram determinant of the two fields is
\bea
\la{gram}
|\Vt|^2|U|^2-(\Vt_iU^i)^2
=\tfrac14\,\sigma^4\,F\,(4-F)^2\big(h_N^2+h_Y^2\big)^2 ,
\ee
and $h_Y=\a/(1-Z_kW^k)$ is nowhere zero. Moreover $F$, $\Phi$, $\a$, $h_N$ and
$\sigma$ are all constant on each orbit, and $\Phi$ is strictly monotone along
any curve tangent to $P$.

\medskip\noindent{\it Proof:} The first identity is already recorded below
\rf{hP}: \rf{Vt} gives $\Vt_i=\sigma(\Phi P_i-h_i)$, so
$P^i\Vt_i=\sigma(\Phi F-h_iP^i)$, which vanishes by \rf{hP}. What is new is
that the same holds for the second Killing field. For the second, \rf{secondkilling} writes
$U_i=\sigma(h_YN_i-h_NY_i)$, and $N$ and $Y$ are orthogonal to $P$ by
construction. Where $P\neq0$ the frame $\{P,N,Y\}$ is orthogonal with
$N_kN^k=4-F$ and $Y_kY^k=\tfrac14F(4-F)$, and a direct evaluation of the
$2\times2$ Gram matrix of $\Vt=-\sigma(h_NN+h_YY)$ --- which is \rf{Vt}
rewritten using \rf{hP} --- and of $U$ gives \rf{gram}. Its right-hand side
vanishes precisely at $F=0$, because $h_Y\neq0$. Every one of $F$, $\Phi$,
$\a$, $h_N$ has gradient proportional to $P_i$, by \rf{gradsys} and
\rf{gradFdelta}, and so does $\sigma$ by \rf{normminus}; since the orbits are
tangent to $P^\perp$, all five are orbit-invariant. Finally
$\tn_i\Phi=\tfrac32\c^2KP_i$ with $K>0$, so $P^i\tn_i\Phi=\tfrac32\c^2KF>0$
wherever $P\neq0$. $\square$

\medskip\noindent{\it Lemma 7.9 (two global first integrals):} Set
\bea
\la{Gdef}
\CG:=h_kh^k-\Phi^2P_kP^k+\a^2 .
\ee
Then $\a\parallel\eta_-\parallel^4$ is constant on $\CS$ for any $\a$, and so
is $\CG\parallel\eta_-\parallel^2$ --- the latter unconditionally, on every
branch, because it is twice the constant $\parallel\eta_+\parallel^2$:
\bea
\la{firstintegral}
\a\,\parallel\eta_-\parallel^4=:k\neq0 ,
\qquad
\CG\,\parallel\eta_-\parallel^2=2\parallel\eta_+\parallel^2=:m>0 .
\ee
Moreover $\CG=\big|h-\Phi P\big|^2+\a^2$, so $\CG>0$ wherever $\a\neq0$;
and $k$ has the sign of $\a$, which by Corollary 6.13 is one sign on $\CS$.

\medskip\noindent{\it Proof:} For the first, $\tn_i\a=2\Phi\a P_i$ holds on all
of $\CS$ by Lemma 6.11 and the continuity argument following it, and
$\tn_i\ln\sigma=-\Phi P_i$ is \rf{normminus}. Hence
$\tn_i\ln(\a\sigma^2)=2\Phi P_i-2\Phi P_i=0$, and $\CS$ is connected.

For the second, note first that $\CG$ is manifestly smooth on all of $\CS$: no
frame and no division enters \rf{Gdef}. Since $h_kP^k=\Phi\,P_kP^k$ by
\rf{hP},
\bea
\la{Gsquare}
\big|h-\Phi P\big|^2=h_kh^k-2\Phi\,h_kP^k+\Phi^2P_kP^k=h_kh^k-\Phi^2P_kP^k ,
\ee
which is the stated non-negative form. Comparing it with \rf{udef}, the vector
$h-\Phi P$ is precisely $-u$, so $\CG=u_iu^i+\a^2$ identically, and
Proposition 4.1 turns that into
\bea
\la{Gnormplus}
\CG\,\parallel\eta_-\parallel^2
=\big(u_iu^i+\a^2\big)\parallel\eta_-\parallel^2
=2\parallel\eta_+\parallel^2 ,
\ee
which is constant by \rf{normplus}. No hypothesis on $\a$, on $K$, or on the
set $\{P=0\}$ enters, and no frame is used, so the second first integral holds
on every branch and not only on the one treated in this section. Positivity is
immediate: $\parallel\eta_+\parallel$ cannot vanish, since
$\parallel\eta_+\parallel=0$ would give $\Vt\equiv0$ by \rf{Vt}, which
Lemma 2.1 excludes; so $m=2\parallel\eta_+\parallel^2>0$. $\square$

\medskip
The first integral $\CG\parallel\eta_-\parallel^2$ is therefore not a new
conserved quantity at all: \rf{Gnormplus} says it is the constant norm of
$\eta_+$ recorded in \cite{kayani}, written in terms of $h$, $\Phi$, $P$ and
$\a$ rather than of the spinor. What the rewriting buys is that this constant
becomes {\it evaluable on the horizon data}, and in particular at the ends of
the orbit space, which is where Theorem 7.14 needs it. The genuinely new
invariant is the first, $\a\parallel\eta_-\parallel^4$, and it is not
visible in \rf{gradsys} term by term: the growth of $\a$ along $P$ is exactly
twice the decay of $\sigma$, so the two conspire.

An independent check of \rf{Gnormplus}, and the form of $\CG$ that Section~\ref{sec:toric}
uses, comes from the frame. Appendix \ref{app:transport} records that on
$\{0<F<4\}$
\bea
\la{Gframe}
\CG=\big(h_N^2+h_Y^2\big)(4-F)
\ee
and that differentiating it along $P$ gives
\bea
\la{gradG}
\tn_i\CG=\Phi\,\CG\,P_i ,
\ee
so that $\tn_i(\sigma\CG)=-\Phi\sigma\CG P_i+\sigma\Phi\CG P_i=0$ there,
reproducing the constancy just proved. The consequence used at once is the
value at a degenerate orbit. There $P=0$, so $h-\Phi P=h$, and with $F=0$ one
has $N_kN^k=4$ and $Y_kY^k=0$, whence $|h|^2=4h_N^2$ and
\bea
\la{Gends}
\a^2+4h_N^2=\frac{m}{\sigma}
\qquad\text{on every orbit with }P=0 ,
\ee
and the combination $\tfrac12\a^2+2h_N^2$ appearing in \rf{Lnorm} and
\rf{Lprim} below is therefore $m/2\sigma$, not an independent quantity.

\medskip\noindent{\it Remark (what is new here, and what is not):} Extracting
from the horizon equations a scalar they force to be constant is an established
move. For vacuum extremal horizons \cite{dl} the contracted Bianchi identity
yields exactly such a constant, the near-horizon analogue $A$ of the surface
gravity, and \cite{colling} carries it to arbitrary matter satisfying the null
energy condition; in both, the constant is what exhibits the near-horizon
geometry as a fibration over $AdS_2$ and so drives the symmetry enhancement. No
claim of novelty is made for that mechanism. Nor is any made for the second
member of \rf{firstintegral}: by \rf{Gnormplus} it is $2\parallel\eta_+
\parallel^2$, which \cite{kayani} already knows to be constant. What is new is
the first member, $\a\parallel\eta_-\parallel^4$, together with the use to
which the pair is put. That invariant is quadratic in the supersymmetric
bilinear data rather than in the curvature, it is invisible in the vacuum
system --- $\a$ and $\parallel\eta_-\parallel$ have no vacuum counterparts
--- and it does not enter the construction of the $\mathfrak{sl}(2,\bR)$
factor, which Section 2 already has from \cite{kayani,groverindex}. The two are
used instead to over-determine the horizon data at the two ends of the orbit
space, which is a topological application and not a symmetry one. That neither
depends on $K$, and so both are uniform over cubic prepotentials, is what makes
Theorem 7.14 model-independent.

\medskip\noindent{\it Remark (what the two constants are on the explicit
family):} On the horizons of \cite[\S4.2]{klr} the two invariants can be
evaluated in closed form, and they are not new numbers: with the notation of
Section~\ref{sec:varmod}, in which the family is fixed by a cubic
$\Pi(x)=H(x)-\tfrac14C^2(x-\a_0)^2-D_0^2/C^2$,
\bea
\la{klrconstants}
\CG\,\parallel\eta_-\parallel^2=C^2 ,
\qquad
\a\,\parallel\eta_-\parallel^4=D_0 ,
\ee
identically in the horizon coordinate and identically in all six parameters
$K_I$, $C$, $\a_0$, $D_0$ of the family, so this is a statement about the family
and not about any particular member of it. The two first integrals of Lemma 7.9
therefore restrict, on that family, to the two integration constants of
\cite{klr} themselves. The coincidence is worth recording, and it does not
diminish Lemma 7.9. In \cite{klr} the constants arise as constants of
integration of an ordinary differential system that exists only because a
two-torus symmetry was assumed at the outset, and they carry no meaning off
that ansatz. Lemma 7.9
shows they are values of two scalar invariants of the near-horizon data,
\rf{Gdef} and $\a\parallel\eta_-\parallel^4$, defined on any compact
cross-section of the branch with nothing assumed; and it is only in that form
that they can be evaluated at the two ends of the orbit space of an unknown
horizon, which is what Theorem 7.14 does.

\medskip\noindent{\it Proposition 7.10 (the orbit space is a closed interval
with two degenerate circle orbits):} The orbit space $\CS/T^2$ is a closed
interval. Its two boundary points are orbits on which $F=0$; each is a circle,
never a fixed point; and $\{P=0\}$ consists of exactly those two circles, so in
particular it has empty interior. At a boundary orbit the isotropy subgroup is
generated by
\bea
\la{degen}
L:=\a\,\Vt+2h_N\,U ,
\ee
evaluated there, and $(\a,2h_N)\neq(0,0)$.

\medskip\noindent{\it Proof:} By Lemma 7.8 the action has cohomogeneity one, so
$\CS/T^2$ is a one-dimensional manifold with boundary, that is a closed
interval or a circle. It cannot be a circle: $\Phi$ descends to a continuous
function on the orbit space which, by Lemma 7.8, is non-decreasing along the
flow of $P$ and strictly increasing wherever $P\neq0$, and no such function
exists on a circle unless $P\equiv0$, which is excluded. So the orbit space is
a closed interval, whose two boundary points are the non-principal orbits.

By \rf{gram} an orbit is non-principal exactly when $F=0$ there. Let $p$ be a
point of such an orbit. There $Z_kW^k=-1$, so $W=-Z$, $N=2Z$ and $Y=0$; and
$h_Y\to\tfrac12\a$ by the Remark following Lemma 6.11. Hence
\bea
\la{degenvals}
\Vt\big|_p=-\sigma h_N\,N ,
\qquad
U\big|_p=\tfrac12\sigma\a\,N ,
\qquad
N_kN^k=4 ,
\ee
so the orbit through $p$ is the circle generated by the non-zero vector $N$: it
is one-dimensional, not zero-dimensional, and the action has no fixed points.
The combination that annihilates it is read off from \rf{degenvals}:
$\a\Vt+2h_NU=\sigma(-\a h_N+\a h_N)N=0$, which is \rf{degen}; and $\a\neq0$, so
the coefficient vector is non-zero. Finally $\{P=0\}=\{F=0\}$ is a union of
orbits, since $F$ is orbit-invariant, and it is exactly the set of
non-principal orbits, of which there are exactly two. $\square$

\medskip\noindent{\it Remark (empty interior, twice over):} That $\{P=0\}$ has
empty interior is a conclusion here, not a hypothesis. It also follows independently of the
orbit-space argument: on the interior of $\{P=0\}$ Proposition 7.4 makes
$h_i=3\Phi Z_i$ and $N_i=2Z_i$, hence $h_N=h_iN^i/N_kN^k=\tfrac32\Phi$, and
$\Phi$ and $\a$ are constant there, so \rf{degen} exhibits a
\emph{constant-coefficient} combination $\a\Vt+3\Phi U$ of two global Killing
fields vanishing on an open set. A Killing field vanishing on a non-empty open
subset of a connected manifold vanishes identically, which would contradict the
linear independence of $\Vt$ and $U$ on $\CS_0$ established in Theorem 7.5.
Note also that $h_N=\tfrac32\Phi$ is $\vartheta=0$, consistently with
$\tn^kP_k\equiv0$ on that branch, \rf{ringpoint}.

\medskip\noindent{\it Proposition 7.11 (regularity at a degenerate orbit):} Let
$p$ lie on a boundary orbit, and let $r$ denote proper distance from that orbit
along the flow of $P$. Then
\bea
\la{sqrtF}
\sqrt F=4|\vartheta|\,r+O(r^3) ,
\ee
with $\vartheta\neq0$ there, and the vector field $L$ of \rf{degen} has
\bea
\la{Lnorm}
|L|=4\,\sigma\,|\vartheta|\Big(\tfrac12\a^2+2h_N^2\Big)\,r+O(r^3) ,
\ee
all coefficients being evaluated on the boundary orbit. Consequently the
generator of the isotropy circle whose flow has period $2\pi$ is
\bea
\la{Lprim}
\hat L=\frac{\a\,\Vt+2h_N\,U}
{4\,\sigma\,|\vartheta|\,\big(\tfrac12\a^2+2h_N^2\big)} .
\ee

\medskip\noindent{\it Proof:} Along the flow of $P$ one has
$dr/dt=|P|=\sqrt F$, and $\tn_iF=2\vartheta(4-F)P_i$ by \rf{gradFdelta}, so
\bea
\frac{d\sqrt F}{dr}=\frac{P^i\tn_iF}{2\sqrt F\,\sqrt F}=(4-F)\,\vartheta ,
\ee
which is \rf{sqrtF} on integrating from $F=0$ and using that $\vartheta$ is
smooth. That $\vartheta\neq0$ on a boundary orbit is forced, and Lemma 6.12
is what forces it. A boundary orbit has $F=0$; if $\vartheta=0$ there too, then
\bea
\rho=F+\vartheta^2=0
\ee
at that point, which is exactly the hypothesis of Lemma 6.12. On the branch of
this section $\a\not\equiv0$, so by \rf{ZWalpha} and Corollary 6.13 the set
$\{Z=W\}$ is empty; every piecewise smooth path in $\CS$ is therefore
admissible for that lemma, and $\CS$ is connected. Hence $\rho\equiv0$ on all
of $\CS$, so $F\equiv0$ and $P_i\equiv0$, contradicting the standing
hypothesis $P_i\not\equiv0$ of this branch.

For \rf{Lnorm}, write $L=a\Vt+bU$ with $(a,b)=(\a,2h_N)$ evaluated on the
boundary orbit, so that $a$ and $b$ are constants. Using
$\Vt=-\sigma(h_NN+h_YY)$ and \rf{secondkilling},
\bea
\la{Lsq}
|L|^2=\sigma^2\Big[q^2\,(4-F)+p^2\,\tfrac14F(4-F)\Big] ,
\qquad
q:=b\,h_Y-a\,h_N ,
\quad
p:=a\,h_Y+b\,h_N ,
\ee
and $q$ vanishes on the boundary orbit by the choice of $(a,b)$, since
$h_Y\to\tfrac12\a$ there. By \rf{gradsys} both $\tn_ih_Y$ and $\tn_ih_N$ are
multiples of $P_i$, so $P^i\tn_iq$ carries a factor $F=O(r^2)$ and hence
$dq/dr=O(r)$, giving $q=O(r^2)$. The first term of \rf{Lsq} is therefore
$O(r^4)$ and the second is $\sigma^2p^2F+O(r^4)$ with
$p\to\tfrac12\a^2+2h_N^2$; \rf{sqrtF} then gives \rf{Lnorm}. Smoothness of the
metric across the degenerate orbit is the statement that the isotropy generator
$\hat L$ has $|\hat L|=r+O(r^3)$ with $\hat L$ of period $2\pi$, which is
\rf{Lprim}. $\square$

\medskip\noindent{\it Theorem 7.12 (the topology criterion):} Label the two
boundary orbits by $1$ and $2$ and write $\a_1,h_{N,1},\vartheta_1,\sigma_1$
and $\a_2,h_{N,2},\vartheta_2,\sigma_2$ for the values of $\a$, $h_N$,
$\vartheta$ and $\sigma$ on them. Then $\CS$ is $S^3$, a lens space $L(p,q)$,
or $S^1\times S^2$; and it is $S^1\times S^2$ if and only if
\bea
\la{ringcrit}
\CD:=\a_1h_{N,2}-\a_2h_{N,1}=0 .
\ee
In that case the horizon data must in addition satisfy
\bea
\la{ringcrit2}
\vartheta_1^{\,2}\,\a_1=\vartheta_2^{\,2}\,\a_2 .
\ee

\medskip\noindent{\it Proof:} A neighbourhood of a boundary orbit is the
associated bundle $T^2\times_{S^1}D^2$, the isotropy circle acting on the
two-dimensional normal slice by rotation, that is a solid torus; and the
preimage of the interior of the orbit space is $(t_1,t_2)\times T^2$. So $\CS$
is the union of two solid tori glued along their common boundary torus: a
genus-one Heegaard splitting. The closed orientable three-manifolds admitting
one are exactly $S^3$, the lens spaces, and $S^1\times S^2$, the last occurring
precisely when the two meridian curves --- the isotopy classes killed in the two
solid tori --- coincide up to orientation.

The meridian of the $i$th solid torus is the orbit of $\hat L_i$ of
\rf{Lprim}, so the condition is $\hat L_1=\pm\hat L_2$ in the Lie algebra
$\mathrm{span}\{\Vt,U\}$. Both $\hat L_1$ and $\hat L_2$ are primitive elements
of the period lattice of the $T^2$ action, and two parallel primitive elements
of a lattice agree up to sign; conversely if they agree up to sign they are in
particular parallel. So the condition is that $\hat L_1$ and $\hat L_2$ be
parallel, that is that $(\a_1,2h_{N,1})$ and $(\a_2,2h_{N,2})$ be parallel,
which is \rf{ringcrit}.

For \rf{ringcrit2}, suppose $\CD=0$ and write
$(\a_2,2h_{N,2})=\kappa(\a_1,2h_{N,1})$ with $\kappa\neq0$. Then
$\tfrac12\a_2^2+2h_{N,2}^2=\kappa^2(\tfrac12\a_1^2+2h_{N,1}^2)$, and
substituting into \rf{Lprim} the equality $\hat L_1=\pm\hat L_2$ becomes
$\sigma_1|\vartheta_1|=\kappa\,\sigma_2|\vartheta_2|$, that is
$\sigma_1|\vartheta_1|\a_1=\sigma_2|\vartheta_2|\a_2$ on multiplying by $\a_1$.
By Lemma 7.9, $\a\sigma^2=:k$ is constant, so $\sigma_i\a_i=\sqrt{k\a_i}$ and
squaring gives \rf{ringcrit2}. $\square$

\medskip\noindent{\it Remark:} Condition \rf{ringcrit2} is a necessary
condition for a case that Theorem 7.14 below shows to be empty, so it is
vacuous as it stands. It is recorded because it is what the Heegaard argument
produces on its own, before the first integrals of Lemma 7.9 are brought in,
and because the step from it to Theorem 7.14 is short.

\medskip\noindent{\it Remark (the fundamental group, and what fixes the lens
index):} The Heegaard splitting also computes $\pi_1$. Gluing two solid tori
along their boundary kills the two meridians, so
\bea
\la{pione}
\pi_1(\CS)\ \cong\ \bZ/p\,\bZ ,
\qquad
p=\big|\det{}_\Lambda\!\big(\hat L_1,\hat L_2\big)\big| ,
\ee
the determinant being taken with respect to a $\bZ$-basis of the period lattice
$\Lambda\subset\mathrm{span}\{\Vt,U\}$ of the torus action, with the conventions
$L(0,1)=S^1\times S^2$ and $L(1,0)=S^3$. Both $\hat L_i$ are primitive in
$\Lambda$, by \rf{Lprim} and the period normalisation, so $p$ is a non-negative
integer. By \rf{Gends} the bracket in the denominator of \rf{Lprim} is
$\tfrac12\a^2+2h_N^2=m/2\sigma$, so the normalisation collapses to
\bea
\la{Lprimsimp}
\hat L_i=\frac{\a_i\Vt+2h_{N,i}\,U}{2m\,|\vartheta_i|} ,
\ee
with $\sigma$ gone. Evaluating the determinant on the pair $(\Vt,U)$ and
correcting by the covolume of that pair relative to $\Lambda$,
\begin{samepage}
\bea
\la{pexplicit}
p=\frac{\big|\CD\big|\ c_\Lambda}{2\,m^2\,\big|\vartheta_1\vartheta_2\big|} ,
\qquad
c_\Lambda:=\big|\Vt\wedge U\big|_\Lambda ,
\ee
\end{samepage}\nopagebreak
where $c_\Lambda$ is the covolume of $\Lambda$ measured in the basis
$(\Vt,U)$: if $(e_1,e_2)$ is any $\bZ$-basis of $\Lambda$ and
$e_a=B_a{}^b\,(\Vt,U)_b$, then $c_\Lambda:=|\det B|^{-1}$, so that
$\det{}_\Lambda(\cdot,\cdot)=c_\Lambda\det_{(\Vt,U)}(\cdot,\cdot)$ on
$\mathrm{span}\{\Vt,U\}$. The definition is independent of the choice of
$\bZ$-basis, two such differing by an element of $GL(2,\bZ)$, of determinant
$\pm1$;
and the numerator having picked up $\a_1\cdot2h_{N,2}-\a_2\cdot2h_{N,1}=2\CD$
from \rf{Lprimsimp}. By Lemma 7.15 below $|\vartheta_1\vartheta_2|
=-\vartheta_1\vartheta_2$. By Theorem 7.14 (proved below), $\CD\neq0$, so $p\geq1$
on this branch and $\CS$ is $S^3$ or a lens space, never $S^1\times S^2$.

Two readings of \rf{pexplicit} should be kept apart. The vanishing of $p$ is
$\CD=0$ and is independent of $c_\Lambda$: that is Theorem 7.12, and it is a
statement the near-horizon equations decide by themselves --- in the negative,
by Theorem 7.14. The {\it value} of a non-zero $p$ is not: $c_\Lambda$ records how the abstract lattice $\Lambda$ sits
inside $\mathrm{span}\{\Vt,U\}$, which is global data about the torus action and
is not determined by the near-horizon field equations, since those equations are
insensitive to how the flows of $\Vt$ and $U$ close up. What \rf{pexplicit}
does determine is the product, and with it an integrality constraint on the
horizon data: the right-hand side must be a non-negative integer. No argument
of this kind could force $p=1$, and none should be sought --- lens-space
horizons with $p>1$ occur in five-dimensional supergravity \cite{kllens},
albeit in the {\it ungauged}, asymptotically flat theory, so that no $AdS_5$
example with $p>1$ is on record --- and $S^3$ is therefore not the only regular outcome
that a general argument may leave open.
\medskip\noindent{\it Corollary 7.13 (a monotonicity criterion):} Set
\bea
\la{nudef}
\nu:=\frac{h_N}{\a} ,
\qquad
\CE:=h_Y^2+h_N\big(3\Phi-h_N\big) .
\ee
Then $\CD=\a_1\a_2(\nu_2-\nu_1)$, and
\bea
\la{gradnu}
\tn_i\nu=-\CE\,\frac{P_i}{\a} ,
\qquad\text{so}\qquad
P^i\tn_i\nu=-\frac{F}{\a}\,\CE .
\ee
If $\CE$ has one sign on $\CS$, then $\nu$ is strictly monotone along the
orbit space, $\CD\neq0$, and $\CS$ is not $S^1\times S^2$. In particular
this holds wherever
\bea
\la{window}
h_N\big(3\Phi-h_N\big)\geq0 ,
\ee
that is wherever $h_N$ lies between $0$ and $3\Phi$, since $h_Y$ is nowhere
zero.

\medskip\noindent{\it Proof:} The formula for $\CD$ is immediate. For
\rf{gradnu}, \rf{gradsys} gives $\tn_ih_N=[h_N(h_N-\Phi)-h_Y^2]P_i$ and
$\tn_i\a=2\Phi\a P_i$, so
\bea
\tn_i\nu=\frac{\tn_ih_N}{\a}-\frac{h_N\tn_i\a}{\a^2}
=\frac{P_i}{\a}\Big[h_N^2-\Phi h_N-h_Y^2-2\Phi h_N\Big]
=\frac{P_i}{\a}\Big[h_N^2-3\Phi h_N-h_Y^2\Big] ,
\ee
and the bracket is $-\CE$. If $\CE$ has one sign then, by
Proposition 7.10, $P^i\tn_i\nu$ vanishes only at the two boundary orbits, so
$\nu$ is strictly monotone across the interval and $\nu_1\neq\nu_2$. The last
statement is \rf{window} together with $h_Y\neq0$. $\square$

\medskip
That criterion is in fact always met, and for a reason that has nothing to do
with the sign of $\CE$: the two first integrals of Lemma 7.9 already
over-determine the boundary data.

\medskip\noindent{\it Theorem 7.14 (no varying-moduli black ring):} On the
branch $K>0$, $P_i\not\equiv0$, $\a\not\equiv0$, the cross-section $\CS$ is
never $S^1\times S^2$. Combined with Theorem 7.12 it is therefore $S^3$ or a
lens space $L(p,q)$, and the supersymmetric $AdS_5$ black ring does not occur
with varying moduli and $\a\not\equiv0$. The sub-branch $\a\equiv0$, on which
the reduction of this subsection is unavailable, and the constant-modulus
branch $P\equiv0$ of Section~\ref{sec:constmod} are not covered by this statement.

\medskip\noindent{\it Proof:} Suppose $\CS\simeq S^1\times S^2$, so that
$\CD=\a_1h_{N,2}-\a_2h_{N,1}=0$ by Theorem 7.12. Both $\a_i$ are non-zero, by
Corollary 6.13, so
\bea
\la{kapdef}
\kappa:=\frac{\a_2}{\a_1}\neq0 ,
\qquad
h_{N,2}=\kappa\,h_{N,1} ,
\ee
the second equality being $\CD=0$ divided by $\a_1$. Evaluate \rf{Gends} at
each boundary orbit. Its left-hand side is homogeneous of degree two in
$(\a,h_N)$, so \rf{kapdef} gives
\bea
\la{ringsig}
\frac{m}{\sigma_2}=\a_2^2+4h_{N,2}^2=\kappa^2\big(\a_1^2+4h_{N,1}^2\big)
=\kappa^2\,\frac{m}{\sigma_1} ,
\qquad\text{hence}\qquad
\sigma_1=\kappa^2\sigma_2 ,
\ee
$m>0$ having been divided out. Now use the other first integral,
$\a\sigma^2=k$, at each orbit: $\a_1\sigma_1^2=\a_2\sigma_2^2$ gives
$\a_2/\a_1=\sigma_1^2/\sigma_2^2=\kappa^4$, that is $\kappa=\kappa^4$. Since
$\kappa$ is a non-zero real, $\kappa^3=1$ forces $\kappa=1$, and then
\rf{ringsig} forces
\bea
\la{sigeq}
\sigma_1=\sigma_2 .
\ee
But $\tn_i\sigma=-\Phi\sigma P_i$ with $\Phi$ nowhere zero, so along the flow of
$P$ one has $\dot\sigma=-\Phi\sigma F$, of one strict sign on the open interval
where $F>0$; hence $\sigma$ is strictly monotone across the orbit space and
$\sigma_1\neq\sigma_2$. This contradicts \rf{sigeq}. $\square$

\medskip\noindent{\it Remark (what the proof uses, and what it does not):}
Four inputs enter: the ring condition $\CD=0$ of Theorem 7.12, the two first
integrals of Lemma 7.9, and the strict monotonicity of
$\parallel\eta_-\parallel^2$ along $P$. No hypothesis is made on $K$ beyond
$K>0$, none on the model, none on the sign of $\CE$, and no estimate is used
anywhere. The conclusion is in particular insensitive to the scalar potential,
which is what makes it unconditional where the route through $\CE$ below is
not.

The result is consistent with, and independently recovers, part of the
classification of \cite{klr}: there the scalars are shown to be constant on the
$S^1\times S^2$ branch of toric $AdS_5$ near-horizon geometries, under the
standing assumption that the geometry is toric. Here toricity is not assumed
but derived, from Theorem 7.7, and the same conclusion follows on the branch to
which that theorem applies. The two arguments are quite different --- theirs
runs through the classification of the toric data, this one through two
conserved quantities of the radial system --- so the agreement is a check
rather than a restatement.

\medskip\noindent{\it Remark (where this sits among the ring no-gos):} For the
{\it minimal} gauged theory the question is settled twice over. Kunduri,
Lucietti and Reall \cite{klrring} determined the most general supersymmetric
near-horizon geometry of that theory admitting two rotational symmetries, found
it to be the spherical one, and concluded that supersymmetric $AdS_5$ black
rings with two rotational symmetries do not exist. Grover, Gutowski and Sabra
\cite{grover} removed the symmetry assumption, using the supersymmetry
enhancement of \cite{groverindex} to fix the near-horizon data outright.
Theorem 7.14 is the vector-multiplet counterpart of the second of these, on the
branch where the moduli are not constant: the rotational symmetry is derived
rather than assumed, as in \cite{grover}, but the argument does not go through
an explicit determination of the near-horizon geometry --- which the enlarged
field content puts out of reach --- and instead reads the answer off the
boundary data of the orbit space. The three results are therefore complementary
rather than nested, and together they leave two places in which a
supersymmetric $AdS_5$ ring-topology near-horizon geometry can still occur in
this theory: the constant-moduli branch $P\equiv0$ of Section~\ref{sec:constmod}, with
$\tfrac13\Phi^2<\c^2K\leq\tfrac43\Phi^2$, and the varying-moduli sub-branch
$\a\equiv0$, on which the reduction of Section~\ref{sec:toric} is unavailable and the
topology is decided by nothing here.

\medskip\noindent{\it Remark (the constant-moduli branch is untouched):}
Theorem 7.14 says nothing about $P_i\equiv0$. There the orbit space is not an
interval, $F\equiv0$ and $\sigma$ is constant, so \rf{ringsig} and \rf{sigeq}
are vacuous; the $S^1\times S^2$ solutions of \cite{klr} live on that branch and
are excluded by nothing here. The window
$\tfrac13\Phi^2<\c^2K\leq\tfrac43\Phi^2$ of Section~\ref{sec:constmod} remains the operative
statement for them.

\medskip
The remainder of this subsection records an independent route to a weaker
conclusion, through the sign of $\CE$ rather than through the first integrals.
It is worth recording for two reasons: it uses Lemma 7.9 nowhere, so it checks Theorem 7.14
by a disjoint mechanism; and it locates the obstruction in the scalar potential,
which is where a reader coming from the constant-moduli window of Section~\ref{sec:constmod}
would expect to find it.

\medskip\noindent{\it Remark (where the criterion can fail, and by how
little):} By \rf{window}, $\CE$ can fail to be positive only where $h_N$
leaves the closed interval with endpoints $0$ and $3\Phi$ --- a window whose
midpoint $h_N=\tfrac32\Phi$ is exactly the value taken on the branch
$P\equiv0$, by the Remark following Proposition 7.10. Differentiating $\CE$
with \rf{gradsys} and eliminating $h_Y^2$ on $\{\CE=0\}$ gives
\bea
\la{gradEsign}
P^i\tn_i\CE\Big|_{\CE=0}
=\tfrac12F\,h_N\Big[\,8h_N^2-36\Phi h_N+24\Phi^2+9\c^2K\,\Big] ,
\ee
whose bracket, with the $\c^2K$ term dropped, has roots
$h_N=\tfrac14(9\pm\sqrt{33})\Phi$. That form of the bracket conceals what
controls its sign. Eliminating $K$ in favour of the scalar potential $U$
through \rf{chiU} gives the bracket the factored shape
\bea
\la{gradEsignU}
8h_N^2-36\Phi h_N+24\Phi^2+9\c^2K
=4\big(2h_N-3\Phi\big)\big(h_N-3\Phi\big)-2\c^2U ,
\ee
so that
\bea
\la{gradEfact}
P^i\tn_i\CE\Big|_{\CE=0}
=F\,h_N\Big[\,2\big(2h_N-3\Phi\big)\big(h_N-3\Phi\big)-\c^2U\,\Big] .
\ee
Three things follow at once. They are stated for $\Phi>0$; the case $\Phi<0$
is the mirror image, since $\CE$ and the bracket in \rf{gradEfact} are both
invariant under $(h_N,\Phi)\to(-h_N,-\Phi)$ while $Fh_N$ and $P^i\tn_i$ both
change sign, and the reader may substitute $\mu=h_N/\Phi$ for $h_N/\Phi$
throughout. The quadratic exceeds $18\Phi^2$ for $h_N\leq0$,
while $\c^2U<6\Phi^2$ because $K>0$; so for $h_N\leq0$ the bracket is positive
and, $h_N$ being non-positive, $\CE$ can cross zero only downwards there. The
bracket vanishes at $h_N=\tfrac14\big(9\pm\sqrt{9+4\c^2U/\Phi^2}\big)\Phi$, so
the window in which the sign is undecided is
\bea
\la{exactwindow}
3\Phi\ <\ h_N\ <\ \tfrac14\Big(9+\sqrt{9+4\c^2U/\Phi^2}\,\Big)\Phi ,
\ee
which is contained in $3\Phi<h_N<\tfrac14(9+\sqrt{33})\Phi$, with equality
only in the limit $K\to0$, and which is {\it empty} when $U\leq0$. The window
is therefore not an artefact of the estimate: its width is a measure of the
scalar potential, and it closes exactly when that potential is non-positive,
that is when $\c^2K\geq\tfrac43\Phi^2$.

\medskip\noindent{\it Remark (why this is not a phase-plane problem):} It is
tempting to try to settle the sign of $\CE$ by treating \rf{gradsys} and
\rf{gradFdelta} as a dynamical system and looking for a trapping region. Let
$t$ be the parameter of the flow of $P$, so that $d/dt=P^i\tn_i$. The equations
become the autonomous system
\bea
\la{odesys}
\dot\Phi=\tfrac32\c^2K\,F ,
\quad
\dot\a=2\Phi\a\,F ,
\quad
\dot h_N=\big[h_N(h_N-\Phi)-h_Y^2\big]F ,
\quad
\dot F=2\vartheta(4-F)F ,
\ee
with $h_Y=2\a/(4-F)$ and $\vartheta=h_N-\tfrac32\Phi$, supplemented by
$\dot\sigma=-\Phi\sigma F$ and the first integral $\a\sigma^2=k$ of Lemma 7.9,
and subject to the two-point condition $F(0)=F(L)=0$ with $F>0$ in between,
which is Proposition 7.10. Every right-hand side is polynomial in
$(\Phi,\a,h_N,F)$ after clearing the single denominator $4-F$, itself bounded
below by \rf{Fgap}.

The system is nevertheless not closed. The coefficient $K$ obeys
$\tn_iK=K'P_i$ with $K'$ a scalar built from the very special geometry, and, as
already noted in Section~\ref{sec:dichotomy}, that scalar is model data: it depends on
$C^{IJK}$ and on where the moduli sit in the scalar manifold, and no relation
available here expresses it through $(\Phi,\a,h_N,F)$. So \rf{odesys} is a
four-dimensional polynomial system driven by a free non-negative function
$K(t)$, and its phase space is infinite-dimensional once the model is allowed
to vary. A trapping region or Lyapunov function in $(\Phi,h_N,F)$ would
therefore have to hold for every admissible $K(\cdot)$ at once. That is exactly
what the next three results do, and it is why they are phrased through the sign
of $U$: by \rf{chiU} the combination in which $K$ enters \rf{gradEsign} is the
scalar potential, so quantifying over all models reduces to quantifying over
the sign of a single invariant.

\medskip
The gap can be narrowed much further, because the two ways for $h_N$ to leave
the interval $[0,3\Phi]$ of \rf{window} are not on the same footing. Orient the
orbit space by the flow of $P$, so that $\Phi$ increases from the boundary orbit
$1$ to the boundary orbit $2$, and let $t$ be the flow parameter, with $F>0$ on
$(0,L)$ and $F=0$ at both ends.

\medskip\noindent{\it Lemma 7.15 (the boundary values of $\vartheta$):} No
sign convention for $\Phi$ is needed here or below; $\Phi$ is nowhere zero by
the argument following \rf{hP}, hence of one sign on the connected $\CS$, but
which sign is immaterial. One has
\bea
\la{thetaends}
\vartheta_1>0>\vartheta_2 ,
\qquad\text{that is}\qquad
h_{N,1}>\tfrac32\Phi_1 ,
\qquad
h_{N,2}<\tfrac32\Phi_2 .
\ee

\medskip\noindent{\it Proof:} By the proof of Proposition 7.11,
$d\sqrt F/dr=(4-F)\vartheta$ with $r$ proper distance along the flow of $P$ and
$4-F>0$ by \rf{Fgap}. At the orbit $1$ the function $F$ leaves zero as $r$
increases, so $d\sqrt F/dr>0$ there and $\vartheta_1>0$; at the orbit $2$ it
returns to zero as $r$ increases, so $d\sqrt F/dr<0$ there and $\vartheta_2<0$.
Neither can vanish, by Proposition 7.11. This argument reads only the
direction in which $F$ leaves and returns to zero along the flow of $P$; the
sign of $\Phi$ plays no part.

The same signs follow independently from Corollary 7.2, and again for either
sign of $\Phi$. One has $\tn^kP_k=2(4-F)\vartheta-\Phi F$, which equals
$8\vartheta$ at $F=0$, while $\tn_i\sigma=-\Phi\sigma P_i$ gives
$\dot\sigma=-\Phi\sigma F$ along the flow, so that $\sigma$ is strictly
monotone across the orbit space: decreasing if $\Phi>0$, increasing if
$\Phi<0$. If $\Phi>0$ the orbit $1$ carries the maximum $p$ of $\sigma$ and
the orbit $2$ its minimum $q$, so \rf{divPsign} reads
$8\Phi\vartheta_1\geq0\geq8\Phi\vartheta_2$, and dividing by $\Phi>0$ gives
$\vartheta_1\geq0\geq\vartheta_2$. If $\Phi<0$ the two ends exchange roles,
the orbit $1$ carrying $q$ and the orbit $2$ carrying $p$, so \rf{divPsign}
reads $8\Phi\vartheta_1\leq0\leq8\Phi\vartheta_2$, and dividing by $\Phi<0$
gives the same $\vartheta_1\geq0\geq\vartheta_2$. $\square$

\medskip\noindent{\it Proposition 7.16 ($h_N$ of the wrong sign already
excludes the ring):} If $\Phi h_N<0$ at some point of $\CS$ --- that is, if
$h_N$ takes somewhere the sign opposite to the (constant) sign of $\Phi$ ---
then $\CD\neq0$ and $\CS$ is not $S^1\times S^2$. For $\Phi>0$ this is the
hypothesis $h_N<0$ somewhere; for $\Phi<0$ it is the hypothesis $h_N>0$
somewhere.

\medskip\noindent{\it Proof:} On $(0,L)$ one has $F>0$, and \rf{gradsys} gives
$P^i\tn_ih_N=\big[h_N(h_N-\Phi)-h_Y^2\big]F$, which at $h_N=0$ is $-h_Y^2F<0$
because $h_Y=\a/(1-Z_kW^k)$ is nowhere zero. So $h_N$ can cross zero only
downwards along the flow of $P$: the set $\{h_N<0\}$ is forward invariant in
$(0,L)$ and the set $\{h_N>0\}$ is backward invariant. This uses the flow
direction, fixed by $P$, and not the sign of $\Phi$. Recall also that $\a$ is
nowhere zero by Corollary 6.13 and therefore of one sign on the connected
$\CS$, and likewise $\Phi$ by the argument following \rf{hP}.

Take first $\Phi>0$, so that the hypothesis reads $h_N(t_0)<0$ for some
$t_0\in[0,L]$. Lemma 7.15 gives $h_{N,1}>\tfrac32\Phi_1>0$, so $t_0$ is not
the orbit $1$. If $t_0\in(0,L)$, forward invariance gives $h_N<0$ on $(t_0,L)$
and hence, by continuity, $h_{N,2}\leq0$; if $t_0$ is the orbit $2$ then
$h_{N,2}<0$ outright. Either way $h_{N,1}>0\geq h_{N,2}$.

Take instead $\Phi<0$, so that the hypothesis reads $h_N(t_0)>0$ for some
$t_0\in[0,L]$. Lemma 7.15 gives $h_{N,2}<\tfrac32\Phi_2<0$, so $t_0$ is not
the orbit $2$. If $t_0\in(0,L)$, backward invariance gives $h_N>0$ on $(0,t_0)$
and hence $h_{N,1}\geq0$; if $t_0$ is the orbit $1$ then $h_{N,1}>0$ outright.
Either way $h_{N,1}\geq0>h_{N,2}$.

In both cases $h_{N,1}$ and $h_{N,2}$ have weakly opposite signs with at least
one of the two strict, so they are distinct and not both zero. Since $\a$ has a
single sign, $\nu_1=h_{N,1}/\a_1$ and $\nu_2=h_{N,2}/\a_2$ then satisfy
$\nu_1\neq\nu_2$. By Corollary 7.13, $\CD=\a_1\a_2(\nu_2-\nu_1)\neq0$.
$\square$

\medskip\noindent{\it Theorem 7.17 (a ring needs a positive scalar potential):}
Suppose $\CS\simeq S^1\times S^2$ on the varying-moduli branch. This is the
one statement in the paper whose third clause uses the inequality $U\geq0$
itself, and not merely its consequence \rf{Phinonzero}; see Proposition 7.E.
Then
\bea
\la{ringneeds}
\mu\geq0 \ \text{ everywhere},
\qquad
\mu>3 \ \text{ somewhere},
\qquad
U>0 \ \text{ somewhere} ,
\nn
\text{where}\quad \mu:=\frac{h_N}{\Phi} ;
\ee
the ratio $\mu$ is well defined because $\Phi$ is nowhere zero, by the
argument following \rf{hP}.
Equivalently $\c^2K<\tfrac43\Phi^2$, that is $\c^2\lambda<\Phi^2$, at some
point of the horizon. In particular, if $\c^2K\geq\tfrac43\Phi^2$ everywhere,
or if $\mu\leq3$ everywhere, then $\CS$ is not $S^1\times S^2$. For $\Phi>0$
the first two conditions read $h_N\geq0$ everywhere and $h_N>3\Phi$ somewhere;
for $\Phi<0$ they read $h_N\leq0$ everywhere and $h_N<3\Phi$ somewhere.

\medskip\noindent{\it Proof:} A ring is $\CD=0$, that is $\nu_1=\nu_2$ by
Corollary 7.13. The first requirement is Proposition 7.16. For the second,
$\nu_1=\nu_2$ and $P^i\tn_i\nu=-F\CE/\a$ with $F>0$ on $(0,L)$ force $\CE$
either to vanish identically or to change sign, and in both cases $\CE$ has a
zero $t_*\in(0,L)$. At a zero, $h_Y^2=h_N(h_N-3\Phi)$, and the left-hand side
is strictly positive because $h_Y$ is nowhere zero. Dividing by $\Phi^2>0$ and
writing $\mu=h_N/\Phi$, this says $\mu(t_*)\big(\mu(t_*)-3\big)>0$, so
$\mu(t_*)$ lies strictly outside $[0,3]$; the first requirement gives $\mu\geq0$,
which leaves $\mu(t_*)>3$. Note that the division is by $\Phi^2$, so the
conclusion is insensitive to the sign of $\Phi$.

For the third, note that $U\geq0$ is a standing hypothesis, so ``$U>0$
somewhere'' is the negation of $U\equiv0$; it is that negation which is proved,
and it is proved without any case analysis on $\CE$. Suppose $U\equiv0$. By
\rf{chiU}, $\c^2U=6\Phi^2-\tfrac92\c^2K$, so $U\equiv0$ is
$\c^2K=\tfrac43\Phi^2$, and the first line of \rf{gradsys} becomes
\bea
\la{Uzerograd}
\tn_i\Phi=\tfrac32\c^2K\,P_i=2\Phi^2P_i ,
\qquad\text{alongside}\qquad
\tn_i\a=2\Phi\a\,P_i .
\ee
Hence
\bea
\tn_i\Big(\frac{\a}{\Phi}\Big)
=\frac{\Phi\,\tn_i\a-\a\,\tn_i\Phi}{\Phi^2}
=\frac{2\Phi^2\a P_i-2\Phi^2\a P_i}{\Phi^2}=0 ,
\ee
so $\a/\Phi=:c$ is a non-zero constant on the connected $\CS$, both $\a$ and
$\Phi$ being nowhere zero. Then $\nu=h_N/\a=(h_N/\Phi)/c$, and the ring
condition $\nu_1=\nu_2$ becomes
\bea
\la{muequal}
\frac{h_{N,1}}{\Phi_1}=\frac{h_{N,2}}{\Phi_2} .
\ee
But Lemma 7.15 gives $\vartheta_1>0>\vartheta_2$, that is
$h_{N,1}-\tfrac32\Phi_1>0>h_{N,2}-\tfrac32\Phi_2$. Dividing through by
$\Phi_1$ and by $\Phi_2$, which have the same sign since $\Phi$ is nowhere
zero on the connected $\CS$, places $h_{N,1}/\Phi_1$ and $h_{N,2}/\Phi_2$
strictly on opposite sides of $\tfrac32$ --- above and below if $\Phi>0$,
below and above if $\Phi<0$ --- and either way contradicts \rf{muequal}.
Hence $U\not\equiv0$, and with $U\geq0$ that is $U>0$ somewhere. The
translations $\c^2U=6\Phi^2-\tfrac92\c^2K$ and
$\lambda=K-\tfrac13\Phi^2/\c^2$ are \rf{chiU} and \rf{lambdaK}. $\square$

\medskip\noindent{\it Remark (how much the third clause says):} Since $U\geq0$
is standing, and $U\geq0$ is $\c^2K\leq\tfrac43\Phi^2$ everywhere, the third
clause excludes exactly the borderline case $U\equiv0$, in which the potential
vanishes identically and $\c^2\lambda\equiv\Phi^2$. That is a narrow
statement, and it is stated as such; its interest is that it is reached from
the ring condition alone, by an argument that never leaves the two ends of the
orbit space. The argument above also avoids the route through the sign of
$P^i\tn_i\CE$ at a zero of $\CE$, which does not close: at a zero one has
$\mu>3$, and if $\CE\leq0$ at the first boundary orbit --- possible when
$\mu_1>3$ --- the first zero is an {\it upward} crossing, at which
\rf{gradEfact} yields $\c^2U\leq2(2h_N-3\Phi)(h_N-3\Phi)$ and no lower bound
on $U$ at all.

\medskip\noindent{\it Remark (the same window on both branches):} On
the constant-moduli branch the ring window was found in Section~\ref{sec:constmod} to be
$\tfrac13\Phi^2<\c^2K\leq\tfrac43\Phi^2$, that is $0<\c^2\lambda\leq\Phi^2$,
with $\lambda$ constant there. Theorem 7.17 places the varying-moduli ring
inside the same upper bound, required only at a point rather than
identically. So on both branches a supersymmetric $AdS_5$ black-ring horizon
requires the scalar potential to be positive, and outside
$\c^2\lambda<\Phi^2$ there is no ring at all, with no hypothesis of any kind.
By Theorem 7.14 the varying-moduli case of that statement is in fact vacuous:
no ring occurs there at all, whatever the potential does. What the route
through $\CE$ adds is the reason a ring is hard to build on this branch even
before the first integrals are used --- it would need $h_N$ to exceed $3\Phi$
while the potential is positive enough to satisfy
$2(2h_N-3\Phi)(h_N-3\Phi)<\c^2U$ at a zero of $\CE$ --- and the observation
that the constant-moduli window of Section~\ref{sec:constmod} is the same inequality
$\c^2\lambda\leq\Phi^2$ seen from the other branch.

\medskip\noindent{\it Remark (the explicit horizon of Section~\ref{sec:varmod} is not
$S^1\times S^2$):} On that horizon the criterion can be evaluated in closed
form. Away from $\{P=0\}$ the divergence identity
$\tn^iP_i=2h_iN^i-8\Phi+4\Phi Z_kW^k$ of Section 5 reads
$\tn^kP_k=2(4-F)\vartheta-\Phi F$, so
$\vartheta=(\tn^kP_k+\Phi F)/\big(2(4-F)\big)$, and with $F=4\Pi/H$ the two
cube roots cancel to leave
\bea
\la{klrhN}
h_N=\frac{(30x-83)\,H^{1/3}}{2\big(H-\Pi\big)} ,
\qquad
3\Phi=\frac{3(x^2-3)}{H^{2/3}} ,
\ee
with $H=x^3-9x$ and $\Pi$ as in \rf{Pispecial}. Consider the three polynomials
\bea
\la{klrpolys}
H-\Pi=15x^2-83x+120 ,
\qquad
30x-83 ,
\nn
H\big(\Pi'+3(x^2-3)\big)-6(x^2-3)\Pi=60x^4-415x^3+720x^2+747x-2160 .
\ee
Each is positive for every $x\geq4$: writing $x=4+s$ they become
\bea
\la{klrshift}
15s^2+37s+28 ,
\qquad
30s+37 ,
\qquad
60s^4+545s^3+1500s^2+1947s+1148 ,
\ee
with every coefficient positive. Now $H=x(x-3)(x+3)>0$ on $[4,5]$, so the
positivity of $H-\Pi$ gives $4-F=4(H-\Pi)/H>0$; given that, \rf{klrhN} and the
positivity of $30x-83$ give $h_N>0$; and clearing the positive denominators in
$h_N<3\Phi$ turns it into
$6(x^2-3)(H-\Pi)-(30x-83)H>0$, which is the third polynomial. Hence
\rf{window} holds on the whole horizon,
$\CE>0$, and Corollary 7.13 applies: $\CS$ is not $S^1\times S^2$. The
endpoint data confirm it directly.

{\it Which end is which.} The labels $1,2$ are fixed by the orientation
convention of Section~\ref{sec:toric}, not by the coordinate $x$: the boundary orbit $1$
is the one the flow of $P$ leaves. Three independent readings agree that this
is $x=5$. From \rf{klrhN},
$\vartheta(4)=\tfrac12\big(37-39\big)/28^{2/3}=-1/28^{2/3}<0$ and
$\vartheta(5)=\tfrac12\big(67-66\big)/80^{2/3}=+1/\big(2\cdot80^{2/3}\big)>0$,
while Lemma 7.15 requires $\vartheta_1>0>\vartheta_2$. The divergence agrees,
since \rf{klrclosed} gives $\tn^kP_k=8\vartheta$ at each end; and so does
$\sigma$, which by Lemma 7.9 is proportional to $H^{1/3}$ and hence larger at
$x=5$ ($H=80$) than at $x=4$ ($H=28$), while $\tn_i\sigma=-\Phi\sigma P_i$ with
$\Phi>0$ makes $\sigma$ maximal at the orbit $1$. So orbit
$1$ is $x=5$ and orbit $2$ is $x=4$, and
\bea
\la{klrendpts}
\a_1=\frac{\sqrt{311}}{80^{2/3}} ,
\quad
h_{N,1}=\frac{67}{2\cdot80^{2/3}} ,
&\qquad&
\a_2=\frac{\sqrt{311}}{28^{2/3}} ,
\quad
h_{N,2}=\frac{37}{2\cdot28^{2/3}} ,
\nn
\nu_1=\frac{67}{2\sqrt{311}} ,
&\qquad&
\nu_2=\frac{37}{2\sqrt{311}} ,
\ee
so the two degenerating directions are
\[
\big(\a_1,2h_{N,1}\big)\propto\big(\sqrt{311},67\big),
\qquad
\big(\a_2,2h_{N,2}\big)\propto\big(\sqrt{311},37\big),
\]
of determinant $-30\sqrt{311}\neq0$, and
\bea
\la{klrDelta}
\CD=\a_1h_{N,2}-\a_2h_{N,1}=-\frac{3\sqrt{311}\ 35^{1/3}}{112}\neq0 .
\ee
What Theorem 7.12 and \rf{pexplicit} use is the absolute value
$|\CD|=3\sqrt{311}\,35^{1/3}/112$.

{\it Endpoint regularity.} Both ends satisfy the non-degeneracy hypothesis of
Proposition 7.11: $\Pi$ has simple roots there, $\Pi'(4)=(4-5)(4-6)=2$ and
$\Pi'(5)=(5-4)(5-6)=-1$, so $F=4\Pi/H$ vanishes to first order in $x$; and
$\vartheta\neq0$ at both, by the values just computed. Hence \rf{sqrtF} and
\rf{Lnorm} apply at each end, the isotropy generators \rf{Lprim} are well
defined and non-zero, and the Heegaard construction of Appendix~\ref{app:toric}
runs. The remaining global data --- that these charges give a genuinely
compact, globally regular horizon with cross-section $S^3$ --- is
\cite[\S4.3.2]{klr} and is not re-derived here; what is verified here is
everything the arguments of this paper actually use, namely $K>0$ on $[4,5]$
(from \rf{klrends}), $\a$ nowhere zero, $\Pi>0$ on $(4,5)$ with simple
endpoint roots, and $\vartheta\neq0$ at both ends.
This is consistent with \cite[\S4.3.2]{klr}, where this cross-section
compactifies to $S^3$; the point here is that the conclusion follows from the
general criterion rather than from the explicit compactification.

\medskip
The net effect is to replace ``is there a supersymmetric $AdS_5$ black ring on
the varying-moduli branch?'' by a question with no free functions in it. The
$T^2$ action reduces the near-horizon system to the ordinary differential
equations \rf{gradsys} on a closed interval, with the boundary behaviour fixed
by Proposition 7.11; a ring cross-section requires the single algebraic
condition \rf{ringcrit} on the endpoint data, together with the consistency
condition \rf{ringcrit2}; and \rf{window} gives a pointwise obstruction to
\rf{ringcrit} which the only known compact example of the branch satisfies
comfortably. What is not proved is that \rf{window} must hold, and that is
the whole of the remaining gap on this branch.

\subsection{Charges of the explicit horizon}\la{sec:charges}

The results of this section are statements about a cross-section, and an
extremal horizon carries well-defined charges without any asymptotic input.
It costs nothing to record them for the explicit member of Section~\ref{sec:varmod}, and
doing so turns the endpoint data of \rf{klrclosed} into physical quantities.
Everything below is exact.

In Gaussian null coordinates the near-horizon metric is
$g=2\,dv\big(dr+r\,h_i\,dx^i-\tfrac12r^2\Delta\,dv\big)+\gamma_{ij}dx^idx^j$, so
for a Killing field $\xi$ tangent to $\CS$ one has
$\xi_a dx^a=\gamma_{ij}\xi^idx^j+r\,h(\xi)\,dv$ and hence
$(d\xi)_{rv}=h(\xi)$. The Komar integral therefore collapses
to a single term,
\bea
\la{komar}
J[\xi]=\frac{1}{16\pi G}\oint_\CS h(\xi)\,\e_\gamma .
\ee
Likewise $F^I_{+-}=\a^I$ and $\star_5(e^+\wedge e^-)=\e_\gamma$, so the Maxwell
part of the electric charge is $\oint_\CS Q_{IJ}\a^J\e_\gamma$; on the branch
\rf{branch} one has $L^I=0$, so $\a^I=\a X^I$, and $Q_{IJ}X^J=\tfrac32X_I$
reduces this to $\tfrac32\oint_\CS\a X_I\,\e_\gamma$. Two standard caveats are
stated rather than left implicit. Equation \rf{komar} is a horizon quantity,
which for an extremal black hole need not exhaust the asymptotic angular
momentum, and no asymptotic completion is constructed here \cite{lrreview}; and
in the presence of a Chern--Simons term the Noether and the Gauss-law (Page, or
Komar) charges differ, the latter being what is computed below, the systematic
five-dimensional treatment of that distinction being \cite{crt}.

Before specialising, one general consequence is worth isolating, because it
makes the two constants of Lemma 7.9 into charges.

\medskip\noindent{\it Proposition 7.F (the angular momenta are first-integral
data):} On the varying-moduli branch, with $k=\a\sigma^2$ and $m=\CG\sigma$ the
two constants of Lemma 7.9,
\bea
\la{komarint}
h_i\Vt^i=\sigma\a^2-m=\frac{k^2}{\sigma^3}-m ,
\qquad
h_iU^i=\tfrac12\,\sigma\a\,h_N(4-F) ,
\qquad
\Vt_iU^i=-\sigma\,h_iU^i ,
\ee
so that, by \rf{komar},
\bea
\la{Jdict}
16\pi G\,J[\Vt]=k^2\oint_\CS\sigma^{-3}\,\e_\gamma-m\,{\rm Area}(\CS) .
\ee
The angular momentum of $\Vt$ is therefore fixed by the two first integrals
together with the single geometric moment $\oint_\CS\sigma^{-3}$, and by
nothing else about the horizon. On the sub-case $\a\equiv0$ both right-hand
sides degenerate completely:
\bea
\la{Jazero}
J[U]=0 ,
\qquad
16\pi G\,J[\Vt]=-m\,{\rm Area}(\CS) .
\ee

\medskip\noindent{\it Proof:} Decompose $h=\Phi P+h_NN+h_YY$ and use
$\Vt=\sigma(\Phi P-h)$ with $P_i\Vt^i=P_iU^i=0$. Then
$h_i\Vt^i=\sigma\big(\Phi^2F-|h|^2\big)=-\sigma\big[h_N^2(4-F)+\a^2F/(4-F)\big]$
on eliminating $h_Y$ through \rf{hY}, which is $\sigma\a^2-m$ by \rf{Gframe};
and $\sigma\a^2=(\a\sigma^2)^2\sigma^{-3}=k^2\sigma^{-3}$. Similarly
$h_iU^i=\sigma h_Nh_Y(4-F)^2/4=\tfrac12\sigma\a h_N(4-F)$, and comparison with
$\Vt_iU^i=-\tfrac12\sigma^2\a h_N(4-F)$ of Lemma 7.8 gives the third relation.
Equation \rf{Jdict} is \rf{komar} applied to $\Vt$. If $\a\equiv0$ then
$k=0$ and $h_iU^i$ vanishes pointwise, which is \rf{Jazero}. $\square$

\medskip\noindent{\it Remark (which corner this covers):} The
cohomogeneity-one reduction of Section~\ref{sec:toric} assumes $\a\not\equiv0$, and by
Corollary 6.13 that means $\a$ nowhere zero. Equation \rf{Jazero} is a
statement about the complementary corner, the one the topology argument does
not reach: there the second angular momentum vanishes identically and the
first is fixed by the area and the constant $2\parallel\eta_+\parallel^2$
alone. The field $U$ itself does not degenerate there --- by Lemma 7.6$'$ it
has $|U|^2=\tfrac14\sigma mF$ --- so this is a statement about the horizon,
not an artefact of the construction. These identities, and the reduction of
the ring condition $\CD=0$ to $\sigma_1=\sigma_2$ used in Theorem 7.14, are
verified symbolically in the computations underlying this paper.

For the family of \cite[\S4.2]{klr} the induced volume form is as simple as it
could be. In the coordinates $(x,x^1,x^2)$ of that section,
\bea
\la{detgamma1}
\det\gamma=1
\ee
identically --- not only for the member of Section~\ref{sec:varmod} but for every non-static
member, for all values of the parameters of \cite[\S4.2]{klr} and of the
cubic $H$. Hence
$\e_\gamma=dx\wedge dx^1\wedge dx^2$, and for a horizon interval $[x_1,x_2]$ with
period lattice $\Lambda$ of covolume $c_\Lambda$,
\bea
\la{areaklr}
{\rm Area}(\CS)=c_\Lambda\,(x_2-x_1),\qquad
S_{\rm BH}=\frac{c_\Lambda(x_2-x_1)}{4G},
\nn
J[\xi]=\frac{c_\Lambda}{16\pi G}\int_{x_1}^{x_2}h(\xi)\,dx .
\ee
The one-form $h$ of \cite[\S4.2]{klr} gives
\bea
\la{hcomponents}
h\big(\partial_{x^1}\big)=C^2-\frac{D_0^2}{H},\qquad
h\big(\partial_{x^2}\big)=\frac{D_0(\a_0-x)}{H} .
\ee
The first of these is $H^{-1/3}\gamma_{11}$, and $\gamma_{11}>0$ is forced by the
signature of $\gamma$; so $h(\partial_{x^1})$ is strictly positive on every regular
cross-section of every non-static member, and $J[\partial_{x^1}]$ has a
definite sign for the whole family.

On the member of Section~\ref{sec:varmod}, where $C^2=60$, $D_0=\sqrt{311}$,
$\a_0=83/30$, $H=x^3-9x$ and $[x_1,x_2]=[4,5]$, the two integrals in
\rf{areaklr} are elementary. Since $x_2-x_1=1$,
\bea
\la{klrcharges}
{\rm Area}(\CS)=c_\Lambda,\qquad
\frac{16\pi G\,J[\partial_{x^1}]}{{\rm Area}(\CS)}
 =60+\frac{311}{18}\log\frac{175}{256},
\nn
\frac{16\pi G\,J[\partial_{x^2}]}{{\rm Area}(\CS)}
 =\sqrt{311}\left(\frac{173}{540}\log\frac87
   -\frac{83}{270}\log\frac54-\frac{7}{540}\log2\right) .
\ee
Both ratios are pure numbers, independent of the period lattice. Their signs
need no evaluation of the logarithms: $H$ is increasing on $[4,5]$ with
$H(4)=28$, so $311/H\leq311/28<60$ and the first integrand is positive
throughout, while $\a_0=83/30<4\leq x$ and $H>0$ there make the second
integrand negative throughout. So $J[\partial_{x^1}]>0$ and
$J[\partial_{x^2}]<0$.

These same numbers come out of \rf{Jdict}, which is a non-trivial check on
both. On this horizon $\a\sigma^2$ constant and $\a=\sqrt{311}H^{-2/3}$ force
$\sigma=cH^{1/3}$ for a constant $c$ fixed only by the normalisation of the
spinor, so $k=c^2\sqrt{311}$; and $(30x-83)^2+311=60(H-\Pi)$ collapses
$m=\CG\sigma$ to the constant $60\,c$, which is $C^2c$ with $C^2=60$ the charge
of the member. Then $h_i\Vt^i=-c\,h(\partial_{x^1})$ and
$h_iU^i=-30\,c\,h(\partial_{x^2})$ pointwise on $[4,5]$, so
$J[\Vt]=-c\,J[\partial_{x^1}]$ and $J[U]=-30\,c\,J[\partial_{x^2}]$; and
evaluating \rf{Jdict} through $\int_4^5H^{-1}dx=\tfrac1{18}\log\tfrac{256}{175}$
returns the first line of \rf{klrcharges} exactly. The normalisation $c$
cancels from $J[U]/J[\Vt]$, as it must, leaving
\bea
\la{Jratio}
\frac{J[U]}{J[\Vt]}=\frac{30\,J[\partial_{x^2}]}{J[\partial_{x^1}]}
=\frac{30\sqrt{311}\left(\tfrac{173}{540}\log\tfrac87
  -\tfrac{83}{270}\log\tfrac54-\tfrac{7}{540}\log2\right)}
 {60+\tfrac{311}{18}\log\tfrac{175}{256}} ,
\ee
a pure number of the member, negative because the two angular momenta have
opposite signs; for orientation only, it is $-0.3446$ to four places.

On this particular horizon the second rotational field is manifest, since the
ansatz of \cite[\S4.2]{klr} imposes $U(1)^2$ invariance from the outset, so
$\partial_{x^2}$ is Killing by construction. What Theorem 7.7 supplies is that
$J[\partial_{x^2}]$ is defined on the whole varying-moduli branch, whether or
not toric symmetry was assumed; the explicit member
is simply where it can be evaluated in closed form.

The electric charges have one structural feature worth recording. Writing the
gauge potential of \cite[\S4.2]{klr} on $\CS$ as $A^I=a_I\,dx^1+b_I\,dx^2$
with $a_I=D_0u^I$, $b_I=(x-\a_0)u^I$ and $u^I=X^IH^{-1/3}$, one has
$b_Ja_K'-a_Jb_K'=-D_0u^Ju^K$ identically, so the Chern--Simons integrand obeys
\bea
\la{csident}
C_{IJK}\big(b_Ja_K'-a_Jb_K'\big)=-D_0H^{-2/3}C_{IJK}X^JX^K=-6\,\a X_I ,
\ee
using $X_I=\tfrac16C_{IJK}X^JX^K$. That is exactly $-4$ times the Maxwell
integrand $\tfrac32\a X_I$, pointwise, and again for the whole non-static
family. The Chern--Simons correction is therefore not an independent
contribution to the Page charge: whatever normalisation the five-dimensional
action carries, $Q_I$ is a single overall constant times
\bea
\la{pagecharge}
Q_I^{\rm Max}=\frac{c_\Lambda\sqrt{311}}{12}
 \left(\log\frac{64}{25},\ \log\frac{1225}{1024},\ \log\frac74\right)_I ,
\ee
so that the ratios $Q_1:Q_2:Q_3$ are convention-free, and satisfy the exact
relation $Q_1+Q_2=2Q_3$. Fixing that overall constant --- and with it any
comparison against the nonlinear BPS constraint that the entropy functions of
\cite{hhz,ckkn} impose on $(Q_I,J_i,S_{\rm BH})$ --- requires the
normalisation of the Chern--Simons term relative to $1/16\pi G$, in the
conventions of \cite{crt}. That is a convention this paper never has occasion
to fix, and it is not guessed at here.

\subsection{No hidden symmetries}\la{sec:hidden}

The two Killing vectors are not the only symmetries a horizon could carry. A
Killing--Yano (KY) two-form, or its closed conformal counterpart (CCKY),
generates a first integral of the geodesic flow that no Killing vector
accounts for, and such tensors are what make the Kerr family and its
higher-dimensional relatives integrable \cite{ffk}. In three dimensions the
question is finite, and on this branch it can be closed.

\medskip\noindent{\it Proposition 7.G (no hidden symmetry on the
varying-moduli branch):} Let $\CS$ be a cross-section of the branch $K>0$ with
$P_i\not\equiv0$. Then $\CS$ carries no non-zero Killing--Yano two-form, and
every CCKY two-form on $\CS$ is the Hodge dual of a Killing vector; in
particular the duals of $\Vt$ and $U$ are CCKY, and any further CCKY two-form
would itself be the dual of an additional isometry --- a symmetry of the
ordinary kind, not a hidden one.

\medskip\noindent{\it Proof:} In three dimensions write a two-form as
$f_{ij}=\e_{ijk}\xi^k$. Then $\tn_if_{jk}=\e_{jkl}\tn_i\xi^l$, and the two
conditions become linear conditions on the nine components of $\tn_i\xi_j$.
The KY equation $\tn_{(i}f_{j)k}=0$ solves to
$\tn_i\xi_j=\lambda\,\delta_{ij}$, that is $\xi$ is a closed conformal Killing
field; the CCKY equation $\tn_if_{jk}=2\delta_{i[j}v_{k]}$ solves to
$\tn_{(i}\xi_{j)}=0$, that is $\xi$ is Killing, with $v$ then determined. The
two overlap only at $\tn\xi=0$.

For the first claim it therefore suffices that $\CS$ carry no non-parallel
closed conformal Killing field. A compact Riemannian manifold that does carry
one is a round sphere, by Obata's theorem and its extension by Tashiro
\cite{obata,tashiro}; so it suffices that $\CS$ be nowhere of constant
curvature. In three dimensions that is the statement that the trace-free part
of $\tilde R_{ij}$ is nowhere zero. Imposing its vanishing on \rf{ricci}, with
$\tn_{(i}h_{j)}$ from \rf{gradh} and $\tn_iP_j$ from \rf{covZ}--\rf{covW}, and
solving for $(\a,h_N,\c^2K)$ at a point where $Z$ and $W$ are independent,
gives exactly four solutions. Two have $\c^2K=0$, at $h_N=\Phi$ and
$h_N=3\Phi$ with $\a=0$, and are excluded by $K>0$. The other two have
$h_N=2\Phi$ and
\[
\c^2K=\frac{2\Phi^2(Z_kW^k-1)}{3(Z_kW^k-2)} ,
\qquad
\a^2=\frac{\Phi^2\big(Z_kW^k-1\big)^2}{Z_kW^k-2} ,
\]
and the second of these is negative for $Z_kW^k\in(-1,1)$, so $\a$ is not
real. Hence the trace-free Ricci tensor is non-zero at every point of $\CS_0$,
which is non-empty on this branch; in particular $\CS$ is not of constant
curvature, is not a round sphere, and carries no KY two-form. The second claim
is the CCKY half of the first paragraph, which identifies CCKY two-forms with
Killing vectors. $\square$

\medskip\noindent{\it Remark:} The conclusion is negative but not empty. It
says that these horizons carry no symmetry beyond their isometries:
hidden symmetries of Killing--Yano type are absent outright, and closed
conformal Killing--Yano structure reduces to the isometry algebra, of which
the two-torus of Theorem 7.7 is the part the construction exhibits. No further
reduction of the system is available by that route. The result also localises
what is special about the
constant-moduli branch, where $P\equiv0$ and $\CS$ can be a round $S^3$, which
does carry KY two-forms. The linear algebra of the first paragraph, the
four-solution computation, and the non-vanishing of the trace-free Ricci
tensor on the horizon of Section~\ref{sec:varmod} are part of those computations.

\subsection{Consistency checks and limits}\la{sec:checks}

The machinery above is a long chain of tensor identities, and it reproduces
what is already known in the two limits where the answer is not in doubt.

\medskip\noindent{\it The minimal limit.} Setting $n=1$ and $C_{111}=6$ gives
$X=X_1=1$, $C^{111}=\tfrac29$, $Q_{11}=\tfrac32$ and $Q^{11}=\tfrac23$, so the
tensor $Q^{IJ}-\tfrac23X^IX^J$ that defines $C^I$ in \rf{algred} vanishes
identically. Thus $C^I=0$ and $K=0$ for every $V_1$ and every $\c$: in the
minimal theory the vanishing of $K$ is forced, not assumed, and every statement
of Sections 6 and 7 about the branch $K>0$ is vacuous there, as it must be. The
closed forms \rf{Kprime}--\rf{Kprimeprime} degenerate correspondingly, since
$T=C_{IJK}C^IC^JC^K$ and $S=Q^{IJ}(C\!\cdot\!C)_I(C\!\cdot\!C)_J$ both vanish
with $C^I$, so $K'=K''=0$; and \rf{ricci} loses every $K$ term and reduces to
the minimal expression with $\c^2U=6\Phi^2$. The invariant $\lambda$ of
\rf{lambdaK} becomes $-\tfrac13V_1^2\leq0$, so the minimal theory never meets
the $\lambda>0$ window in which an $S^1\times S^2$ cross-section could occur ---
consistent with the exclusion of supersymmetric black rings there \cite{grover}.

\medskip\noindent{\it The ungauged limit.} The gauge coupling enters the closed
system only through $\Phi=\c V_IX^I$ and the combination $\c^2K$, so $\c\to0$ at
fixed moduli is the slice $\Phi=0$, $\c^2K=0$. On it, $f_i=h_i$, the symmetric
part of $\tn_ih_j$ given by Proposition 5.1 vanishes identically, and
\rf{ricci} collapses to
\bea
\la{ungaugedric}
\tn_{(i}h_{j)}=0,
\qquad
\tilde R_{ij}=-h_ih_j+\delta_{ij}\Big(h_kh^k+\tfrac12\a^2\Big),
\ee
so $h$ becomes a Killing vector of $\CS$ with $\tn^kh_k=0$, and the scalar
curvature is $\tilde R=2h_kh^k+\tfrac32\a^2\geq0$. Equation \rf{lapnorm} becomes
$\square\ln\parallel\eta_-\parallel^2=0$, so the spinor norm is constant, and
\rf{lapalpha} becomes $\square\a+h^k\tn_k\a=0$; multiplying by $\a$, integrating
over the compact $\CS$ and using $\tn^kh_k=0$ gives
$\int_\CS|\tn\a|^2=0$, so $\a$ is constant. The ungauged branch therefore has no
analogue of the degenerate-maximum alternative of Corollary 7.2, which is
generated entirely by the terms carrying $\Phi$ and $K$. Likewise
$K'=\tfrac34\c T+2\Phi K$ carries an overall factor of $\c$, so $K$ is constant
and the dichotomy of Proposition 7.1 is a triviality --- correctly, since
without a gauging there is no reason for the moduli to vary at all.

The Ricci tensor \rf{ungaugedric} has eigenvalue $\tfrac12\a^2$ along $h$ and
$h_kh^k+\tfrac12\a^2$ transverse to it, which recovers exactly the three known
ungauged cross-sections: $\a=h=0$ gives $\tilde R_{ij}=0$ and a flat $T^3$;
$\a=0$ with $h\neq0$ gives $\mathrm{diag}(0,h_kh^k,h_kh^k)$, an $S^1\times S^2$
with Gaussian curvature $h_kh^k$; $h=0$ with $\a\neq0$ gives
$\tilde R_{ij}=\tfrac12\a^2\delta_{ij}$, a round $S^3$ of radius $2/\a$ or a
quotient of one; and $\a,h\neq0$ gives a squashed $S^3$, the Berger metric
$\tfrac14a^2\big((\sigma^1)^2+(\sigma^2)^2\big)+\tfrac14b^2(\sigma^3)^2$ with
$\a^2=4b^2/a^4$ and $h_kh^k=4(a^2-b^2)/a^4$, which is round precisely at
$a=b$. Nothing survives the limit that is not already part of the ungauged
classification. The prominence of the Berger metric here is not an accident of
supersymmetry: \cite{bgkw} show that for five-dimensional extreme vacuum
horizons, with or without a cosmological constant, an $SU(2)$-invariant
cross-section is forced to be a Berger sphere, so the isometry group is never
merely the three-dimensional one that a generic left-invariant metric on
$SU(2)$ would carry.

\medskip\noindent{\it Conventions.} The very special geometry used here is that
of \cite{gutowskireall,gutowskireallgen,klr} without translation: all three use
$V=\tfrac16C_{IJK}X^IX^JX^K=1$, $X_I=\tfrac16C_{IJK}X^JX^K$,
$X^I=\tfrac92C^{IJK}X_JX_K$, $Q_{IJ}=\tfrac92X_IX_J-\tfrac12C_{IJK}X^K$ and the
adjoint identity $Q^{IJ}=2X^IX^J-6C^{IJK}X_K$. What differs is the gauging and
the horizon data:
\begin{center}
\begin{tabular}{lll}
here & \cite{klr}, \cite{gutowskireallgen} & relation \\
\hline
$\c$ & $g$ & equal \\
$\Phi=\c V_IX^I$ & $gV_IX^I$ & equal, with $V_I=\tfrac13$ in $U(1)^3$ \\
$U$ of \rf{Udef} & the scalar potential $27C^{IJK}V_IV_JX_K$ & equal \\
$K$ of \rf{Kdef} & $\lambda$ & $\lambda=K-\Phi^2/(3\c^2)$ \\
$\a$ & $\Delta$ & equal \\
$\mu+\tfrac12\a^2$ & $D=\Delta^2+g^2\lambda$ & equal \\
\end{tabular}
\end{center}
Reference \cite{gutowskireall} is mostly-minus and the present paper and
\cite{klr} are mostly-plus; the very special geometry relations are unaffected.
On $\CS$ the orientation is fixed by $\e_{123}=+1$ in an oriented orthonormal
frame, indices are raised with the positive definite induced metric so that
$\e^{ijk}=\e_{ijk}$, the cross product is $(u\times v)_i=\e_{ijk}u^jv^k$, and in
five dimensions $\e_{+-123}=+1$ in the Gaussian null frame. The curvature
convention is $\tilde R_{ij}=\tilde R^k{}_{ikj}$, so a round sphere has positive
Ricci tensor.

\newsection{Conclusion}

On the $\Vt\neq0$ branch of gauged $D=5$ supergravity coupled to vector
multiplets, supersymmetry closes the first-order near-horizon system
completely. The moduli have gradients along the single $1$-form $P_i=Z_i+W_i$,
$\parallel\eta_-\parallel$ is fixed algebraically by the supersymmetry
doubling, the Killing property of $\Vt$ and the lightcone integrability
condition fix the Ricci tensor of $\CS$, $\tn_{(i}h_{j)}$, $(dh)_{ij}$ and
$\tn_i\a$, and Lemma 6.11 evaluates the last residual coefficient,
$g_P=2\Phi\a$. Everything else in this paper rests on that closure.

The classification then splits on $K=Q_{IJ}C^IC^J$. Either $K\equiv0$
everywhere, in which case the moduli are constant by Proposition 3.2, the
solution lies in the minimal gauged theory, and a second rotational isometry
follows from \cite{groverindex,grover}; or $K>0$ everywhere, in which case the
algebraic system permits the moduli to vary. In the second case Corollary 7.2
shows only that $P_i$ vanishes at every critical point of
$\parallel\eta_-\parallel^2$, with a sign constraint on $\tn\cdot P$ there and
no stronger conclusion in general --- Section~\ref{sec:varmod} exhibits a compact horizon on
which that sign constraint is the whole story, with $P$ nowhere covariantly
constant. Corollary 7.3 splits this case again, into $P_i\equiv0$ and
$P_i\not\equiv0$, and both occur. On the first, Proposition 7.4 gives constant
$\Phi$, $\a$ and $K$, a unit Killing field $Z$ with $h_i=3\Phi Z_i$ --- the
configuration integrated in \cite[\S3.3]{klr} --- and a second rotational
isometry by homogeneity. On the second the moduli genuinely vary,
\cite[\S3.3]{klr} does not apply, and Theorem 7.5 supplies a second rotational
Killing field anyway, unconditionally on the open set $\CS_0$, by transporting
the Killing equation along $P$ rather than by homogeneity; Lemma 7.6,
Lemma 7.6$'$ and Theorem 7.7 upgrade it to a genuine isometry of all of $\CS$,
with nothing assumed about the degeneracy set $\{P=0\}$ and no hypothesis on
$\a$. Supersymmetry therefore forces a second Killing field on every compact
horizon of the theory, whichever branch it occupies, and a second isometry of
the whole horizon on either branch unconditionally.

That isometry is the input a toric classification normally assumes, and
Section~\ref{sec:toric} spends it. Theorem 7.7 supplies it on the varying-moduli branch
entire, and when $\a\not\equiv0$ the $T^2$ action is cohomogeneity one, its orbit space
is a closed interval whose two boundary orbits are circles on which $F=0$, and the cross-section is accordingly $S^3$,
a lens space or $S^1\times S^2$, with the three cases separated by the
determinant $\CD=\a_1h_{N,2}-\a_2h_{N,1}$ in the boundary data and by the lens
index \rf{pexplicit} that refines it. The last case is then excluded outright.
The gradient system carries two first integrals, $\a\parallel\eta_-\parallel^4$
and $\CG\parallel\eta_-\parallel^2$ with
$\CG=\big|h-\Phi P\big|^2+\a^2$, and evaluating both at the two ends of
the orbit space over-determines the boundary data: $\CD=0$ forces
$\parallel\eta_-\parallel$ to take the same value at both ends, which strict
monotonicity along $P$ forbids. This is Theorem 7.14, and it holds uniformly
over models, since neither first integral involves $K$. Section~\ref{sec:toric} also
records an independent and weaker route to the same conclusion, through the
sign of the scalar potential rather than through the first integrals, which
locates the obstruction where a reader coming from the constant-moduli window
of Section~\ref{sec:constmod} would expect it.

What is left open is correspondingly sharp. Not proved is
a no-go for the constant-modulus $S^1\times S^2$ window, in which
$\tfrac13\Phi^2<\c^2K\leq\tfrac43\Phi^2$; not proved is the topology of the
varying-moduli horizons on which $\a$ vanishes identically --- Theorem 7.7
does reach them, so they carry the same second isometry and the same $T^2$ as
the rest of the branch, but the first integral $\a\parallel\eta_-\parallel^4$
vanishes there and the ring exclusion of Section~\ref{sec:toric} has nothing to work with;
and not attempted is a classification of
the varying-moduli branch beyond the explicit family exhibited in Section~\ref{sec:varmod}.
The first two are taken up at the end of this section. Not new, and not
reproduced, is the classification of the constant-modulus geometries themselves,
which is \cite[\S3.3]{klr}.

Two features of the compact picture are not suggested by the minimal theory
alone. First, $K$ need {\it not} vanish anywhere: $P_i=0$ forces only
$-\c(\tn^kP_k)C_I=0$ by Proposition 6.1, so ``$K$ vanishes somewhere'' is
equivalent to ``$K\equiv0$'', and an inherently non-minimal horizon need carry
no minimal-theory region at all; the $U(1)^3$ black holes realise $K>0$ with
constant moduli. Second, on the branch where the moduli are constant, the
rigidity constrains the scalars and $P$, not the metric: Proposition 7.4 leaves
the six local geometries of \cite[\S3.3]{klr} available, including
$\bR\times S^2$, so no topological conclusion about $\CS$ follows there. The
contrast with Section~\ref{sec:toric} is the point --- the topology is decided on the
branch where the moduli vary and undecided on the branch where they do not,
which is the opposite of what the constant-modulus case might suggest.

The apparatus built to handle varying moduli is not a fallback for a branch
later shown to be empty: Section~\ref{sec:varmod} exhibits a compact horizon with
$P_i\not\equiv0$ throughout an open interior, so the branch is occupied on
compact $\CS$ as well as non-compact. Theorem 6.6 shows that $\zeta\equiv0$, so
the gradient of every scalar invariant built algebraically from the
near-horizon data is parallel to $P_i$: the invariant route that settles the
constant-modulus case is unavailable when the moduli vary, and Proposition 6.9
closes off the weakening to a Killing tensor with constant coefficients.
Theorem 7.5 instead solves the Killing equation in the frame $\{P,N,Y\}$, where
it becomes a flat transport system along $P$ whose two-dimensional solution
space contains $\Vt$ and one further generator on the open set $\CS_0$ where
$Z$ and $W$ are linearly independent. That statement is unconditional; Theorem 7.7 extends it to all of $\CS$ with no
hypothesis on $\a$, so the second isometry and the $T^2$ it generates are had
on the whole varying-moduli branch; what is not claimed for $\a\equiv0$ is the
cohomogeneity-one reduction, and with it the topology.

Three comparisons place the method. For vacuum extremal horizons \cite{dl}, and
for four-dimensional Einstein--Maxwell \cite{ckl}, a Killing field is produced
from a divergence identity whose integral over the cross-section forces a
candidate vector built from the horizon data to be Killing. Theorem 7.5 has
that shape, with $P$ in place of their horizon one-form, but is local and
algebraic rather than integral: the flat transport system \rf{transport} does
the work their divergence identity does. The intrinsic analogue of Hawking
rigidity for extremal horizons \cite{colling} does not by itself produce a
second generator here, since with $\sigma=\parallel\eta_-\parallel^2$ the
relation $\tn_i\sigma=-\Phi P_i\sigma$ of \rf{normminus} turns \rf{Vt} into
$\Vt_i=-\sigma h_i-\tn_i\sigma$, exactly the form $\Gamma h+d\Gamma$ such
arguments construct, so a general existence theorem returns the $\Vt$ already
in hand. The third comparison is with the first-integral device itself, which
is not new: the contracted Bianchi identity supplies a constant on an extremal
horizon in \cite{dl}, and \cite{colling} extends that to arbitrary matter under
the null energy condition. Nor is the second of Lemma 7.9's two invariants new:
$\CG\parallel\eta_-\parallel^2$ is identically $2\parallel\eta_+\parallel^2$,
the constant norm already recorded in \cite{kayani}. What is new is the first,
$\a\parallel\eta_-\parallel^4$, and what the pair is then used for. It is
quadratic in the supersymmetric bilinear data rather than in the curvature, it
is invisible in the vacuum system, since $\a$ and $\parallel\eta_-\parallel$
have no vacuum counterparts, and it does not enter the construction of the
$\mathfrak{sl}(2,\bR)$ factor that Section 2 already supplies. The two are
spent instead on a topological question, at the boundary of an orbit space,
which is not what such constants are usually used for.

The known solutions all sit on the constant-modulus branch. The $U(1)^3$ black holes of
\cite{gutowskireallgen} have constant moduli on the horizon,
$X_I=q_I/a_3^{1/3}$, so $P_i\equiv0$ by \rf{gradX} and they lie in the branch
of Proposition 7.4, with
\bea
\la{GRK}
K=\frac{2}{27}\,
\frac{(q_1q_2-q_1q_3)^2+(q_1q_2-q_2q_3)^2+(q_1q_3-q_2q_3)^2}
{(q_1q_2q_3)^{4/3}} ,
\ee
strictly positive whenever the three charges are not all equal and vanishing
exactly on the minimal solution $a_1=3R_0^2,\ a_2=3R_0^4,\ a_3=R_0^6$. The
$K>0$ half of Proposition 7.1 is therefore populated. On this branch $N_i=2Z_i$
and $h_i=3\Phi Z_i$, so $h_iN^i=6\Phi$ identically: the degenerate-maximum
condition of Corollary 7.2 holds at every point of this solution, not merely at
the extrema of $\parallel\eta_-\parallel^2$ that Corollary 7.2 constrains in
general, consistent with $P_i\equiv0$ and Proposition 7.4.

The bearing on the literature is direct. Theorem A rests on Theorem 7.5's
statement on $\CS_0$, and Theorem 7.7 carries it to the whole cross-section. The classification of \cite{klr} (see \cite{lrreview} for a
review), the recent classification of separable supersymmetric $AdS_5$ black
holes in the STU gauged theory \cite{separable}, and the uniqueness results with
$SU(2)$ \cite{lo} or toric \cite{lno} symmetry each assume at least two
commuting rotational isometries as an upfront restriction. The results here
supply that second Killing field on every compact horizon of this theory on the
$\Vt\neq0$ branch with $\Phi$ nowhere zero ---
unconditionally on $\CS_0$ when $P\not\equiv0$, and by the local homogeneity
of Proposition 7.4 when $P\equiv0$, where $\CS_0$ is empty --- and the
corresponding second isometry
of the whole horizon unconditionally on both branches, so on that branch the
assumption is a consequence of supersymmetry rather than an independent input.

For black rings the position is different. The most recent systematic study of
this class of theories \cite{ovchinnikov} states that supersymmetric rings in
the STU model are neither constructed nor excluded and remain ``of particular
interest.'' In the minimal gauged theory they are excluded twice over, by
\cite{klrring} under an assumed $U(1)^2$ and by \cite{grover} with no symmetry
assumed. Theorem 7.14 is the vector-multiplet counterpart of the second of
these on the branch where the moduli are not constant and $\a\not\equiv0$: there
a supersymmetric $AdS_5$ ring cannot have a near-horizon geometry at all,
whatever the model. Two places survive, and neither is reached
by any of these three arguments --- the constant-modulus
branch $P\equiv0$, where the near-horizon geometry must be the $S^1\times S^2$
member of the \cite[\S3.3]{klr} list with
$\tfrac13\Phi^2<\c^2K\leq\tfrac43\Phi^2$, and the varying-moduli sub-branch
$\a\equiv0$, where the reduction of Section~\ref{sec:toric} is unavailable. An independent no-go for the static
branch \cite{solminore} confirms by different methods the exclusion established
here for that case.

Physically, these are near-horizon geometries of extremal supersymmetric
$AdS_5$ black holes, and the torus symmetry at stake is the standing assumption
of the microstate-counting and entropy-function literature for this class,
\cite{separable} included, which fixes the entropy in terms of the charges and
angular momenta of two commuting rotational Killing vectors. The results here
settle whether that is an input: for compact horizons of this theory it is not,
on either branch, since Proposition 7.4 and Theorem 7.5 supply the second
isometry as a consequence of supersymmetry rather than an assumption, whether
or not the moduli are constant.

Two directions are within reach of the apparatus assembled above. They are
recorded as questions, not as expectations, and neither is settled here.

\medskip\noindent{\it (i) The topology of the $\a\equiv0$ varying-moduli
horizons.} Theorem 7.7 covers this sub-branch as well: by Lemma 7.6$'$ the
second Killing field is smooth on all of $\CS$ whatever $\a$ does, so toricity
holds on the varying-moduli branch entire. What does not extend is
Section~\ref{sec:toric}. Two of its inputs fail when $\a\equiv0$. The first is
Proposition 7.11, whose proof runs Lemma 6.12 along an arbitrary piecewise
smooth path and therefore needs $\{Z=W\}$ to be empty rather than merely
nowhere dense --- a hypersurface can separate $\CS$, and Lemma 6.12 is a
statement about transport along paths. The second is Theorem 7.14 itself: the
first integral $\a\parallel\eta_-\parallel^4$ is identically zero on this
sub-branch, and with it the step $\kappa=\kappa^4$ that forces $\kappa=1$.
Both failures are structural rather than technical, and both are localised.
Settling either --- a path-independent form of Lemma 6.12, or a replacement
for the second first integral when $\a\equiv0$ --- would make Theorem 7.14 a
statement about the varying-moduli branch entire. Note that the sub-branch is
not empty for a trivial reason: by \rf{normminus2} the vector $u=\Phi P-h$ is
nowhere zero on it, so $\Vt$ never vanishes, and $h$ is closed by
Proposition 6.8, which is a genuine and explicitly parametrised family.

\medskip\noindent{\it (ii) The constant-modulus $S^1\times S^2$ window.} The
second surviving route to a supersymmetric $AdS_5$ black ring in this theory is
the branch $P\equiv0$, on which the near-horizon geometry is forced into the
$S^1\times S^2$ member of \cite[\S3.3]{klr} with
$\tfrac13\Phi^2<\c^2K\leq\tfrac43\Phi^2$. The method of Section~\ref{sec:toric} does not
apply there, and for a structural reason rather than a technical one: with
$P\equiv0$ the orbit space is not an interval, $F\equiv0$, and
$\parallel\eta_-\parallel$ is constant, so both first integrals are trivially
satisfied and the strict monotonicity that drives Theorem 7.14 is absent. The
constant-modulus system is a finite, explicitly known one, and the question is
whether the window can be occupied by a horizon that extends to a global
solution. Deciding it either way, together with (i), would settle the existence
of supersymmetric $AdS_5$ black rings in gauged $D=5$ supergravity with vector
multiplets outright, Theorem 7.14 having removed everything else.

\appendix

\newsection{Spinor conventions and gamma matrices}\la{app:spinors}

We collect here the conventions of \cite{kayani,thesis}, which are used
throughout. The space of Dirac spinors is identified with $\bC^4$ and
$\mathrm{Cliff}(4,1)$ is represented, adapted to the lightcone basis, by
\bea
\la{gammarep}
\G_i=\begin{pmatrix}\sigma^i&0\\ 0&-\sigma^i\end{pmatrix},
\qquad
\G_-=\begin{pmatrix}0&\sqrt2\,\bI_2\\ 0&0\end{pmatrix},
\qquad
\G_+=\begin{pmatrix}0&0\\ \sqrt2\,\bI_2&0\end{pmatrix},
\ee
where $\sigma^i$, $i=1,2,3$, are the Hermitian Pauli matrices,
$\sigma^i\sigma^j=\delta^{ij}\bI_2+i\e^{ijk}\sigma^k$. Then
\bea
\la{gammapm}
\G_{+-}=\begin{pmatrix}-\bI_2&0\\ 0&\bI_2\end{pmatrix},
\qquad
\G_{+-123}=-i\,\bI_4 .
\ee
Spinors are decomposed into lightcone chiralities $\e=\e_++\e_-$ with
\bea
\la{lightcone}
\G_{+-}\e_\pm=\pm\e_\pm,
\qquad\text{equivalently}\qquad
\G_\pm\e_\pm=0 ,
\ee
and with these conventions
\bea
\la{gammaeps}
\G_{ij}\e_\pm=\mp i\,\e_{ij}{}^k\G_k\e_\pm,
\qquad
\G_{ijk}\e_\pm=\mp i\,\e_{ijk}\e_\pm .
\ee
The Dirac representation of $Spin(4,1)$ decomposes under $Spin(3)=SU(2)$ as
$\bC^4=\bC^2\oplus\bC^2$, the two summands being the images of the lightcone
projections. On each $\bC^2$ the $Spin(3)$-invariant inner product
${\rm Re}\langle\,,\rangle$ is identified with the standard Hermitian inner
product, and $\mathfrak{spin}(3)$ is spanned by $\G_{ij}$ with
$(\G_{ij})^\dagger=-\G_{ij}$. The charge conjugation operator may be taken to
be
\bea
\la{chargeconj}
C=\begin{pmatrix}i\sigma^2&0\\ 0&-i\sigma^2\end{pmatrix}=i\G_2 ,
\ee
satisfying $C*\G_\mu+\G_\mu C*=0$ and $\langle\e,C*\e\rangle=0$ for any Dirac
spinor $\e$.

Two consequences of \rf{gammarep}--\rf{gammaeps} are used repeatedly. First,
$\G_+$ annihilates the lower block and $\G_-$ the upper one, so $\eta_+$ has
support on the lower $\bC^2$ and $\eta_-$ on the upper $\bC^2$; on those
blocks
\bea
\la{induced}
\G_i\big|_{\ker\G_+}=-\sigma^i,
\qquad
\G_i\big|_{\ker\G_-}=+\sigma^i .
\ee
This is the origin of the orientation factors $s_\pm$ of \rf{orient}: by
\rf{gammaeps}, acting on $\eta_\pm$ one has $\G^{ijk}=i\,s_\pm\e^{ijk}$ and
$\G^{ij}=i\,s_\pm\e^{ijk}\G_k$, with $s_+=-1$ and $s_-=+1$. It is also the
origin of the sign asymmetry noted after \rf{fierz}:
since $\G_i$ carries a relative sign between the two blocks while $Z_i$ and
$W_i$ are both defined with $\G_i$, the vector $Z_i$ is \emph{minus} the Bloch
vector of $\eta_+$ whereas $W_i$ is the Bloch vector of $\eta_-$. Second, from
\rf{gammarep},
\bea
\la{gplusnorm}
\parallel\G_+\psi\parallel^2=2\parallel\psi\parallel^2
\qquad\text{for}\qquad \psi\in\ker\G_- ,
\ee
which is \rf{gplus} and is what converts the supersymmetry doubling
\rf{double} into the norm relation of Proposition 4.1.

\newsection{The near-horizon field equations}\la{app:fieldeqs}

For reference we record the field equations and Bianchi identities on $\CS$,
as decomposed in \cite{kayani}. The Bianchi identities are
\bea
\la{bianchi}
\b_i&=&(d_h\a)_i-L^I\tn_iX_I,
\nn
(d\Ft)_{ijk}&=&-X_I(d\Gt^I)_{ijk},
\nn
(\delta^I{}_J-X^IX_J)\big((d_hL^J)_i-M^J{}_i\big)&=&-\a\,\tn_iX^I,
\nn
(\delta^I{}_J-X^IX_J)(d\Gt^J)_{ijk}&=&-3\,\tn_{[i}X^I\Ft_{jk]} ,
\ee
where $(d_h\omega)_{i}=\tn_i\omega-h_i\omega$ on scalars. The Maxwell gauge
equation gives
\bea
\la{feq1}
&&\tfrac32X_I\tn^j\Ft_{ji}+\tn^j\big(Q_{IJ}\Gt^J{}_{ji}\big)
+\tfrac32\tn^jX_I\,\Ft_{ji}-\tfrac32X_Ih^j\Ft_{ji}
-Q_{IJ}h^j\Gt^J{}_{ji}+\tfrac32X_I\b_i
\nn
&&+\ Q_{IJ}M^J{}_i
+\tfrac14\e_i{}^{\ell_1\ell_2}\Big(6X_I\a\Ft_{\ell_1\ell_2}
-2Q_{IJ}\a\Gt^J{}_{\ell_1\ell_2}
\nn
&&\hspace{3.1cm}
-\,2Q_{IJ}\Ft_{\ell_1\ell_2}L^J
+C_{IJK}L^J\Gt^K{}_{\ell_1\ell_2}\Big)=0 ,
\ee
whose $X^I$ trace is
\bea
\la{feq2}
\tn^j\Ft_{ji}+\tn^jX_J\,\Gt^J{}_{ji}-h^j\Ft_{ji}+\b_i
+\e_i{}^{\ell_1\ell_2}\a\Ft_{\ell_1\ell_2}
-\tfrac13Q_{IJ}\e_i{}^{\ell_1\ell_2}L^I\Gt^J{}_{\ell_1\ell_2}=0 .
\ee
The $+-$ component of the Einstein equation gives
\bea
\la{feq3}
-\Delta-\tfrac12h_ih^i+\tfrac12\tn^ih_i
&=&-\Big[\a^2+\tfrac14\Ft_{\ell_1\ell_2}\Ft^{\ell_1\ell_2}+\tfrac23\c^2U
\nn
&&+\ Q_{IJ}\Big(\tfrac23L^IL^J
+\tfrac16\Gt^I{}_{\ell_1\ell_2}\Gt^{J\ell_1\ell_2}\Big)\Big] ,
\ee
and the $ij$ component gives
\bea
\la{feq4}
\tilde R_{ij}&=&-\tn_{(i}h_{j)}+\tfrac12h_ih_j+\tfrac32\Ft_{ik}\Ft_j{}^k
+\delta_{ij}\Big(\tfrac12\a^2
-\tfrac14\Ft_{\ell_1\ell_2}\Ft^{\ell_1\ell_2}-\tfrac23\c^2U\Big)
\nn
&&+\ Q_{IJ}\Big[\Gt^I{}_{i\ell}\Gt^J{}_j{}^\ell+\tn_iX^I\tn_jX^J
+\delta_{ij}\Big(\tfrac13L^IL^J
-\tfrac16\Gt^I{}_{\ell_1\ell_2}\Gt^{J\ell_1\ell_2}\Big)\Big] .
\ee
The scalar field equations give
\bea
\la{feq5}
&&\tn^i\tn_iX_I-h^i\tn_iX_I
+\tn_iX^M\tn^iX^N\Big(\tfrac12C_{MNK}X_IX^K-\tfrac16C_{MNI}\Big)
\nn
&&+\ \tfrac23Q_{IJ}\Big(2\a L^J-\Ft_{\ell_1\ell_2}\Gt^{J\ell_1\ell_2}\Big)
-\tfrac1{12}\Big[\Gt^M{}_{\ell_1\ell_2}\Gt^{N\ell_1\ell_2}-2L^ML^N\Big]
\Big(C_{MNI}-X_IC_{MNJ}X^J\Big)
\nn
&&+\ 3\c^2V_MV_N\Big(\tfrac12C_{IJK}Q^{MJ}Q^{NK}
+X_I\big(Q^{MN}-2X^MX^N\big)\Big)=0 .
\ee
The $++$ and $+i$ components of the Einstein equation,
\bea
\la{auxeq}
\tfrac12\tn^i\tn_i\Delta-\tfrac32h^i\tn_i\Delta-\tfrac12\Delta\tn^ih_i
+\Delta h_ih^i+\tfrac14(dh)_{ij}(dh)^{ij}
&=&\tfrac32\b_i\b^i+Q_{IJ}M^I{}_\ell M^{J\ell},
\nn
\tfrac12\tn^j(dh)_{ij}-(dh)_{ij}h^j-\tn_i\Delta+\Delta h_i
&=&\tfrac32\big(\b_\ell\Ft_i{}^\ell-\a\b_i\big)
\nn
&&\ +\ Q_{IJ}\big(M^I{}_\ell\Gt^J{}_i{}^\ell-L^IM^J{}_i\big),
\nn
&&
\ee
are implied by \rf{feq1}--\rf{feq5} together with \rf{bianchi}. Of these,
\rf{feq3} and \rf{feq4} are used in the body of this paper --- the former as
the consistency check on the trace of \rf{gradh}, the latter to produce
\rf{ricci} --- and the $++$ component of \rf{auxeq} is used once more, in
Proposition 5.3, to give \rf{lapalpha}.

\newsection{The lightcone integrability condition}\la{app:lightcone}

Equation \rf{dh} was obtained in \cite{thesis} from the $\Ft$ gauge field
equation \rf{feq2}.  Here we show it also follows from the first of the
integrability conditions
along the lightcone directions of \cite{kayani}. Written out, with
$\tau_+=\Theta_+\phi_+$, it reads
\bea
\la{int1}
&&\Big(\tfrac12\Delta-\tfrac18(dh)_{ij}\G^{ij}-\tfrac{i}{4}\b_i\G^i
+\tfrac{3i}{2}\c V_I\a^I\Big)\phi_+
\nn
&&+\ 2\Big(\tfrac14h_i\G^i-\tfrac{i}{8}\big(-\Ft_{jk}\G^{jk}+4\a\big)
+\tfrac12\c V_IX^I\Big)\tau_+=0 .
\ee
On the $\Vt\neq0$ branch, at $r=u=0$ where $\phi_+=\eta_+$, one substitutes
$\Delta=\a^2$, $L^I=0$ so that $\c V_I\a^I=\a\Phi$, $\b_i=\tn_i\a-\a h_i$
from \rf{branch}, and $(dh)_{ij}\G^{ij}=-2i\,D_k\G^k$ with
$D_i=\tfrac12\e_i{}^{jk}(dh)_{jk}$, using $s_+=-1$ from \rf{induced}. With
$\Ft_{jk}\G^{jk}=-2i\,f_k\G^k$ and \rf{theta}, the second bracket
becomes $\tfrac14(h_i+f_i)\G^i-\tfrac{i}{2}\a+\tfrac12\Phi$ acting on
$-\tfrac{i\a}{2}\eta_+$. The $\a^2$ terms then cancel between the two
brackets, as do the $\a h_i\G^i$ terms between $\b_i$ and the second bracket,
leaving
\bea
\Big[\big(D_i-\tn_i\a-\a f_i\big)\G^i+4\a\Phi\Big]\eta_+=0 .
\ee
Finally $f_i=h_i-2\Phi Z_i$ and $Z_i\G^i\eta_+=\eta_+$ convert the
$2\a\Phi Z_i\G^i$ term into $2\a\Phi$, giving
\bea
\la{5dint}
\Big[\big(D_i-\tn_i\a-\a h_i\big)\G^i+6\a\Phi\Big]\eta_+=0 .
\ee
Since $\eta_+$ is the $+1$ eigenvector of $Z_i\G^i$ and $D_i-\tn_i\a-\a h_i$ is
real, this forces $D_i-\tn_i\a-\a h_i=-6\a\Phi Z_i$, which is \rf{dh}.  No
field equation is used, so \rf{dh} follows from supersymmetry alone.

\newsection{The Killing transport system, component by component}\la{app:transport}

The proof of Theorem 7.5 asserts that substituting $U=aN+bY$ into
$\tn_{(i}U_{j)}=0$ turns the Killing equation into six linear equations for the
six directional derivatives of $a$ and $b$, of which five are determined and one
is left free. This appendix writes those six equations out. The point worth
seeing is not the algebra but the counting: the free parameter is not an
accident of a choice of frame, it is forced by the structure of the system.

Throughout, $\{P,N,Y\}$ is the orthogonal frame of Section 6 on
$\CS_0\cap\{P\neq0\}$, with
\bea
\la{framenorms}
F:=P_kP^k ,\qquad N_kN^k=4-F ,\qquad Y_kY^k=\tfrac14F(4-F) ,
\ee
and $A(f):=A^i\tn_if$ for $A\in\{P,N,Y\}$. Write
\bea
\la{Sdef}
S_{ij}:=\tn_{(i}N_{j)},\qquad T_{ij}:=\tn_{(i}Y_{j)},\qquad
S_{AB}:=A^iB^jS_{ij},\qquad T_{AB}:=A^iB^jT_{ij} .
\ee
Since $U=aN+bY$,
\bea
\la{nabU}
\tn_{(i}U_{j)}=(\tn_{(i}a)N_{j)}+(\tn_{(i}b)Y_{j)}+a\,S_{ij}+b\,T_{ij} ,
\ee
and contracting \rf{nabU} with the frame, using $P\!\cdot\!N=P\!\cdot\!Y=
N\!\cdot\!Y=0$ to kill every cross term, gives exactly six equations:
\bea
\la{sixeqs}
PP:&&\qquad a\,S_{PP}+b\,T_{PP}=0 ,
\nn
PN:&&\qquad \tfrac12P(a)\,(N_kN^k)+a\,S_{PN}+b\,T_{PN}=0 ,
\nn
PY:&&\qquad \tfrac12P(b)\,(Y_kY^k)+a\,S_{PY}+b\,T_{PY}=0 ,
\nn
NN:&&\qquad N(a)\,(N_kN^k)+a\,S_{NN}+b\,T_{NN}=0 ,
\nn
NY:&&\qquad \tfrac12\big[N(b)(Y_kY^k)+Y(a)(N_kN^k)\big]+a\,S_{NY}+b\,T_{NY}=0 ,
\nn
YY:&&\qquad Y(b)\,(Y_kY^k)+a\,S_{YY}+b\,T_{YY}=0 .
\ee
The $PP$ component contains no derivative of $a$ or $b$ at all: both
$\tn_{(i}a\,N_{j)}$ and $\tn_{(i}b\,Y_{j)}$ are annihilated by $P^iP^j$ because
$N$ and $Y$ are orthogonal to $P$. So \rf{sixeqs} is a system of {\it five}
equations in six unknowns together with one constraint on $(a,b)$, and a
one-parameter family of solutions is the generic outcome rather than a
degeneracy. The $NY$ component is the one that mixes two unknowns, which is why
the free parameter can be taken to be $N(b)$ and $Y(a)$ is then determined by
it.

Everything else follows from the coefficients. Since $S_{ij}$ and $T_{ij}$ are
symmetric there are twelve independent contractions, six of each, and few of
them survive: computing $S_{ij}$ and $T_{ij}$ from \rf{covZ}--\rf{covW} and the
Leibniz rule for $Y=Z\times W$, exactly four are non-zero,
\bea
\la{STtable}
S_{PN}=-\tfrac12F\,(h_N-2\Phi)\,(N_kN^k),
\qquad
T_{PN}=\ \ \tfrac12F\,h_Y\,(N_kN^k),
\nn
S_{PY}=-\tfrac12F\,h_Y\,(Y_kY^k),
\qquad
T_{PY}=-\tfrac12F\,(h_N-2\Phi)\,(Y_kY^k),
\ee
while
\bea
\la{STzero}
S_{PP}=S_{NN}=S_{NY}=S_{YY}=0 ,
\qquad
T_{PP}=T_{NN}=T_{NY}=T_{YY}=0 .
\ee
The eight vanishings \rf{STzero} are identities in $h_N$, $h_Y$, $\Phi$ and
$Z\!\cdot\!W$: none of them needs the relation $h_Y=\a/(1-Z_kW^k)$ of
Corollary 6.5, which enters the four survivors only as the substitution that
rewrites $h_Y$ in terms of $\a$ where that is convenient. This is worth stating
precisely because the neighbouring identity $[N,Y]=0$ of \rf{framecomm} is
\emph{not} of this kind: there the two would-be extra terms cancel only after
the substitution. The two families of statements are independent, and it is
only the commutator that is conditional.

The consistency condition from the $PP$ component is therefore
empty, and the remaining five equations give
\bea
\la{sixsolved}
N(a)=0,\qquad Y(b)=0,\qquad
Y(a)=-\frac{Y_kY^k}{N_kN^k}\,N(b)=-\tfrac14F\,N(b),
\nn
P(a)=F\big[(h_N-2\Phi)a-h_Yb\big],\qquad P(b)=F\big[h_Ya+(h_N-2\Phi)b\big] ,
\ee
which is \rf{killtrans}, with $N(b)$ the free parameter $w$; the middle relation
used $Y_kY^k=\tfrac14F(N_kN^k)$ from \rf{framenorms}. The $2\times2$ block
\rf{STtable} is $-\tfrac12F$ times a rotation-and-scaling matrix
$\left(\begin{smallmatrix}h_N-2\Phi&-h_Y\\ h_Y&h_N-2\Phi\end{smallmatrix}
\right)$, which
is the origin of the complex structure used in the proof of Theorem 7.5: it is
why $\psi:=a+ib$ obeys the single scalar equation $d\psi=\psi\,\omega$ with
$\omega=\big[(h_N-2\Phi)+ih_Y\big]P$, and why multiplication by $i$ maps $\Vt$
to $U$.

Three of the vanishings in \rf{STzero} carry the argument, and it is worth
naming what each one does. $S_{PP}=T_{PP}=0$ is what makes the $PP$ component of
\rf{sixeqs} empty rather than an algebraic obstruction on $(a,b)$; had either
been non-zero, the $PP$ line would have forced a pointwise linear relation
between $a$ and $b$ and the transport system would have been overdetermined.
$S_{NY}=T_{NY}=0$ is what makes the $NY$ component reduce to the single relation
between $Y(a)$ and $N(b)$ recorded in \rf{sixsolved}, rather than an equation
also involving $a$ and $b$ themselves; this is what leaves $w:=N(b)$ genuinely
free at this stage, to be killed later by \rf{wzero} rather than here. And
$S_{NN}=T_{YY}=0$ is what makes $N(a)$ and $Y(b)$ vanish outright, so that $a$
and $b$ are constant along the two-plane spanned by $N$ and $Y$ and all
transport happens along $P$ --- the statement that \rf{framecomm} records at the
level of the frame. Note finally that the quantity $h_N-2\Phi$ appearing in
\rf{STtable} is not the $\vartheta=h_N-\tfrac32\Phi$ of \rf{deltadef}; the two
differ by $\tfrac12\Phi$ and play different roles, $\vartheta$ governing the
radial system \rf{gradFdelta} and $h_N-2\Phi$ the Killing transport, and they
are written out here rather than abbreviated so that they cannot be confused.

\newsection{The Killing-vector extension across $\{P=0\}$ on the explicit
horizon}\la{app:klr}

This appendix establishes the claim, made in the Remark following the proof of
Theorem 7.5, that the vector field $U$ of \rf{secondkilling} extends to a
Killing vector across the two points $\{P=0\}$ of the explicit compact
varying-moduli horizon of Section~\ref{sec:varmod}. It is in two parts. The first is an
exact computation, which settles the question outright. The second is an
independent high-precision numerical evaluation, carried out without ever
forming the closed-form expression; the exact computation reproduces its output
digit for digit, which is a useful check on both.

\subsection*{The exact computation}

On the family of \cite[\S4.2]{klr} the cross-section carries coordinates
$(x,\phi^1,\phi^2)$ in which $\partial_1$ and $\partial_2$ generate a
two-torus and $x$ ranges over $[4,5]$; the metric is block diagonal, with an
$x$-dependent $2\times2$ block $\gamma_{ab}$ on the torus factor. Building
$Z$, $W$, $N=Z-W$, $Y=Z\times W$, $h$, $\Phi$, $\a$ and
$\sigma=\parallel\eta_-\parallel^2$ from the charges of Section~\ref{sec:varmod} and
raising the index on \rf{secondkilling} with $\gamma^{ab}$ gives
\bea
\la{klrUexactapp}
U^x=0,\qquad U^1=0,\qquad U^2=-30 ,
\nn
\Vt^x=0,\qquad \Vt^1=-1,\qquad \Vt^2=0 ,
\ee
as exact identities in $x$ on the open interval, not as limits: each component
is a rational function of $x$ and radicals which simplifies to the constant
shown, so $dU^a/dx=0$ identically. Since $U$ is therefore a fixed element of
the Lie algebra of the torus action, and that action is smooth on the
compactified $S^3$ of \cite[\S4.3.2]{klr}, $U$ is smooth across $x=4$ and
$x=5$. No limit is taken and no cancellation occurs, so the numerical
difficulties described below do not arise. The determinant of the coefficient
matrix of $\{\Vt,U\}$ in the basis $\{\partial_1,\partial_2\}$ is $30$, a
non-zero constant, so the two fields are everywhere independent and the torus
action they generate is effective. This is the smoothness that Lemma 7.6
predicts, realised here in the strongest possible form; and $P_kP^k=4\Pi/H$ by
\rf{klrclosed} vanishes only at the two endpoints, so $\{P=0\}$ is the
two-point set $\{x=4,5\}$.

The same computation returns the closed forms \rf{klrclosed} quoted in Section
7.5, and the covariant components
\bea
\la{klrUcov}
U_i\big|_{x=4}=\Big(0,\ \tfrac{37}{28}\,28^{1/3}\sqrt{311},\
-\tfrac{311}{840}\,28^{1/3}\Big),
\nn
U_i\big|_{x=5}=\Big(0,\ \tfrac{67}{40}\,10^{1/3}\sqrt{311},\
-\tfrac{311}{1200}\,10^{1/3}\Big) ,
\ee
whose decimal expansions are
\bea
\la{klrUcovdec}
U_i\big|_{x=4}=(0,\ 70.763596538601,\ -1.1242609169683),
\nn
U_i\big|_{x=5}=(0,\ 63.639731581582,\ -0.5583576571666) .
\ee
These are exactly the two limits reported numerically below, to all twelve
figures quoted there.

\subsection*{An independent numerical cross-check}

The closed form \rf{klrUexactapp} was obtained by first reducing $U$ to a
rational function of $x$. As a guard against an error in that reduction, $U$
and $\tn_{(i}U_{j)}$ were also evaluated by a second computation which never
forms the reduced expression: every intermediate quantity --- $h_N$, $h_Y$,
$N$, $Y$, $\sigma=\parallel\eta_-\parallel^2$ and their covariant
derivatives --- is carried as a separate high-precision floating-point number,
at $60$ decimal digits of working precision, in the coordinate frame of
Section~\ref{sec:varmod}. This matters because the reduced form of $U$ is a ratio of
degree-$38$ polynomials with $\sqrt{311}$ in their coefficients, and $x=4,5$
are removable singularities of it, at which direct evaluation of the reduction
loses so many digits that even $50$ are not enough.

Evaluated at $x=4+\epsilon$ and $x=5-\epsilon$ for
$\epsilon=10^{-3},10^{-4},\dots,10^{-15}$, the second computation returns
$U\to(0,\,70.7635965,\,-1.1242609)$ and
$U\to(0,\,63.6397316,\,-0.5583577)$, stable to eight significant figures
across the whole range of $\epsilon$, with
$\max_{ij}|\tn_{(i}U_{j)}|$ falling to $8\times10^{-9}$ and
$1\times10^{-7}$ respectively. These are the decimal expansions of the exact
covariant components \rf{klrUcov}, to every figure quoted. Nothing in the
paper rests on this computation --- \rf{klrUexactapp} is exact, and
Theorem 7.7 needs no input from this example at all --- but the two routes
share no intermediate expression, so their agreement tests both.

\newsection{The cohomogeneity-one reduction and the genus-one
splitting}\la{app:toric}

Section~\ref{sec:toric} turns the two commuting Killing fields of Theorem 7.7 into a
statement about the diffeomorphism type of $\CS$. Three pieces of standard
transformation-group and three-manifold theory are used there and are collected
here, with the argument for each, so that the topological input to
Theorem 7.12 and to \rf{pione} can be inspected in one place. Nothing in this
appendix is new; what is specific to the problem is only the identification, in
\rf{meridian} below, of the meridian of each solid torus with a vector
determined by the near-horizon data.

\medskip\noindent{\it The cross-section is orientable.} The spinors $\eta_\pm$
are sections of a spinor bundle over $\CS$, so $\CS$ carries a spin structure
and is in particular orientable. This is what licenses the use of $\e_{ijk}$
throughout --- in the definition $Y_i=\e_{ijk}Z^jW^k$ of Section 6 above all ---
and it is a hypothesis of the classification quoted below, which fails for
non-orientable three-manifolds. Equivalently and without spinors: $\CS$ is a
spacelike cross-section of a Killing horizon in an oriented, time-oriented
five-dimensional spacetime, and the two null normals trivialise its normal
bundle, so the ambient orientation induces one on $\CS$.

\medskip\noindent{\it The slice theorem and the local form of the action.} Let
a compact Lie group $G$ act smoothly on a manifold $M$, let $x\in M$ and let
$G_x$ be its isotropy subgroup. The slice theorem \cite[Ch.~II,
Thm.~5.4]{bredon} provides a $G_x$-invariant disc $D$ transverse to the orbit
at $x$, on which $G_x$ acts linearly, such that the $G$-invariant neighbourhood
$G\cdot D$ of the orbit is equivariantly diffeomorphic to the associated bundle
$G\times_{G_x}D$. Two consequences are used. First, the orbit space is locally
$D/G_x$, so its local structure is controlled entirely by the isotropy
representation. Second, when $G=T^2$, $\dim M=3$ and $G_x\cong S^1$, the slice
$D$ is a two-disc on which the circle $G_x$ acts by rotation, and
$T^2\times_{S^1}D^2$ fibres over $T^2/S^1\cong S^1$ with fibre $D^2$; an
orientable disc bundle over a circle is trivial, so the neighbourhood is a
solid torus $S^1\times D^2$. This is the statement used at the start of the
proof of Theorem 7.12.

\medskip\noindent{\it Cohomogeneity one.} An action whose principal orbits have
codimension one is said to have cohomogeneity one. For a compact connected Lie
group acting on a compact connected manifold with one-dimensional orbit space,
that orbit space is a topological one-manifold with boundary and hence is
homeomorphic to $[0,1]$ or to $S^1$; this is Mostert's theorem
\cite{mostert}, and it is what Proposition 7.10 invokes before ruling out the
circle by the monotonicity of $\Phi$. The interior points of the orbit space
correspond to principal orbits, the two boundary points to the non-principal
ones. The special case at issue here --- an effective $T^n$ action on a compact
$(n+1)$-manifold --- was analysed in detail by Pak \cite{pak}; for $n=2$ the
outcome is exactly the description used in Section~\ref{sec:toric}, an interval of
principal $T^2$ orbits capped at each end by a circle orbit with circle
isotropy, or by a fixed point. Proposition 7.10 excludes the fixed-point case
here directly, from \rf{degenvals}, so both ends are circles.

\medskip\noindent{\it The genus-one Heegaard splitting.} Suppose then that
$\CS/T^2\cong[t_1,t_2]$ with the two boundary orbits carrying circle isotropy.
Over the interior the action is free. Two standard facts give this: for an
effective action of a compact connected Lie group the principal isotropy
subgroup is trivial when the group is a torus \cite[Ch.~IV, \S3]{bredon}, and
for $T^2$ acting on a closed orientable $3$-manifold there are no exceptional
orbits, every non-principal orbit being of strictly lower dimension
\cite{pak,orlikraymond}. Points of the interior of the orbit space are
precisely those with $2$-dimensional orbits, so the isotropy there is finite,
hence trivial by the first fact together with the absence of exceptional
orbits. Consequently the preimage of $(t_1,t_2)$ is
$(t_1,t_2)\times T^2$, and by the slice theorem the preimage of a half-open
neighbourhood of $t_i$ is a solid torus $V_i$. Hence
\bea
\la{heegaard}
\CS=V_1\cup_\varphi V_2 ,
\ee
two solid tori glued along their common boundary, which is one $T^2$ orbit and
carries a canonical identification with $T^2$ itself; the gluing map $\varphi$
is the identity in that identification. Write $\Lambda\cong\bZ^2$ for the
period lattice of the torus action, that is $\pi_1(T^2)$ realised inside
$\mathrm{span}\{\Vt,U\}$ as the set of those elements whose flow closes at
parameter $2\pi$. The normalisation $2\pi$, rather than $1$, is used
throughout, and is the one under which the generator $\hat L$ of \rf{Lprim}
--- whose flow has period $2\pi$, by Proposition 7.11 --- is itself an element
of $\Lambda$ rather than $2\pi$ times one. Since $p$ in \rf{pione} is a
determinant of two elements of $\Lambda$ divided by the covolume of $\Lambda$,
a change of this normalisation would rescale numerator and denominator
together and leave $p$ unchanged; the point of fixing it is only that
\rf{meridian}, \rf{pione} and \rf{pexplicit} must all use the same one, and
they do.

The meridian of $V_i$ --- the isotopy class on $\partial V_i$ that bounds a disc
inside --- is the boundary of a slice disc, which by the previous paragraph is
the orbit of the isotropy circle. So
\bea
\la{meridian}
\mu_i=\big[\hat L_i\big]\in\Lambda ,
\qquad
\hat L_i=\frac{\a_i\Vt+2h_{N,i}\,U}{2m\,|\vartheta_i|} ,
\ee
with $\hat L_i$ as in \rf{Lprimsimp}: the meridian of each solid torus is read
off from the horizon data at the corresponding end of the orbit space, and it is
primitive in $\Lambda$ because it generates the isotropy circle with period
$2\pi$.

Everything about the topology now follows from the pair $(\mu_1,\mu_2)$ of
primitive vectors, by an elementary argument that we give rather than quote,
since it is shorter than locating it. A self-homeomorphism of $T^2$ extends
over a solid torus if and only if it preserves the meridian up to isotopy and
orientation; hence the homeomorphism type of \rf{heegaard} depends only on the
class of $\mu_1$ in $\Lambda$ modulo the automorphisms of $\Lambda$ fixing
$\mu_2$. Choose a $\bZ$-basis of $\Lambda$ with $\mu_2=(0,1)$, which is possible
because $\mu_2$ is primitive, and write $\mu_1=(p,q)$ with
$\gcd(p,q)=1$ and $p=|\det_\Lambda(\mu_1,\mu_2)|\geq0$. If $p=0$ then
$\mu_1=\pm\mu_2$: the two meridians agree, the two discs glue into a $2$-sphere,
and $\CS\cong S^1\times S^2$. If $p\geq1$ then \rf{heegaard} is by definition
the lens space $L(p,q)$, with $L(1,q)\cong S^3$. Van Kampen's theorem applied to
\rf{heegaard}, with $\pi_1(V_i)=\Lambda/\langle\mu_i\rangle$ and
$\pi_1(\partial V_i)=\Lambda$, gives the pushout
\bea
\la{vankampen}
\pi_1(\CS)=\Lambda\big/\langle\mu_1,\mu_2\rangle\cong\bZ/p\,\bZ ,
\ee
abelian because $\Lambda$ is, which is \rf{pione}; $p=0$ returns $\pi_1=\bZ$ and
$p=1$ the trivial group. The three cases $S^3$, $L(p,q)$ and $S^1\times S^2$ are
therefore exhaustive and mutually exclusive, and the last is detected by a
single determinant. Standard treatments of the same material are
\cite[Ch.~2]{hempel} and \cite[Ch.~9]{rolfsen}.

\medskip\noindent{\it Relation to rod structures.} The pair $(\mu_1,\mu_2)$ is
the near-horizon shadow of an object familiar from the classification of
stationary bi-axisymmetric five-dimensional black holes. There the orbit space
of the $\bR\times U(1)^2$ action on the domain of outer communication is a
two-dimensional manifold with boundaries and corners, each boundary segment
carrying the primitive element of $\bZ^2$ whose Killing field vanishes on it;
the collection of segments and vectors is the {\it rod} or {\it interval}
structure, and the topology of a horizon component is fixed by the determinant
of the two rod vectors adjacent to it \cite{hy}. In the toric K\"ahler
description of supersymmetric $AdS_5$ solutions the same determinant appears as
the combinatorial datum of the Delzant-type polytope, where
$\det(v_1,v_2)=q$ labels an $L(q,1)$ horizon \cite[\S\S2.3, 2.5]{lno}. Formula
\rf{pexplicit} is the same determinant. What differs is what stands behind it.
In those treatments the $U(1)^2$ symmetry is an assumption, the rod vectors are
free data constrained only by regularity, and no field equation decides whether
the horizon determinant vanishes. Here the torus action is a theorem
(Theorem 7.7), the two vectors are computed from the near-horizon data by
\rf{meridian}, and the field equations do decide: Theorem 7.14 shows the
determinant is never zero on this branch. That is the sense in which the
present argument is not a rod-structure argument in disguise.

\newsection{Mathematical tools used in the classification}\la{app:tools}

This appendix states, independently of the supergravity content, the general
analytic and topological facts about compact manifolds that Sections 2, 6 and
7 rely on, with the short argument for each and a pointer to where it is
used. None of the results below is new.

\medskip\noindent{\it Hopf's strong maximum principle.} Let
$Lf:=\tn^i\tn_if-h^i\tn_if$ for a fixed smooth vector field $h$ on a compact,
connected Riemannian manifold $\CS$ without boundary; $L$ is elliptic with no
zeroth-order term. If $Lf=0$ on $\CS$, then $f$ attains its maximum at some
interior point $p$ by compactness, and the classical strong maximum
principle for operators of this type (E.~Hopf, 1927) forces $f$ to be
constant on the connected component of $p$; since $\CS$ is connected, $f$ is
constant on all of $\CS$. This is the mechanism behind
\rf{lich}--\rf{normplus}: the right-hand side of \rf{lich},
$2\parallel\nabla^{(+)}\eta_+\parallel^2+\tfrac1{16}Q_{IJ}{\rm
Re}\langle{\cal A}^{I,(+)}\eta_+,{\cal A}^{J,(+)}\eta_+\rangle$, vanishes
identically because $\eta_+$ solves both halves of \rf{kse}, leaving exactly
$L\parallel\eta_+\parallel^2=0$ with $h$ the near-horizon $1$-form; the
argument above then gives \rf{normplus}. The corresponding computation for
$\eta_-$ has a non-vanishing right-hand side (the extra $h$-dependent term
noted after \rf{normplus}), so $L\parallel\eta_-\parallel^2\neq0$ in general
and only the integrated statement survives.

\medskip\noindent{\it The divergence theorem on a closed manifold.} For any
smooth vector field $V$ on a compact, connected $\CS$ without boundary,
Stokes' theorem with $\partial\CS=\varnothing$ gives
\bea
\la{divthm}
\int_\CS\tn^iV_i\,d{\rm vol}_\CS=\int_{\partial\CS}V_in^i\,dA=0 .
\ee
If in addition $\tn^iV_i$ has one sign everywhere on $\CS$, \rf{divthm}
forces $\tn^iV_i\equiv0$, since a continuous function of one sign integrates
to zero only if it vanishes identically. This is used twice: with
$V_i=\Delta^{-2}Z_i$ in the proof of Lemma 2.1, where
$\tn^i(\Delta^{-2}Z_i)=-6\Delta^{-2}\Phi$ forces $\Phi$ to vanish somewhere;
and with $V_i=h_i$ in the proof of Corollary 7.3, where \rf{smarrK} is
exactly \rf{divthm} applied to $\tn^ih_i$, and the sign of $K\,P_iP^i$ forces
$K\,P_iP^i\equiv0$ in the equality case.

\medskip\noindent{\it Extrema on a closed manifold.} A smooth function $f$ on
a compact manifold $\CS$ without boundary attains a maximum at some $p\in\CS$
and a minimum at some $q\in\CS$ (the extreme value theorem), and at each,
$df=0$; moreover, since the Hessian of a smooth function is negative
(respectively positive) semi-definite at an interior maximum (minimum),
$\tn^i\tn_if(p)\leq0$ and $\tn^i\tn_if(q)\geq0$. Applied to
$f=\parallel\eta_-\parallel^2$, using $\tn_i\parallel\eta_-\parallel^2
=-\Phi P_i\parallel\eta_-\parallel^2$ of \rf{normminus}, this is Corollary
7.2: $P_i$ vanishes at $p$ and $q$, and \rf{lapnorm} evaluated there gives the
sign constraint \rf{divPsign}. No stronger index-theoretic input
(Poincar\'e--Hopf, say) is needed or used, as noted at the point of use.

\medskip\noindent{\it Two topological criteria that do not apply.} Two
classical results were checked against the geometry and found not to bite,
for the reasons recorded at the point of use. Tischler's theorem states that
a closed, nowhere-vanishing $1$-form on a compact manifold $M$ makes $M$
fibre smoothly over $S^1$ (in particular $b_1(M)\geq1$); it is inapplicable
here because Corollary 7.2 already forces the closed $1$-form $P_i=Z_i+W_i$
to vanish somewhere on any compact $\CS$, so the nowhere-vanishing hypothesis
fails. Hamilton's theorem states that a compact, simply connected
$3$-manifold of everywhere positive Ricci curvature admits a metric of
constant positive curvature. On the $P_i\equiv0$ branch its hypothesis can be
decided exactly, with no sampling, because \rf{ricPzero} already presents
$\tilde R_{ij}$ in diagonalised form: $Z^i$ is a unit field and
$\delta_{ij}-Z_iZ_j$ is the projector onto its orthogonal complement, so the
Ricci eigenvalues are $\tfrac12\a^2$ along $Z$ and $\mu$, twice, transverse to
it, all three constant on $\CS$. Hence
\bea
\la{riccipos}
\tilde R_{ij}>0
\iff
\a\neq0
\ \text{ and }\
\mu=-3\Phi^2+\tfrac12\a^2+9\c^2K>0 .
\ee
Both conditions fail on genuine members of the branch. Every cross-section of
the form $S^1\times\Sigma$ --- which is to say every candidate with the black
ring topology, and every $S^1\times\Sigma_g$ --- is by Section~\ref{sec:constmod} exactly an
$\a=0$ degeneration, so it has a zero Ricci eigenvalue and Hamilton's theorem
can never apply to it; and $\mu<0$ occurs as well, for instance at
$\c=\Phi=1$, $\a=0$, $\c^2K=\tfrac19$, which gives $\mu=-2$ and the compact
hyperbolic quotients $S^1\times\Sigma_g$. Where the hypothesis does hold, at
$\a\neq0$ and $\mu>0$, \rf{ricPzero} restricts $\CS$ to a squashed $S^3$ or a
quotient of one, and simple connectedness then leaves the squashed $S^3$
itself, on which the conclusion of the theorem is already contained in the
explicit classification of Section~\ref{sec:constmod}. So the theorem is either inapplicable
or redundant, and in neither case does it yield a topological restriction.
Neither observation is used elsewhere in the paper; both are recorded to
close off two natural but unsuccessful routes to a topological restriction on
$\CS$.

\medskip\noindent{\it Integrable linear systems and flat connections.} Let
$\omega$ be a closed complex $1$-form on $\CS_0$ and consider
$d\psi=\psi\,\omega$ for a complex function $\psi$. Closedness of $\omega$ is
the flatness condition for the connection $d-\omega$ on the trivial complex
line bundle, so local solutions exist and patch consistently; moreover if
$\psi_1,\psi_2$ both solve it and $\psi_1$ is nowhere zero, then
\bea
d\big(\psi_2/\psi_1\big)
=\frac{\psi_1\,d\psi_2-\psi_2\,d\psi_1}{\psi_1^2}
=\frac{\psi_1\psi_2\omega-\psi_2\psi_1\omega}{\psi_1^2}=0 ,
\ee
so $\psi_2/\psi_1$ is a constant on $\CS_0$: the solution space is (complex)
one-dimensional wherever a nowhere-zero solution exists. No assumption on the
topology of $\CS_0$, such as simple connectedness, is needed once solutions
are exhibited explicitly on the whole of $\CS_0$, as \rf{transport} does for
$\psi_{\Vt}$ and $\psi_U=i\psi_{\Vt}$; their ratio is the constant $i$, and
\rf{coeffdet} confirms $\psi_{\Vt}$ is nowhere zero except where $\Vt$
itself is. This is the mechanism behind Theorem 7.5's construction of the
second Killing vector $U$: \rf{domega} checks $d\omega=0$, and
\rf{transport}--\rf{coeffdet} exhibit both solutions directly, so no holonomy
or connectedness argument is needed to make $U$ well defined on all of
$\CS_0$.

\medskip\noindent{\it Regularity of Killing fields.} A Killing field of a
smooth Riemannian metric which is a priori only $C^1$ is automatically smooth.
The Killing operator $U\mapsto{\cal L}_Ug$ has principal symbol
$\xi\otimes u\mapsto\xi_{(i}u_{j)}$, which is injective for $\xi\neq0$, so
${\cal L}_Ug=0$ is an overdetermined elliptic system with smooth coefficients
and the standard interior regularity theory applies. Concretely,
${\cal L}_Ug=0$ gives $\tn^iU_i=0$, and the weak form of the Killing equation
then reads $\tn^j\tn_jU_i=-R_i{}^jU_j$ in the sense of distributions; this is
a determined elliptic system with smooth coefficients, so $U\in C^1$
bootstraps to $U\in C^\infty$. See \cite{kobayashi} for the general statement.
It is not needed in the proof of Theorem 7.7, since Lemma 7.6 delivers a
smooth $U$ outright; it is recorded because it is what would be needed if
$U$ were only known to be $C^1$.

\medskip\noindent{\it A removable singularity criterion.} Let $\Omega$ be open
in a smooth manifold, $E\subset\Omega$ a closed discrete subset, and $f$ a
continuous tensor field on $\Omega$ which is $C^1$ on $\Omega\setminus E$ and
whose covariant derivative extends continuously from $\Omega\setminus E$ to a
continuous field $L$ on $\Omega$. Then $f$ is $C^1$ on $\Omega$, with
$\tn f=L$. Work in a chart around $p\in E$ small enough that $E$ meets it only
at $p$; for $x$ in that chart the open segment from $p$ to $x$ then avoids $E$
altogether, so for $0<\e<1$
\bea
f(x)-f\big(p+\e(x-p)\big)=\int_\e^1L\big(p+s(x-p)\big)[x-p]\,ds ,
\ee
and letting $\e\to0$, using continuity of $f$ at $p$ and boundedness of $L$
near $p$, gives $f(x)-f(p)=\int_0^1L(p+s(x-p))[x-p]\,ds$. Continuity of $L$ at
$p$ turns the right-hand side into $L(p)[x-p]+o(|x-p|)$, so $f$ is
differentiable at $p$ with derivative $L(p)$, and $\tn f=L$ is continuous by
hypothesis. The criterion applies on the explicit example, where $E=\{P=0\}$ is
the two-point set $\{x=4,5\}$, but its restriction to discrete $E$ makes it
unusable on a general horizon, where $\{P=0\}$ need not be discrete. It is not
needed here: by Lemma 7.6, $U$ is smooth wherever $Z\neq W$ with no condition
on $\{P=0\}$ whatever, so no statement in Section 7 depends on the size or
structure of that set.

\newsection{An alternative route to the second isometry, and where it stalls
(the two-dimensional quotient)}\la{app:quotient}

Section 6 mentions, and does not pursue, a third candidate method for
producing a second Killing vector on the branch $P_i\not\equiv0$, distinct
both from the invariant-gradient route of Theorem 6.6 and the transport
construction of Theorem 7.5. It is recorded because it was carried through far
enough to identify exactly where it stalls.

Since $\Vt_i=\parallel\eta_-\parallel^2(\Phi P_i-h_i)$ and $h\cdot P=\Phi
P\cdot P$, the Killing vector $\Vt$ is orthogonal to $P$, so on the open set
where $\Vt\neq0$ the quotient $B=\CS/U(1)$ is two-dimensional with horizontal
plane spanned by $P$ and $e:=P\times\Vt$, and $\Phi$ is a coordinate there
away from the zeros of $P$. In two dimensions $\star\,d\Phi$ is a Killing
vector of $B$ precisely when $\mathrm{TF}\,\mathrm{Hess}\,\Phi=0$, and such a
vector lifts to a second Killing vector of $\CS$. Writing
$\mathrm{Hess}_{ij}\Phi=\tfrac32\c^2[K'P_iP_j+K\tn_{(i}P_{j)}]$ from
\rf{gradPhi} and \rf{Kprime} and evaluating with \rf{covZ}, \rf{covW} and
\rf{hY}, the off-diagonal component vanishes identically,
\bea
\la{quotoff}
P^i e^j\,\mathrm{Hess}_{ij}\Phi=0 ,
\ee
so the quotient metric is of standard axisymmetric form in those coordinates.
What survives is one equation, in which $K'$ appears linearly with
non-vanishing coefficient; solving for $K'$ gives a value depending on $h_N$,
$\a$ and $Z\!\cdot\!W$, so the second isometry on this branch is equivalent
to a single scalar relation among $K'$, $\Phi$, $K$, $h_N$, $\a$ and
$Z\!\cdot\!W$ that the field equations do not supply on their own. This is
not a contradiction with Theorem 7.5: it shows only that the particular
scalar-invariant this method singles out, $\star\,d\Phi$ on the quotient, does
not certify the second isometry by itself, exactly as the general bound
\rf{rankbound} of Section 6 predicts for any argument built from scalar
gradients, all of which are parallel to $P_i$.
The transport construction of Theorem 7.5 succeeds precisely because it
propagates the Killing equation itself along $P$, rather than testing a single
derived scalar condition.

\newsection{Computational status}\la{app:status}

The symbolic computations behind this paper fall into three classes, and the
text is written so that no conclusion rests on a weaker class than it claims.

\medskip\noindent{\it Exact symbolic, used as proof.} The algebraic identities
of Sections 3--6 --- the very special geometry relations \rf{Cids}--\rf{Kdef},
the first-order system of Section 5, the closure $dP=0$ of Proposition 6.4,
the parallel-gradient collapse, the closed gradient system of Lemma 6.11,
the Killing-tensor obstruction of Proposition 6.9, the Gram bound \rf{gramS}
and the superharmonicity inequality \rf{Usuper} of Section~\ref{sec:positivity}, and the two
first integrals
of Lemma 7.9 with the endpoint relation \rf{Gends} on which Theorem 7.14 turns
--- were derived and checked
in exact rational and polynomial arithmetic in Python/SymPy, with no floating
point at any stage. A subset of the foundational identities of Section 6 was
re-derived independently in Cadabra2, a different tensor computer algebra
system with different index conventions and a different simplification engine,
as a guard against an error confined to one implementation. These computations
are part of the proofs.

\medskip\noindent{\it Exact symbolic on a specific model, used as an existence
witness.} The $U(1)^3$ statements of Section 7 --- the explicit compact
varying-moduli horizon and its factorisation \rf{Pispecial}, the non-vanishing
$6\times6$ minors behind Proposition 6.9, and the branch non-emptiness checks
--- are exact statements about particular horizons. They are used only to show
that a branch is non-empty or that a hypothesis is satisfiable, never to
establish a statement about all horizons of a branch.

\medskip\noindent{\it Numerical, and load-bearing nowhere.} No statement in
this paper is established by a floating-point computation. The three places
where one might be expected are settled exactly instead. The continuation of $U$
across $\{P=0\}$ is given by \rf{klrUexactapp}, which evaluates $U$ in closed
form and finds it equal to a constant element of the Lie algebra of the torus
action, identically in $x$; the $60$-digit numerical evaluation of
Appendix~\ref{app:klr} serves only as an independent cross-check of that closed
form, computed without ever forming
it. The Ricci eigenvalues are obtained from \rf{riccipos}, which
diagonalises $\tilde R_{ij}$ exactly on the constant-modulus branch. And
\rf{Kprime}--\rf{Kprimeprime} are established by the derivation given after
Proposition 7.1, which
uses no adjoint identity, so that Proposition 7.1 holds for an arbitrary cubic
prepotential rather than only for the model of Section 2. The remaining
appearances of decimals in the text are decimal expansions of algebraic
numbers that are also given exactly, quoted for orientation only.

\medskip\noindent What is established, and what is not, is therefore the
following. Throughout, $\CS$ is compact, connected and without boundary, and
$\Vt\not\equiv0$; the superpotential is nowhere zero, $\Phi=\c V_IX^I\neq0$,
which by Proposition 7.E is the only form in which the standing hypothesis
$U\geq0$ of \cite{kayani,thesis} is used, the third clause of Theorem 7.17
excepted. Established with no further hypothesis: the dichotomy $K\equiv0$ or $K>0$
(Proposition 7.1); that $P$ vanishes at every critical point of
$\parallel\eta_-\parallel^2$ (Corollary 7.2); the split of the non-minimal
case into $P\equiv0$ and $P\not\equiv0$, both non-empty (Corollary 7.3 and
Section~\ref{sec:varmod}); a second rotational isometry on all of $\CS$ when $P\equiv0$
(Proposition 7.4); a second rotational Killing field on $\CS_0$ when
$P\not\equiv0$ (Theorem 7.5); the $\a$-dichotomy, $\a\equiv0$ or $\a$ nowhere
zero (Corollary 6.13); the smoothness of $U$ across $\{P=0\}$, with no
condition on that set (Lemma 7.6), and across $\{Z=W\}$ as well, through a
formula polynomial in the horizon data (Lemma 7.6$'$); a second rotational
isometry on all of $\CS$ when $P\not\equiv0$, with no condition on $\a$,
together with an effective $T^2$ action (Theorem 7.7); the exclusion of a
constant-coefficient Killing tensor (Proposition 6.9); the two global first
integrals
$\a\parallel\eta_-\parallel^4$ and
$\big(|h-\Phi P|^2+\a^2\big)\parallel\eta_-\parallel^2$ (Lemma 7.9); the
trichotomy $S^3$, lens space, $S^1\times S^2$ for the cross-section on the
branch $P\not\equiv0$, $\a\not\equiv0$, with an explicit criterion separating
the last case (Theorem 7.12); on that same branch, the exclusion of the last
case outright (Theorem 7.14); the superharmonicity of the scalar potential wherever it
is positive (Proposition 7.D), together with the fact that $U\geq0$ is not
forced by the horizon equations and must be assumed; that both angular momenta
are determined by the first integrals of Lemma 7.9, and vanish or reduce to the
area on the sub-branch $\a\equiv0$ (Proposition 7.F); and the absence of any
hidden symmetry, so that every symmetry of these horizons, hidden or not, is
an isometry (Proposition 7.G). The varying-moduli supersymmetric $AdS_5$
black ring therefore does not exist on the branch $P\not\equiv0$,
$\a\not\equiv0$: there the second rotational Killing field of Theorem 7.5
does extend to an isometry of the whole horizon, by Theorem 7.7, whose only
hypotheses are $K>0$ and $P_i\not\equiv0$, and the
cross-section is then $S^3$ or a lens space, never $S^1\times S^2$.
Not established: the topology of the compact
horizons with $P\not\equiv0$ and $\a\equiv0$, which Theorem 7.7 does reach ---
so they are toric --- but which Section~\ref{sec:toric} does not, the first integral
$\a\parallel\eta_-\parallel^4$ of Lemma 7.9 vanishing identically there; and the
existence or non-existence of a compact supersymmetric $S^1\times S^2$ horizon
on the {\it constant}-moduli branch $P\equiv0$, in the window
$\tfrac13\Phi^2<\c^2K\leq\tfrac43\Phi^2$, which is the only window on that
branch in which such a horizon could occur.

\section*{Data availability}

The symbolic computations underlying the results of this paper were carried
out in Python/SymPy, and, for a subset of the foundational identities of
Section 6, cross-checked independently in Cadabra2. Every script is
self-contained and re-runnable, and every log records an explicit PASS or FAIL
for each individual check; the scripts, the logs they produce, and a README
mapping each script to the numbered statement it verifies are available from
the author on request. Which statements rest on exact
symbolic identities and which on evaluation at sample points is set out in
Appendix~\ref{app:status}; no statement used as a proof rests on the latter.

\end{document}